\documentclass[twocolumn,aps,prd,10pt,superscriptaddress,longbibliography]{revtex4-1}

\usepackage{glossaries}
\usepackage{amssymb,amsmath}
\usepackage{bm}
\usepackage{dcolumn}
\usepackage{graphicx}
\usepackage[version=4]{mhchem}
\usepackage[dvipsnames,svgnames]{xcolor}
\usepackage{multirow}
\usepackage{booktabs}
\usepackage{siunitx}
\usepackage{makecell}

\usepackage{float}
\usepackage{hyperref}
\hypersetup{
pdfnewwindow=true,      
colorlinks=true,        
linkcolor=Blue,         
citecolor=Blue,         
filecolor=Blue,         
urlcolor=Blue           
}

\newacronym{vdw}{vdW}{van der Waals}
\newacronym{2d}{2D}{two-dimensional}
\newacronym{3d}{3D}{three-dimensional}
\newacronym{hbn}{$h$-BN}{hexagonal boron nitride}
\newacronym{gr}{Gr}{graphene}
\newacronym{mos2}{\ce{MoS2}}{molybdenum disulfide}
\newacronym{dft}{DFT}{density functional theory}
\newacronym{lj}{LJ}{Lennard-Jones}
\newacronym{rebo}{REBO}{reactive empirical bond order}
\newacronym{sw}{SW}{Stillinger-Weber}
\newacronym{reaxff}{ReaxFF}{reactive force field}
\newacronym{airebo}{AIREBO}{adaptive intermolecular reactive bond order potential}
\newacronym{kc}{KC}{Kolmogorov-Crespi}
\newacronym{hnemd}{HNEMD}{homogeneous nonequilibrium molecular dynamics} 
\newacronym{pes}{PES}{potential energy surface}
\newacronym{ilp}{ILP}{interlayer potential}
\newacronym{mlp}{MLP}{machine-learned potential}
\newacronym{gnnp}{GNNP}{graph neural network potential}
\newacronym{smlp}{\textit{s}MLP}{single-layer MLP}
\newacronym{nep}{NEP}{neuroevolution potential}
\newacronym{md}{MD}{molecular dynamics}
\newacronym{nn}{NN}{neural network}
\newacronym{tmd}{TMD}{transition metal dichalcogenide}
\newacronym{rmse}{RMSE}{root mean square error}
\newacronym{sm}{SM}{Supplemental Material}
\newacronym{tbg}{TBG}{twisted bilayer graphene}
\newacronym{snes}{SNES}{separable natural evolution strategy}
\newacronym{lri}{LRI}{local registry index}
\newacronym{mbd_nl}{MBD-NL}{nonlocal many-body dispersive}
\newacronym{pbe}{PBE}{Perdew-Burke-Ernzerhof}
\newacronym{eos}{EOS}{equations of state}
\newacronym{dp}{DP}{deep potential}
\newacronym{nequip}{NequIP}{neural equivariant interatomic potential}
\newacronym{gap}{GAP}{gaussian approximation potential}
\newacronym{cg}{CG}{conjugate gradient} 
\newacronym{aimd}{AIMD}{\emph{ab initio} molecular dynamics} 
\newacronym{bnnr}{BNNR}{\textit{h}-BN nanoribbon}
\newacronym{hbnnr}{H-BNNR}{hydrogen passivated \textit{h}-BN nanoribbon}

\DeclareSIUnit\angstrom{\text{Å}}
\DeclareSIUnit{\atom}{atom}
\DeclareSIUnit{\step}{step}
\DeclareSIUnit{\atomstepsecond}{\atom\step\per\second}
\DeclareSIUnit{\evperangstrom}{\electronvolt\per\angstrom}

\begin{document}

\title{Modular Hybrid Machine Learning and Physics-based Potentials for Scalable Modeling of van der Waals Heterostructures}
\author{Hekai Bu}
\thanks{These authors contributed equally to this work.}
\affiliation{Department of Engineering Mechanics, School of Civil Engineering, Wuhan University, Wuhan, Hubei, 430072, China}

\author{Wenwu Jiang}
\thanks{These authors contributed equally to this work.}
\affiliation{Department of Engineering Mechanics, School of Civil Engineering, Wuhan University, Wuhan, Hubei, 430072, China}

\author{Penghua Ying}
\email{penghua@xjtu.edu.cn}
\affiliation{Department of Physical Chemistry, School of Chemistry, Tel Aviv University, Tel Aviv, 6997801, Israel}
\affiliation{Laboratory for multiscale mechanics and medical science, SV LAB, School of Aerospace, Xi’an Jiaotong University, Xi’an, Shaanxi, 710049, China}

 \author{Ting Liang}
\affiliation{Department of Electronic Engineering and Materials Science and Technology Research Center, The Chinese University of Hong Kong, Shatin, N.T., Hong Kong SAR, 999077, China}

\author{Zheyong Fan}
\affiliation{College of Physical Science and Technology, Bohai University, Jinzhou, 121013, China}

\author{Wengen Ouyang}
\email{w.g.ouyang@whu.edu.cn}
\affiliation{Department of Engineering Mechanics, School of Civil Engineering, Wuhan University, Wuhan, Hubei, 430072, China}
\affiliation{State Key Laboratory of Water Resources Engineering and Management, Wuhan University, Wuhan, Hubei, 430072, China}

\date{\today}

\begin{abstract}
Accurately modeling the structural reconstruction and thermodynamic behavior of van der Waals (vdW) heterostructures remains a significant challenge due to the limitations of conventional force fields in capturing their complex mechanical, thermal, electronic, and tribological properties. To address these limitations, we develop a hybrid framework that combines single-layer machine-learned potential (\textit{s}MLP) with physics-based anisotropic interlayer potential (ILP), effectively decoupling intralayer and interlayer interactions. This \textit{s}MLP+ILP approach modularizes the modeling of vdW heterostructures like assembling LEGOs, reducing the required training configurations by at least an order of magnitude compared to the pure MLP approach, while retaining predictive accuracy and computational efficiency. We validate our framework by accurately reproducing the mechanical and thermal transport properties of graphite and bulk \gls{hbn}, and by resolving intricate Moir{\'e} patterns in graphene/$h$-BN bilayer and graphene/graphene/$h$-BN trilayer heterostructures, achieving excellent agreement with experimental observations. Leveraging the developed \textit{s}MLP+ILP approach, we reveal the stacking order-dependent formation of Moir{\'e} superlattice in trilayer graphene/$h$-BN/MoS$_2$ heterostructures, demonstrating its ability to accurately model large-scale vdW systems comprising hundreds of thousands of atoms with near \emph{ab initio} precision. Moreover, applications to nanotribology of $h$-BN nanoribbons demonstrate the framework's unique ability to capture the critical influence of edge chemistry on sliding dynamics, specifically revealing that hydrogen passivation stabilizes zigzag edges against out-of-plane buckling, thereby inducing a pronounced enhancement of stick--slip friction. These findings demonstrate that the hybrid \textit{s}MLP+ILP framework remarkably outperforms existing pure machine-learned or empirical potentials, offering a scalable and transferable solution for accurately and extensively modeling complex vdW materials across diverse applications, including sliding ferroelectricity, thermal management, resistive switching, and superlubric nanodevices.

\end{abstract}

%
%
\maketitle

\section{Introduction\label{intro}}
\Gls{vdw} homojunctions or heterojunctions, formed by vertically stacking \gls{2d} layered materials, exhibit unique mechanical \citep{zelisko2017determining, fatemeh2017thermal, ren2025interlayer, hou2025probing, liu2025homogenization, zhang2018structural, choi2024elasticb, sung2022torsional, park2025unconventional}, thermal transport \citep{kim2021extremelya, jiang2025moiredriven, ouyang2020controllable, jiang2023twistdependent, fredrik2023tuning, liang2025probing, zhang2023effect}, electronic \citep{koren2016coherent, wang2023deducing, zhang2018structural}, and tribological \citep{hod2018structural, ying2025scaling, wang2023kinetic, xue2022peeling, lyu2024graphene, zhang2025homochiral, liu2023situ, yao2025unraveling, yan2024moire, yan2024shapedependent} properties. Accurate modeling of their structural reconstruction and thermodynamic behaviors at the atomic scale is crucial for understanding these properties. However, theoretical investigation at this scale presents a non-trivial challenge, as heterostructure superlattices typically contain hundreds of thousands of atoms due to intrinsic lattice mismatches between monolayers (e.g., $\sim$ 1.8\% between \gls{hbn} and \gls{gr}), making them computationally intractable for first-principles approaches such as \gls{dft}.

Due to the anisotropic nature of \gls{vdw} layered materials, atomic interactions in layered materials can be divided into intralayer interactions, which involve strong covalent bonding, and weak \gls{vdw} interlayer interactions, which include short-range electronic repulsion, long-range attractive dispersion and electrostatic interactions. Building on these physical insights, various empirical potentials, such as the \gls{rebo} potential \citep{brenner2002second}, the Tersoff potential \citep{tersoff1988empirical}, and the \gls{sw} potential \citep{stillinger1985computer} combined with \gls{lj} terms \citep{stuart2000reactive}, have been developed over the past two decades \citep{los2003intrinsic, los2005improved} to model both homogeneous and heterogeneous layered materials. However, the two-body \gls{lj} terms predict an overly shallow sliding \gls{pes} \citep{reguzzoni2012potential, lebedeva2011interlayer}, as they fail to capture the anisotropic nature of \gls{vdw} layered materials. To address this limitation, registry-dependent \gls{ilp} approaches \citep{kolmogorov2005registrydependent} have been proposed, incorporating not only the interatomic distance (as in \gls{lj}) but also interatomic relative lateral displacement to accurately describe the overlap of electronic clouds, i.e., the origin of short-range interlayer repulsion. To better capture the long-range interactions, further improvements \citep{leven2014interlayer, ouyang2018nanoserpents} incorporate the long-range dispersion term from the Tkatchenko-Scheffler correction scheme \citep{tkatchenko2009accurate} along with the electrostatic interaction term. To date, \gls{ilp} approaches \citep{gao2025anisotropici} have been widely used to model the structural reconstruction \citep{vizner2021interfacial, mandelli2019princess}, mechanical properties,\citep{ouyang2020mechanical, ouyang2021registrydependent, wu2024facetgoverned} nanotribological properties \citep{ouyang2018nanoserpents, yan2023origin, yan2024shapedependent, yao2025unraveling, huang2022origin}, thermal transport \citep{ouyang2020controllable, jiang2025moiredriven, jiang2023twistdependent}, growth behaviors \citep{lyu2024graphene, zhang2025homochiral, lyu2022catalytic}, and wetting properties \citep{feng2023registrydependent, feng2023anisotropic} of layered materials and other \gls{vdw} heterostructures \citep{ouyang2021registrydependent, yao2024semianisotropic, liang2023anisotropic}, aligning well with experimental observations.

Although \gls{ilp} approaches have significantly advanced the modeling of \gls{vdw} heterostructures, they are typically combined with empirical potentials to describe intralayer covalent bonding, leading to the following two main limitations: (i) the existing empirical potential models are primarily parameterized for pristine periodic crystal structures, making them less transferable to the systems with structural defects such as edges, vacancies, and grain boundaries \citep{qian2021comprehensive}, or to new \gls{2d} allotropes \citep{ying2023atomistic}; (ii) the accuracy of empirical potentials often falls short of the level of first-principles calculations. To overcome these limitations, one may resort to the emerging \gls{mlp} approaches \citep{behler2007generalized, friederich2021machinelearned}. Instead of distinguishing the intralayer and interlayer interactions based on physical insights, \glspl{mlp} learn directly from \emph{ab initio} data, mapping the atomic positions into corresponding site energies and their gradients (atomic forces) using invariant \citep{bartok2010gaussian, fan2021neuroevolution} or equivariant \citep{batatia2022mace, batzner2022e3equivariant} feature representations for the atomic environments. Dedicated \glspl{mlp} for \gls{2d} monolayers and bilayers can achieve high accuracy in describing short-range interactions \citep{rowe2018development, ying2024effect, liang2025probing, gao2025spontaneous}, which effectively captures the sliding \gls{pes} or structural reconstruction. 

However, most \glspl{mlp} impose an effective cutoff of several angstroms to reduce computational costs, making them insufficient for capturing London dispersion interactions that extend over several nanometers \citep{mandelli2019princess, ying2024combining}. Several efforts have been proposed to address this limitation, including incorporating \gls{lj}-like corrections \citep{wen2019hybrid} or DFT-D3 dispersion \citep{ying2024combining} corrections. Additionally, the \gls{gnnp}, based on message-passing architectures, offers a promising solution by allowing interlayer interactions to be propagated along the atomic graph through multiple convolutional layers. However, current \gls{gnnp}s are highly computationally demanding and remain impractical for modeling large heterostructures with tens of thousands of atoms \citep{ying2024effect}. Another limitation of \gls{mlp} approaches is their reliance on sufficiently large reference datasets to accurately represent local atomic environments. Capturing interlayer interactions using these \gls{mlp} approaches becomes particularly challenging for complex heterostructures containing three or more distinct layers. A comprehensive \gls{mlp} for such systems must account for various stacking orders, surface and interface conditions, and bulk properties—along with different interlayer distances and sliding PES across various temperatures and pressures for each configuration. This complexity may explain why existing \glspl{mlp} for \gls{vdw} heterostructures are limited to bilayer systems, with no models yet developed for heterostructures containing three or more distinct component layers.

\begin{figure*}[htb]
\begin{center}
\includegraphics[width=2\columnwidth]{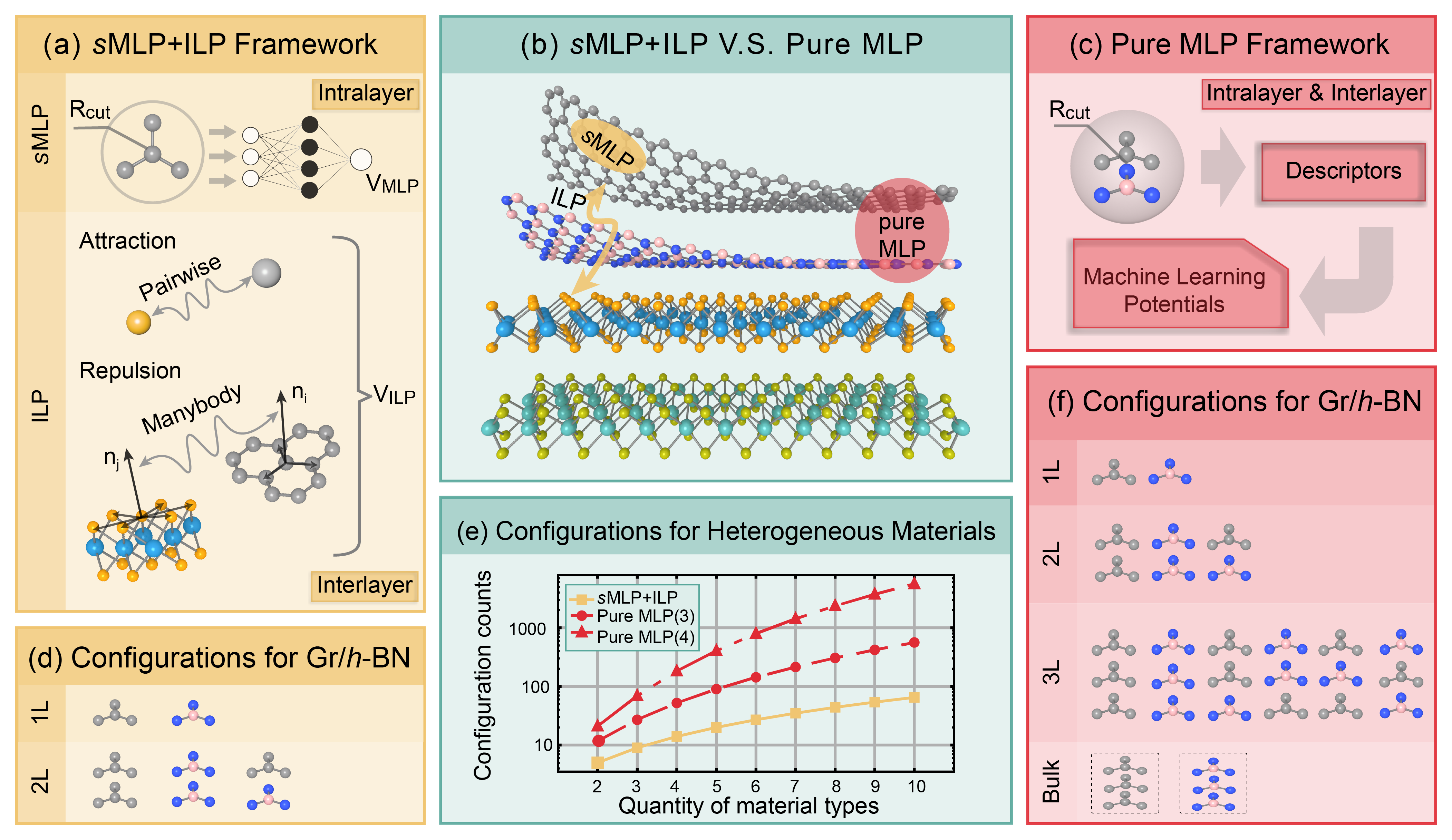}
\caption{Schematic architecture of the \gls{smlp}+\gls{ilp} and pure \gls{mlp} frameworks. (a)-(c) Schematic illustration of the architecture of (a) \gls{smlp}+\gls{ilp} and (c) pure \gls{mlp} as well as (b) their comparison. In the \gls{smlp}+\gls{ilp} framework, the \gls{mlp} is designed to capture the \gls{2d} local chemical environment within a monolayer, while the \gls{ilp} accounts for long-range pairwise and many-body interactions across different layers. In contrast, the pure \gls{mlp} framework learns descriptors from a more complex \gls{3d} local chemical environment across layers, and thus must consider more atoms within the same layer and complex combinations of \gls{vdw} materials. (d)--(f) Comparison of the ideal training configuration numbers for both frameworks. (d) and (f) show the number of training configurations for the \gls{gr}/\gls{hbn} heterostructure in the \gls{smlp}+\gls{ilp} framework and the pure \gls{mlp}(3) framework, respectively. (e) shows the number of required configurations as a function of the number of distinct monolayer components. For the pure \gls{mlp} approach, $N_{\mathrm{MLP(3)}}$ and $N_{\mathrm{MLP(4)}}$ represent the number of required configurations based on the assumption that up to three layers (see \autoref{equation:mlp3}) and four layers (see \autoref{equation:mlp4}) are necessary to accurately describe the surface and interface environment.}
\label{fig:art}
\end{center}
\end{figure*}

In the present study, we propose a LEGO-like hybrid framework that combines \gls{smlp} and physics-based \gls{ilp} using different cutoffs, effectively decoupling short-range intralayer and long-range interlayer interactions (see \autoref{fig:art}(a)). The advantages are twofold: \gls{smlp} with more parameters reduces the required reference dataset by focusing on single layers while improving accuracy and applicability compared to empirical potentials, while \gls{ilp} with fewer parameters ensures efficiency despite using a large cutoff. This approach enables the construction of potential models for \gls{vdw} heterostructures like assembling LEGOs, eliminating the need for extensive reference datasets to capture complex chemical environments, including stacking orders, surface and interface boundary conditions, interlayer spacings, and corresponding sliding \gls{pes}. Here, we chose the highly efficient \gls{nep} framework \citep{fan2022gpumd, song2024generalpurpose} to train the \gls{smlp} models for monolayer \gls{2d} materials, such as \gls{gr}, \gls{hbn} and \gls{mos2}. Note that this workflow is readily adaptable to other \gls{mlp} frameworks for broader applications. By training \gls{smlp} and \gls{ilp} models separately on a small dataset of state-of-the-art \gls{dft} calculations, we demonstrate that this hybrid framework achieves \emph{ab initio} accuracy while maintaining computational efficiency comparable to existing empirical potentials. Furthermore, we validate its transferability by applying it to bilayer, trilayer and bulk complex \gls{vdw} heterostructures, investigating their mechanical, thermal transport properties and structural reconstruction accompanied by complex Moir\'e superlattices. Crucially, applying the framework to the sliding dynamics of \gls{hbn} nanoribbons reveals a distinctive edge-controlled mechanism: hydrogen passivation stabilizes zigzag edges against out-of-plane buckling, thereby promoting pronounced stick--slip friction and significantly enhanced lateral forces in aligned contacts. We implement our \gls{smlp}+\gls{ilp} framework into open-source \gls{md} packages, including \textsc{LAMMPS} \citep{thompson2022lammps} and \textsc{GPUMD} \citep{fan2017efficient, xu2025gpumd}. Notably, an exceptional computational speed exceeding \SI{2e6}{\atom\cdot\step\per\second} is achieved in \textsc{GPUMD} using a single NVIDIA RTX 4090 consumer desktop GPU for \gls{gr}/\gls{mos2}/\gls{hbn} heterostructures comprising 423,552 atoms, enabling sub-micron scale simulations of complex \gls{2d} \gls{vdw} heterostructures at near \emph{ab initio} accuracy for the first time.

\section{\textit{s}MLP+ILP model for vdW structures\label{result}}

\subsection{Theoretical framework}
In our hybrid \gls{smlp}+\gls{ilp} approach, the total potential energy $V_{\mathrm{tot}}^{s\mathrm{MLP+ILP}}$ for a configuration with $N$ atoms and $M$ layers is computed as:
\begin{equation}
V_{\mathrm{tot}}^{s\mathrm{MLP+ILP}} = \sum_{k=1}^{M} \sum_{l=1}^{M}\left[
 V^{s\mathrm{MLP}}_{kl}\delta_{kl} + V^{\mathrm{ILP}}_{kl}(1-\delta_{kl})\right],
\end{equation}
where $V^{s\mathrm{MLP}}_{kl}$ and $V^{\mathrm{ILP}}_{kl}$ represent the potential energy of the \gls{smlp} model and the \gls{ilp} model of the $k^{\mathrm{th}}$ layer with neighbors in the $l^{\mathrm{th}}$ layer, respectively. $\delta_{kl}$ is the Kronecker delta function. 

For these two site potential terms, we construct two independent neighbor lists for atom $i$: $\Omega^{s\mathrm{MLP}}_i$ and $\Omega^{\mathrm{ILP}}_i$. The intralayer neighbors $j$ ($j\in \Omega^{s\mathrm{MLP}}_i$) are searched within the same layer as atom $i$ using a cutoff of $R_{\mathrm{cut}}^{\mathrm{sMLP}}$, while the interlayer neighbors $k$ ($k\in\Omega^{\mathrm{ILP}}_i$) are identified in different layers from atom $i$ using a cutoff of $R_{\mathrm{cut}}^{\mathrm{ILP}}$. In practical implementation, atoms with different molecule IDs in \textsc{LAMMPS} or group labels in \textsc{GPUMD} are used to distinguish atoms belonging to different layers \citep{ouyang2018nanoserpents}.

The \gls{smlp} term focuses on the intralayer interactions, where $k$ equals $l$, for covalent bonds, which can be expressed using a \gls{nn}: 
\begin{equation}
V^{s\mathrm{MLP}}_{kl}\delta_{kl} = V_{kk}^{s\mathrm{MLP}}=\sum_{i\in \Omega^{k}}f_{\mathrm{NN}}\left(q_{1}^{i}, q_{2}^{i}, \dots, q_{N_{\mathrm{des}}}^{i}\right),
\end{equation}
where $\Omega^k$ is the set of atoms in the $k^{\mathrm{th}}$ layer, $f_{\mathrm{NN}}(\cdot)$ represents the \gls{nn} operations and $\{q_{\nu}^{i}\}$ is a set of descriptors that encode local environment messages information, including atomic positions and elements from the neighbors of atom $i$. This transformation can be expressed as:
\begin{equation}
q_{\nu}^{i}=\sum_{j\in \Omega_i^{s\mathrm{MLP}}} g_{\nu}(\boldsymbol{r}_{ij})
\end{equation}
where $\{g_\nu\}$ is a set of descriptor functions, and $\boldsymbol{r}_{ij}$ is the position vector from atom $i$ to atom $j$. Since the intralayer interactions typically require a small cutoff, many \glspl{mlp} can accurately describe them. In this work, we adopt the \gls{nep} framework, which provides an efficient and accurate description of covalent bond interactions in monolayer \gls{2d} materials within our \gls{smlp}+\gls{ilp} approach. Specifically, following the pioneering Behler and Parrinello approach \citep{behler2007generalized}, the \gls{nep} uses a single hidden \gls{nn} layer to compute the site energy $V_{kk}^{s\mathrm{MLP}}$ \citep{fan2021neuroevolution}:
\begin{equation}
\begin{split}
V^{s\mathrm{MLP}}_{kk}= 
\sum_{i\in \Omega ^k}\left[\sum_{\mu=1}^{N_{\mathrm{neu}}} w^{[1]}_{\mu}\tanh{\left(\sum_{\nu=1}^{N_{\mathrm{des}}}w^{[0]}_{\mu\nu}q_{\nu}^{i}-b_{\mu}^{[0]}\right)}-b^{[1]}\right],
\label{equation:site_energy}
\end{split}
\end{equation}
where $w^{[0]}_{\mu\nu}$, $w^{[1]}_{\mu}$, $b^{[0]}_{\mu}$, $b^{[1]}$ are trainable weights and biases of the \gls{nn}, $N_{\mathrm{neu}}$ is the number of neurons in the hidden layer, $N_{\mathrm{des}}$ is the number of descriptors and $\tanh{(\cdot)}$ is the activation function. The \gls{nep} descriptors consist of radial and angular components. The two-body radial descriptors $q_n^i (0 \leq n \leq n_{\mathrm{max}}^{\mathrm{R}})$ are constructed as
\begin{equation}
q_n^i=\sum_{j\in\Omega_i^{s\mathrm{MLP}}}g_n(r_{ij}),
\label{equation:raidial}
\end{equation}
where $r_{ij}$ is the distance between atom $i$ and atom $j$, $g_n$ is the $n$th expansion radial descriptor function. The many-body angular descriptors $q_{nl}^i$ can be expressed as
\begin{equation}
q_{nl}^i={\sum\sum}_{j,k\in\Omega_i^{s\mathrm{MLP}}}g_n(r_{ij})g_n(r_{ik})P_l(\theta_{ijk}),
\label{equation:angular}
\end{equation}
where $P_l(\cdot)$ is the Legendre polynomial of order $l$ and $\theta_{ijk}$ is the angle for the triplet $(ijk)$ with atom $i$ in the center. 

The interlayer interactions are represented by \gls{ilp} term, which includes a long-range dispersion attraction term and Pauli repulsion term \citep{leven2016interlayer}:
\begin{equation}
\begin{split}
&V_{kl}^{\mathrm{ILP}}(1-\delta_{kl})= \\&\sum_{i\in\Omega^k}\sum_{j\in\Omega^l\cap\Omega_i^{\mathrm{ILP}}}^{l\ne k}\mathrm{Tap}(r_{ij}) \left[V^{\mathrm{att}}(r_{ij})+V^{\mathrm{rep}}(r_{ij}, \boldsymbol{n}_i, \boldsymbol{n}_j)\right],
\end{split}
\label{equation:ilp term}
\end{equation}
where $\mathrm{Tap}(\cdot)$ is a long-range taper function of distance, $\mathrm{Tap}(r_{ij})=20{\left({r_{ij}}/{R_{\mathrm{cut}}}\right)}^7-70{\left({r_{ij}}/{R_{\mathrm{cut}}}\right)}^6+84{\left({r_{ij}}/{R_{\mathrm{cut}}}\right)}^5-35{\left({r_{ij}}/{R_{\mathrm{cut}}}\right)}^4+1$. 

The attraction term adopts a dispersion correction similar to Tkatchenko and Scheffler's approach \citep{tkatchenko2009accurate} for augmenting standard DFT exchange-correlation functionals:

\begin{equation}
V^{\mathrm{att}}(r_{ij})=
-\frac{1}{1+\exp \left( {-d_{ij}\left[r_{ij}/(S_{\mathrm{R},ij}\cdot r_{ij}^{\mathrm{eff}})-1\right] }\right)}\frac{C_{6,ij}}{r_{ij}^{6}}.
\end{equation}
The repulsion term follows Kolmogorov and Crespi's approach \citep{kolmogorov2005registrydependent, leven2014interlayer}, using a Morse-like exponential isotropic term with an anisotropic correction:
\begin{equation}
\begin{split}
&V^{\mathrm{rep}}(r_{ij}, \boldsymbol{n}_i, \boldsymbol{n}_j)=\\
&\exp\left(\alpha_{ij}\left(1-{\gamma_{ij}}/{\beta_{ij}}\right)\right) \\
&\left\{\epsilon_{ij}+C_{ij}\left[\exp\left({-{\rho_{ij}}^2/{\gamma_{ij}}^2}\right)+\exp\left(-{\rho_{ji}}^2/{\gamma_{ij}}^2\right)\right]\right\}.
\end{split}
\end{equation}
Here $\rho_{ij}$ and $\rho_{ji}$ are the lateral interatomic distance, which can be expressed as:
\begin{equation}
\left\{
\begin{aligned}
&\rho_{ij}^{2}= r_{ij}^2-{(\boldsymbol r_{ij} \cdot \boldsymbol n_i)}^2\\
&\rho_{ji}^{2}= r_{ij}^2-{(\boldsymbol r_{ij} \cdot \boldsymbol n_j)}^2
\end{aligned}
\right.
\end{equation}
where $\boldsymbol n_{i}$ is the local normal vector of atoms $i$, defined as the normal to the plane formed by its nearest directly bonded neighbors, with the remaining parameters being fitting parameters. 

We note that the definition of the normal vector may vary depending on the specific \gls{vdw} interface. \autoref{fig:art}(a) illustrates the normal vector ($n_i$) definitions for \gls{gr} and \gls{mos2}. For \gls{gr} and \gls{hbn}, which have single-atom thickness, the normal vector is defined based on up to three nearest in-plane neighbors \citep{leven2014interlayer, kolmogorov2005registrydependent}. For \gls{mos2}, which has a sandwich-like layered structure, it is determined using up to six nearest neighbors within each molybdenum or sulfur sublayer \citep{ouyang2021anisotropic}. In \textsc{GPUMD}, we adopt the algorithm from the optimized version of \gls{ilp} in \textsc{LAMMPS} to couple the neighbor lists of \gls{ilp} and \gls{nep} \citep{gao2021lmff}. As \gls{ilp} only considers the nearest neighbor atoms when computing normal vectors, we utilize the radial neighbor list of \gls{nep} to reduce computational cost.

In the present \gls{ilp} formulation (\autoref{equation:ilp term}), we omit explicit Coulomb terms. This simplification is justified because interlayer Coulomb interactions are typically much weaker than \gls{vdw} forces in neutral, weakly polar systems, as further quantified in \gls{sm} Section S5. Consequently, for consistency, we refitted the original \gls{hbn} \gls{ilp} model (which included a Coulomb term) without it; updated parameters are provided in \gls{sm} Section S5.

We explicitly note, however, that Coulomb contributions can become significant in non-neutral or strongly polar systems, such as doped, defective, or electrostatically gated heterostructures. For such cases, the framework remains extensible: a screened Coulomb correction can be added, as demonstrated in earlier \gls{ilp}-based studies \citep{jiang2023anisotropic}. The specific functional form of the Coulomb correction should be tailored to the system under study.

To evaluate the core assumption of the \gls{smlp}+\gls{ilp} framework, the linear separability of intralayer and interlayer energies, we quantified the magnitude of cross-layer coupling in representative layered materials such as graphite and \gls{hbn} (see \gls{sm} Section S6 for details). Our analysis demonstrates that when the interlayer spacing exceeds approximately \SI{3}{\angstrom}, corresponding to normal pressures below about \SI{5}{\giga\pascal}, the cross-layer term contributes less than 5\% of the total interlayer binding energy. In this regime, its influence on key physical properties, including thermal and mechanical responses, is negligible. 

Cross-layer coupling becomes significant only under high compressive loading or substantial structural distortion, consistent with earlier reports \citep{ni2021stronger, martins2017raman}. At even higher pressures (>\textasciitilde\SI{10}{\giga\pascal}), interlayer covalent bonding can emerge, as exemplified by the graphite-to-diamond transition \citep{khaliullin2011nucleation}. Such conditions involve drastic bond rearrangement and strong interlayer coupling. For these regimes, the \gls{smlp}+\gls{ilp} approach, like any method based on a strict separation of interactions, is not suitable for quantitative prediction.

However, typical experimental studies of \gls{2d} layered materials operate at normal pressures well below these thresholds \citep{segura2021tuning, proctor2009highpressure}. Consequently, the combined \gls{smlp}+\gls{ilp} framework remains valid and quantitatively reliable for the vast majority of experimentally relevant conditions.

\subsection{Comparison with existing approaches}

In this section, we discuss various existing potential approaches for \gls{vdw} materials and highlight the necessity of \gls{smlp}+\gls{ilp} approach, particularly for accurately modeling complex multilayer heterostructures. Numerous \glspl{mlp} have now been developed to simulate the \gls{vdw} materials to achieve (near) \emph{ab initio} accuracy, aiming to capture both interlayer and intralayer interactions within a single \gls{nn} framework (see \autoref{fig:art}(c)). While most \glspl{mlp} can theoretically simulate \gls{vdw} materials in this manner, two significant challenges arise, as discussed below. 

The first challenge is how to determine an optimal cutoff for \glspl{mlp} that capture long-range \gls{vdw} interactions while maintaining comparable efficiency, as these \gls{vdw} interactions can extend through several interlayer spacings and have significant impacts on structural reconstructions \citep{mandelli2019princess, zhang2024impact, espanol2023adiscrete} and corresponding physical properties \citep{wang2023deducing}. For example, the interlayer distance of $\mathrm{MoS_2}$ is \SI{6.0}{\angstrom}, so a cutoff of at least \SI{12.0}{\angstrom} should be set to include the forces from the next-nearest layers. However, larger cutoffs result in reduced computational efficiency while not necessarily guaranteeing improved accuracy. This is shown in the \gls{nep} model case for bilayer \gls{gr} \citep{ying2024combining}, where the \gls{nep} using a radial cutoff of \SI{10}{\angstrom} exhibited a larger force \gls{rmse} than that of \SI{4.5}{\angstrom} with identical model setup, while achieving only about half the computational speed. As pure \gls{mlp} approaches don't inherently distinguish between intralayer and interlayer interactions from a physical perspective, implementing them with longer cutoffs forces these \gls{mlp}s to incorporate numerous additional features into local atomic environments, significantly complicating the training process. Consequently, extending cutoffs beyond \SI{1}{\nano\meter} for pure \glspl{mlp} to capture long-range \gls{vdw} interactions from next-nearest layers not only dramatically reduces computational efficiency but can paradoxically lead to decreased accuracy rather than improvement. To resolve this dilemma, the \gls{dft}-D3 dispersion energy correction term \citep{grimme2010consistent} has been combined with \glspl{mlp} \citep{ying2024combining}, allowing the \glspl{mlp} to focus on local chemical environments while the \gls{dft}-D3 handles long-range dispersion interactions.

The second key issue concerns large number of training configurations required for heterostructures composed of various monolayer components. \glspl{mlp} rely on the invariant or equivariant descriptors to transform the spatial coordinates to local chemical information, allowing them to learn from discrete training points and predict a smooth, high-dimensional \gls{pes} in an interpolative manner based on electronic data obtained from \emph{ab initio} calculations. Theoretically, achieving high accuracy and reliability requires the training dataset to comprehensively cover all configurations that may arise during \gls{md} simulations. Furthermore, for \gls{vdw} heterostructures, local chemical environment can extend beyond the nearest layers and vary with different stacking configurations. For the \gls{gr}/\gls{hbn} heterostructure within the pure \gls{nep} framework, it has been demonstrated that at least 13 types of configurations \citep{liang2025probing}, including \gls{2d} structures with up to three layers and bulk structures, are required to accurately capture the surface, interface, and bulk chemical environment of this system, as shown in \autoref{fig:art}(f). Building on the assumption that at least three layers are necessary to capture the interface and surface environment, the total number of configurations required for simulating \gls{vdw} heterostructures with $n$ distinct monolayer components can be determined as follows. The number of monolayer configurations is $\mathrm{C}_n^1$. The number of bilayer homostructures is also $\mathrm{C}_n^1$, while the number of bilayer heterostructures, considering all component pairs, is $\mathrm{C}_n^2$. For trilayer configurations, the total count is $\mathrm{C}_n^1 + 4\mathrm{C}_n^2 + 3\mathrm{C}_n^3$ (see \gls{sm} Section S2 for details). Finally, the number of bulk configurations is also $\mathrm{C}_n^1$. Summing all these contributions, the minimum number of required configurations is given by

\begin{equation}
N_{\mathrm{MLP(3)}} = \frac{1}{2} n^3 + n^2 + \frac{5}{2} n \sim \mathrm O(n^3)
\label{equation:mlp3}
\end{equation}
Furthermore, if we assume that up to four layers are necessary to fully cover the surface and interface environment, the total number of required configurations increases to
\begin{equation}
N_{\mathrm{MLP(4)}} = \frac{1}{2} n^4 + \frac{1}{2} n^3 + \frac{3}{2} n^2 + \frac{5}{2} n \sim \mathrm O(n^4)
\label{equation:mlp4}
\end{equation}
where the detailed derivation is provided in \gls{sm} Section S2.

However, for \gls{smlp}+\gls{ilp} framework, which fully decouples the interlayer and intralayer interactions (see \autoref{fig:art}(b)), it is sufficient to prepare only $\mathrm{C}_n^1$ monolayer configurations for the \gls{mlp}, whereas $\mathrm{C}_n^1$ bilayer homogeneous and $\mathrm{C}_n^2$ bilayer heterogeneous configurations are required for \gls{ilp}. This results in a total number of required configurations given by
\begin{equation}
N_{\mathrm{ILP}} = \frac{1}{2} n^2 + \frac{3}{2} n \sim \mathrm O(n^2)
\end{equation}

In \autoref{fig:art}(d) we show that only five configurations are required in the \gls{smlp}+\gls{ilp} framework for \gls{gr}/\gls{hbn} heterostructure system. \autoref{fig:art}(e) compares the number of required training configurations between existing pure \gls{mlp} approach and the proposed \gls{smlp}+\gls{ilp} framework. For an extremely complex \gls{vdw} heterostructure containing 10 distinct monolayer components, the number of required reference configurations for $N_{\mathrm{MLP(4)}}$ is approximately 100 times larger than that for \gls{smlp}+\gls{ilp}. Moreover, the pure \gls{mlp} dataset includes configurations with significantly more atoms, leading to a computational cost for electronic data calculations that could be several orders of magnitude higher. Additionally, for a given set of fixed configurations, the relatively sliding \glspl{pes} in multilayer systems become significantly more complex, necessitating extensive sampling of reference structures.

\begin{figure*}[htb]
\begin{center}
\includegraphics[width=2\columnwidth]{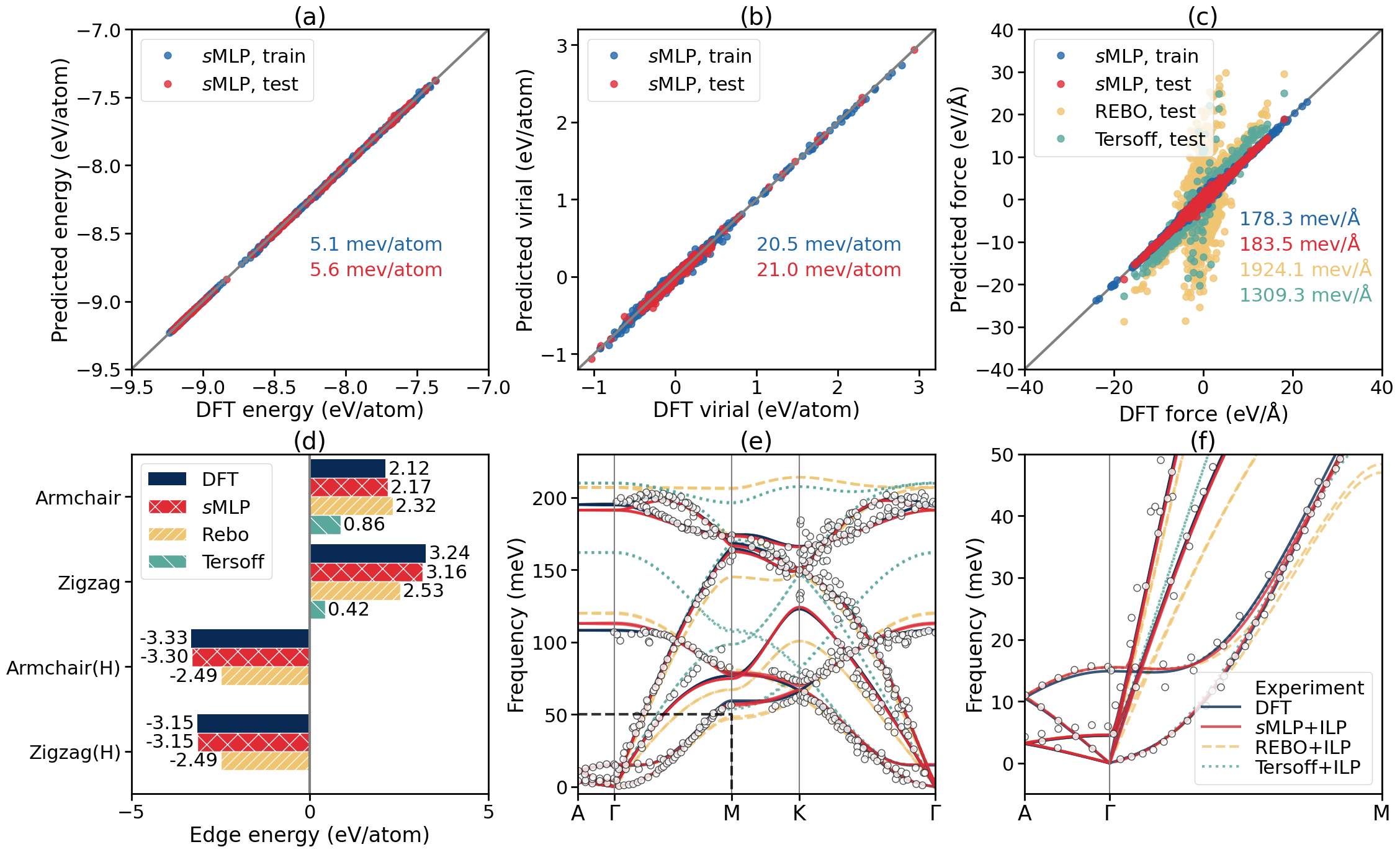}
\caption{(a)-(c) Parity plots for energy, virial, and force comparing \gls{dft} reference data and \gls{smlp} predictions for the whole training and testing dataset. In panel (c), \gls{rebo} and Tersoff predictions for forces across the testing dataset are also provided for comparison. (d) Edge energy calculated by \gls{dft}, \gls{smlp}, \gls{rebo}, Tersoff for monolayer \gls{gr} structures shown in \gls{sm} Figure S15. (e)-(f) Phonon spectra of bulk graphite. The solid red lines, dashed blue lines, and dashed-dotted green lines are dispersion curves calculated using the \gls{smlp}+\gls{ilp}, \gls{rebo}+\gls{ilp} and Tersoff+\gls{ilp} force fields, respectively. Experimental result~\citep{wirtz2004phonon} of bulk graphite is given by open black circles. (f) shows a zoom-in of the low-energy phonon modes around the $\Gamma$-point.
}
\label{fig:benchmark}
\end{center}
\end{figure*}

\section{Validation and benchmark}

\subsection{High accuracy of \textit{s}MLP models}

In this work, we employed the \gls{smlp}+\gls{ilp} framework to investigate homogeneous and heterogeneous \gls{vdw} systems composed of one, two, or all three components from the monolayers \gls{gr}, \gls{hbn}, and \gls{mos2}. Interlayer interactions between \gls{gr}, \gls{hbn}, and their heterostructures were modeled using the \gls{ilp} approach from Ref.~\citep{ouyang2018nanoserpents}. Similarly, interactions for both \gls{gr}/\gls{mos2} and \gls{hbn}/\gls{mos2} were described by the \gls{ilp} model from Ref.~\citep{jiang2025anisotropic}. These \gls{ilp} models are grounded in the electronic nature of weak interlayer interactions and are fitted to state-of-the-art \gls{dft} data augmented by \gls{mbd_nl} corrections \citep{hermann2020density}. As a result, they accurately capture binding energy curves and sliding \glspl{pes}, achieving first-principles accuracy with energy \glspl{rmse} on the order of \SI{0.1}{\milli\electronvolt\per\atom} \citep{ouyang2018nanoserpents, jiang2025anisotropic}. As the necessary \gls{ilp} models are already available, in this section, we focus solely on developing three \glspl{smlp}, corresponding to the individual monolayer components.

To train the \glspl{mlp}, we first prepared a reference dataset consisting of energies, atomic forces, and virials obtained from \gls{dft} calculations at the exchange-correlation \gls{pbe} level \citep{perdew1996generalized} (see \gls{sm} Section S1 for details). Both extended and edge configurations were included, with hydrogen passivation considered for \gls{gr} and \gls{hbn}, and no passivation for \gls{mos2}. For each monolayer, reference structures were generated through random perturbations and \gls{md} simulations at finite temperatures (see \gls{sm} Section S3 for details). 

Once the reference dataset was prepared, the third generation of \gls{nep} framework \citep{fan2022gpumd}, as implemented in \textsc{GPUMD}, was used to train \gls{gr} \gls{smlp} and the fourth generation of \gls{nep} framework \citep{song2024generalpurpose} was used to train \gls{hbn} and \gls{mos2} \glspl{smlp}. The \gls{nep} framework employs a simple feedforward \gls{nn} to represent the site energy of an atom (see \autoref{equation:site_energy}), using descriptor vectors composed of a number of radial (see \autoref{equation:raidial}) and angular (see \autoref{equation:angular}) components. The \gls{snes} \citep{schaul2011high} was used for the optimization of the parameters in \gls{nep} models. For all \gls{smlp} models, a consistent cutoff of \SI{5}{\angstrom} was used for both radial and angular descriptor components. Full details of the \gls{nep} training procedure are provided in \gls{sm} Section S4. 

\autoref{fig:benchmark}(a)–(c) show the energy, virial, and force predictions for \gls{gr} obtained from the \gls{smlp} model, compared against the \gls{dft} reference values. Each panel also reports the corresponding \glspl{rmse} for both the training and test datasets. The \gls{smlp} model demonstrates high accuracy in predicting the energy, virial and forces, with \gls{rmse} values below \SI{6.0}{\milli\electronvolt\per\atom}, \SI{21}{\milli\electronvolt\per\atom}, and \SI{184}{\milli\electronvolt\per\angstrom}, respectively, for both datasets. A comparable level of accuracy is achieved for both \gls{hbn} and \gls{mos2} (see \gls{sm} Figure S4). For comparison, in \autoref{fig:benchmark}(c) we also evaluate the accuracy of two widely used empirical potentials, i.e., the Tersoff potential \citep{tersoff1988empirical} and the \gls{rebo} potential \citep{brenner2002second}, by comparing their predicted forces on the test dataset against the \gls{dft} reference values. It is clear that the \gls{smlp} model, with an \gls{rmse} of \SI{183.5}{\milli\electronvolt\per\angstrom}, exhibits significantly higher accuracy than the other two empirical potentials, whose predictions deviate substantially from the parity line, with corresponding \gls{rmse} values exceeding \SI{1000}{\milli\electronvolt\per\angstrom}.

Beyond the \gls{rmse} values, we further evaluated the predictive accuracy of the \gls{smlp} and the two empirical potentials by comparing their predicted edge energies of \gls{gr} and \gls{hbn} monolayers against \gls{dft} reference results. The edge energy of a monolayer, $V_{\mathrm{EE}}$, is calculated as
\begin{equation}
 V_{\mathrm{EE}} = V_{\mathrm{PBC}} - V_{\mathrm{nPBC}}
\end{equation}
where $V_{\mathrm{PBC}}$ and $V_{\mathrm{nPBC}}$ represent the energies of the monolayer without edges under periodic boundary conditions and that with edges under non-periodic boundary conditions, respectively (see \gls{sm} Section S7 for details). Since the Tersoff potential does not support hydrogen atoms, it is only applied to non-hydrogenated structures. As shown in \autoref{fig:benchmark}(d) and \gls{sm} Figure S15, the edge energies of both \gls{gr} and \gls{hbn} computed using the \gls{smlp} model align closely with the \gls{dft} reference values, exhibiting errors on the order of \SI{0.01}{\electronvolt\per\atom}, which are one to two orders of magnitude smaller than those obtained with two empirical potentials. The high accuracy in predicting energy, atomic forces, virials, and edge energies demonstrates the reliability of the developed \glspl{smlp} in describing intralayer interactions in all three monolayers.

\begin{figure}[htb]
\begin{center}
\includegraphics[width=\columnwidth]{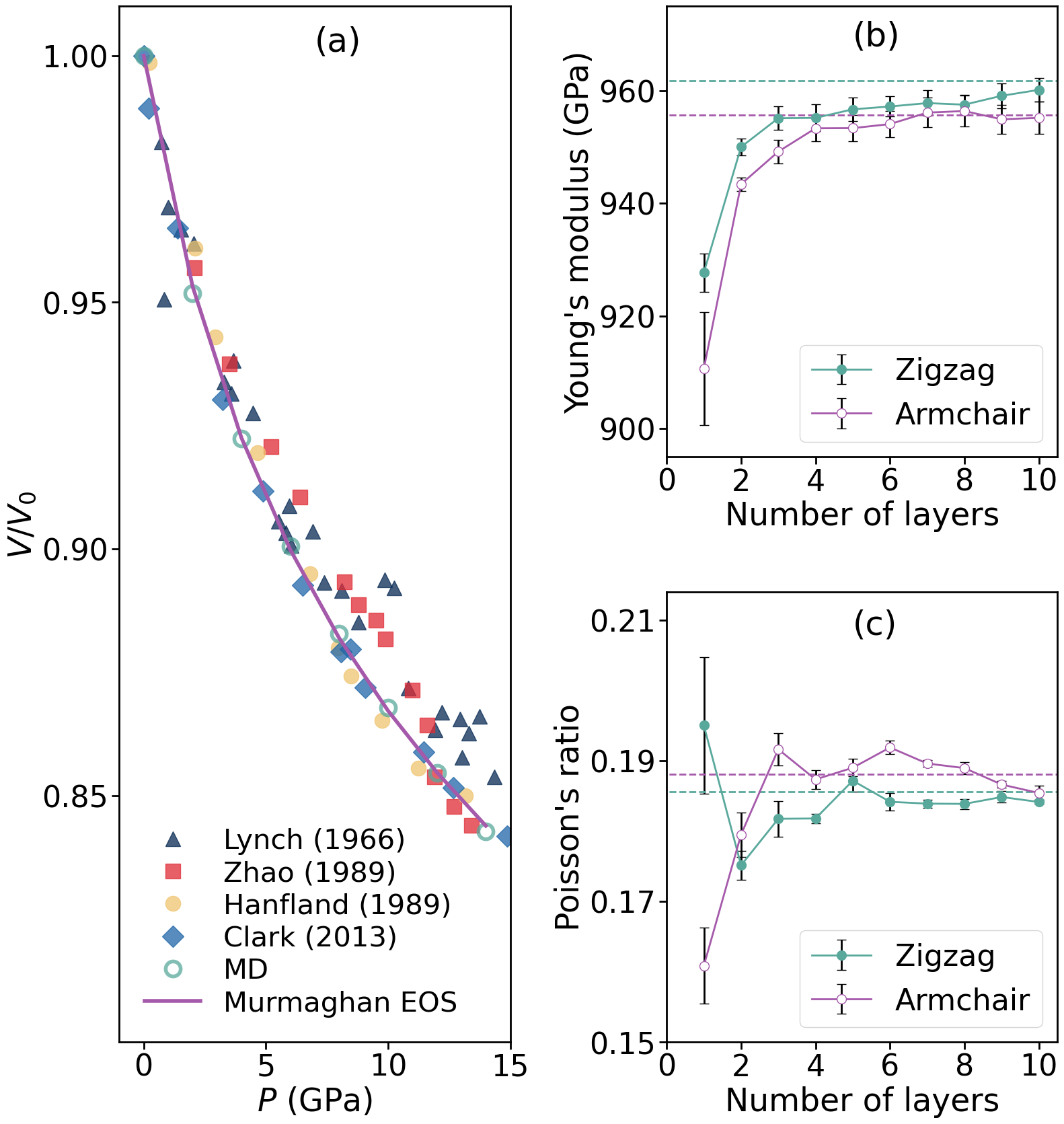}
\caption{(a) Pressure ($P$) dependence of the normalized volume ($V/V_0$) of bulk graphite, where $V$ and $V_0$ represent the pressure-dependent and initial volumes, respectively. The green hollow circles represent \gls{md} simulation results using the \gls{smlp}+\gls{ilp} model. The solid purple line is the fitted curve using the Murnaghan \gls{eos}; the fitted curves from the two additional \glspl{eos} coincide closely and are therefore not shown. 
Blue triangles, red squares, yellow circles, and blue diamonds represent experimental data obtained from Ref.~\citep{lynch1966effect}, Ref.~\citep{zhao1989xray}, Ref.~\citep{hanfland1989graphite}, and Ref.~\citep{clark2013fewlayer}, respectively. (b) Young’s modulus and (c) Poisson’s ratio of the \gls{gr} system as functions of the number of layers, predicted by \gls{smlp}+\gls{ilp}-based \gls{md} simulations. Results (dashed line) for bulk graphite are also included for comparison.}
\label{fig:mechanic}
\end{center}
\end{figure}

\subsection{Accuracy of the \textit{s}MLP+ILP models in predicting phonon and mechanical properties of graphite}

After demonstrating the accuracy of \glspl{smlp} in describing intralayer interactions in monolayers, we now take bulk graphite as an example to further benchmark the accuracy of the \gls{smlp}+\gls{ilp} approach in describing both phonon and mechanical properties. 

\begin{table}[ht]
\setlength{\abovecaptionskip}{1.2cm} 
\caption{Bulk modulus ($B$), zero-pressure derivative ($B'$), intralayer lattice constant ($a_0$), and interlayer lattice constant ($c_0$) of bulk graphite, calculated from \gls{md} simulations using the \gls{smlp}+\gls{ilp} model and compared with experimental data from X-ray diffraction \citep{lynch1966effect, hanfland1989graphite, zhao1989xray}. Values in parentheses represent statistical errors. For the \gls{md} results, three different \glspl{eos} \citep{hanfland1989graphite,murnaghan1944thecompressibility,birch1952elasticity,vinet1987temperature} (see \gls{sm} Section S7 for details) were used to fit $B$ and $B'$. $a_0$ and $c_0$ 
 of \gls{md} are computed at \SI{300}{\kelvin}.}
\centering
\label{tab:bulk_modulus}
\begin{tabular}{lcccc}
\hline
Method & $B$ (GPa) & $B'$ & $a_0$ (\AA) & $c_0$ (\AA) \\
\hline
Experiment (Ref.~\citep{lynch1966effect}) & 32(2) & 12.3(0.7) & 2.4612 & 6.7078 \\
Experiment (Ref.~\citep{hanfland1989graphite}) & 33.8(0.3) & 8.9(0.1) & \makecell{2.603 \\(0.004)} & \makecell{6.706 \\ (0.003)} \\
Experiment (Ref.~\citep{zhao1989xray}) & 30.8(2.0) & -- & 2.462 & 6.707 \\
\hline
MD (Murnaghan) & 32.9(1.7) & 9.6(0.6) & \multirow{3}{*}{\makecell{2.4671\\(0.0002)}} & \multirow{3}{*}{\makecell{6.726\\(0.004)}} \\
MD (Birch-Murn.) & 28.9(0.6) & 15.2(0.5) & \\
MD (Vinet) & 31.8(1.1) & 11.4(0.5) & \\
\hline
\end{tabular}
\end{table}

To this end, we used \gls{smlp}+\gls{ilp} to compute the phonon dispersion curves of bulk graphite at zero pressure and temperature by diagonalizing the dynamical matrix in \textsc{LAMMPS}. The results were compared with those obtained from \gls{rebo}+\gls{ilp}, Tersoff+\gls{ilp}, \gls{dft} calculations, as well as experimental measurements \citep{wirtz2004phonon}. As shown in \autoref{fig:benchmark}(e)-(f), the \gls{smlp}+\gls{ilp} approach accurately reproduces the dispersion of both the low-energy out-of-plane branches near the $\Gamma$ point and the high-energy branches along the $\Gamma \rightarrow \mathrm{M} \rightarrow \mathrm{K} \rightarrow \Gamma$ path. In contrast, significant discrepancies observed in the high-energy modes for both the \gls{rebo}+\gls{ilp} and Tersoff+\gls{ilp} models stem from limitations in their intralayer potential descriptions. Notably, the \gls{smlp}+\gls{ilp} curve closely matches the \gls{dft} results, further confirming that this approach can achieve \gls{dft}-level accuracy. Similar behavior has also been observed in bulk \gls{hbn} systems (see \gls{sm} Section S7 for details). We further employed the \gls{hnemd} method implemented in \textsc{GPUMD} \citep{fan2019homogeneous} to compute the cross-plane thermal conductivity of bulk graphite and \gls{hbn} with different thicknesses, obtaining out-of-plane thermal conductivities in the range of $6.7\pm0.9$ to $8.1\pm0.1$~\unit{\watt\per\metre\per\kelvin} for bulk graphite and $5.8\pm1.4$ to $6.7\pm0.4$~\unit{\watt\per\metre\per\kelvin} for bulk \gls{hbn}, summarized in \autoref{stab:TC}, in excellent agreement with experimental measurements and thereby further confirming the accuracy of the \gls{smlp}+\gls{ilp} framework (see \gls{sm} Section S13 for details).

\begin{table}[htb]
\caption{Cross-plane thermal conductivity $\kappa$ of bulk \gls{gr} and \gls{hbn} at room temperature. Experimental values correspond to $c$-axis transport in highly ordered graphite and bulk \gls{hbn}, as reported in Refs.~\citep{yang2025ultrahigh,han2018submm,jaffe2023thicknessdependent,yuan2019modulating,jiang2018anisotropic,salihoglu2020energy}. The \gls{md} data are obtained in this work from \gls{hnemd} simulations for 20-, 40-, and 60-layer stacks.}
\begin{center}
\begin{tabular}{ccc}
\hline
 Material & Method & $\kappa$ (\unit{\watt\per\meter\per\kelvin})\\
\hline
 \gls{gr} & Experiment \citep{yang2025ultrahigh} & 7.4\\
 & Experiment \citep{han2018submm} & $6.08(0.6)$ \\
 & \gls{md}, \gls{smlp}+\gls{ilp} (20L) & $6.7(0.9)$\\
 & \gls{md}, \gls{smlp}+\gls{ilp} (40L) & $8.1(0.1)$\\
 & \gls{md}, \gls{smlp}+\gls{ilp} (60L) & $7.4(0.5)$\\
\hline
 \gls{hbn} & Experiment \citep{jaffe2023thicknessdependent} & $8.1(0.5)$\\
 & Experiment \citep{yuan2019modulating} & $3.5(0.8)$\\
 & Experiment \citep{jiang2018anisotropic} & $4.8(0.6)$\\
 & Experiment \citep{salihoglu2020energy} & $5.4(0.5)$\\
 & \gls{md}, \gls{smlp}+\gls{ilp} (20L) & $5.8(1.4)$\\
 & \gls{md}, \gls{smlp}+\gls{ilp} (40L) & $6.5(0.7)$\\
 & \gls{md}, \gls{smlp}+\gls{ilp} (60L) & $6.7(0.4)$\\
\hline
\end{tabular}
\label{stab:TC}
\end{center}
\end{table}

To further evaluate the \gls{smlp}+\gls{ilp} performance in describing mechanical properties under hydrostatic pressure, we performed \gls{md} simulations to compute its bulk modulus and compared the results with available experimental data \citep{lynch1966effect, hanfland1989graphite, zhao1989xray, clark2013fewlayer}. As shown in \autoref{fig:mechanic}(a), the computed pressure–volume ($P$–$V$) relationship from our simulations aligns well with experimental measurements. The bulk moduli ($B$) and zero-pressure derivatives ($B'$) of bulk graphite, obtained by fitting three different \glspl{eos} (see Section S4 for details), respectively, fall within the range of \qtyrange{28}{33}{\giga\pascal} and \qtyrange{9}{16}{} (see \autoref{tab:bulk_modulus}). Moreover, our \gls{md} simulations show good agreement with the experimental values of the intra- and interlayer lattice constants. The deviations between simulation and experiment are approximately \SI{0.01}{\angstrom} for the intralayer and \SI{0.02}{\angstrom} for the interlayer lattice constants, except for one reported experimental value of \SI{2.603}{\angstrom}, which significantly overestimates the intralayer lattice constant. Overall, these results demonstrate that the \gls{smlp}+\gls{ilp} approach provides a robust and reliable description of the bulk properties of graphite across a wide range of external pressures.

Next, we validate the \gls{smlp}+\gls{ilp} approach by calculating the elastic constants of the graphite system with varying number of layers. The Young's modulus and Poisson's ratio for monolayer \gls{gr} have been experimentally measured to be approximately \SI{1}{\tera\pascal} and 0.17, respectively \citep{lee2008measurement}. Similarly, Blakslee et al. reported values for bulk graphite of \SI{1.02+-0.03}{\tera\pascal} for the Young's modulus and \num{0.16+-0.06} for the Poisson’s ratio, based on static tensile and compression measurements \citep{blakslee1970elastic}. Here, we calculated Young's modulus and Poisson's ratio by applying uniaxial tension along armchair and zigzag directions using \gls{smlp}+\gls{ilp} model (see \gls{sm} Section S1 for details). For a monolayer \gls{gr}, the Poisson’s ratio was predicted to be 0.161 in the armchair direction and 0.195 in the zigzag direction, while the corresponding Young’s moduli were \SI{0.911}{\tera\pascal} and \SI{0.928}{\tera\pascal}, respectively, indicating a weak elastic anisotropy. As the number of layers increases, the predicted Young's modulus rises and gradually converges to the bulk graphite value. Meanwhile, the anisotropy in the Poisson's ratio progressively decreases with increasing layer thickness. Notably, our \gls{smlp}+\gls{ilp} predictions for both monolayer \gls{gr} and bulk graphite are consistent with experimental values \citep{lee2008measurement, blakslee1970elastic}. Additional results and comparisons with other potential models are provided in \gls{sm} Section S7.

\begin{figure*}
\begin{center}
\includegraphics[width=1.4\columnwidth]{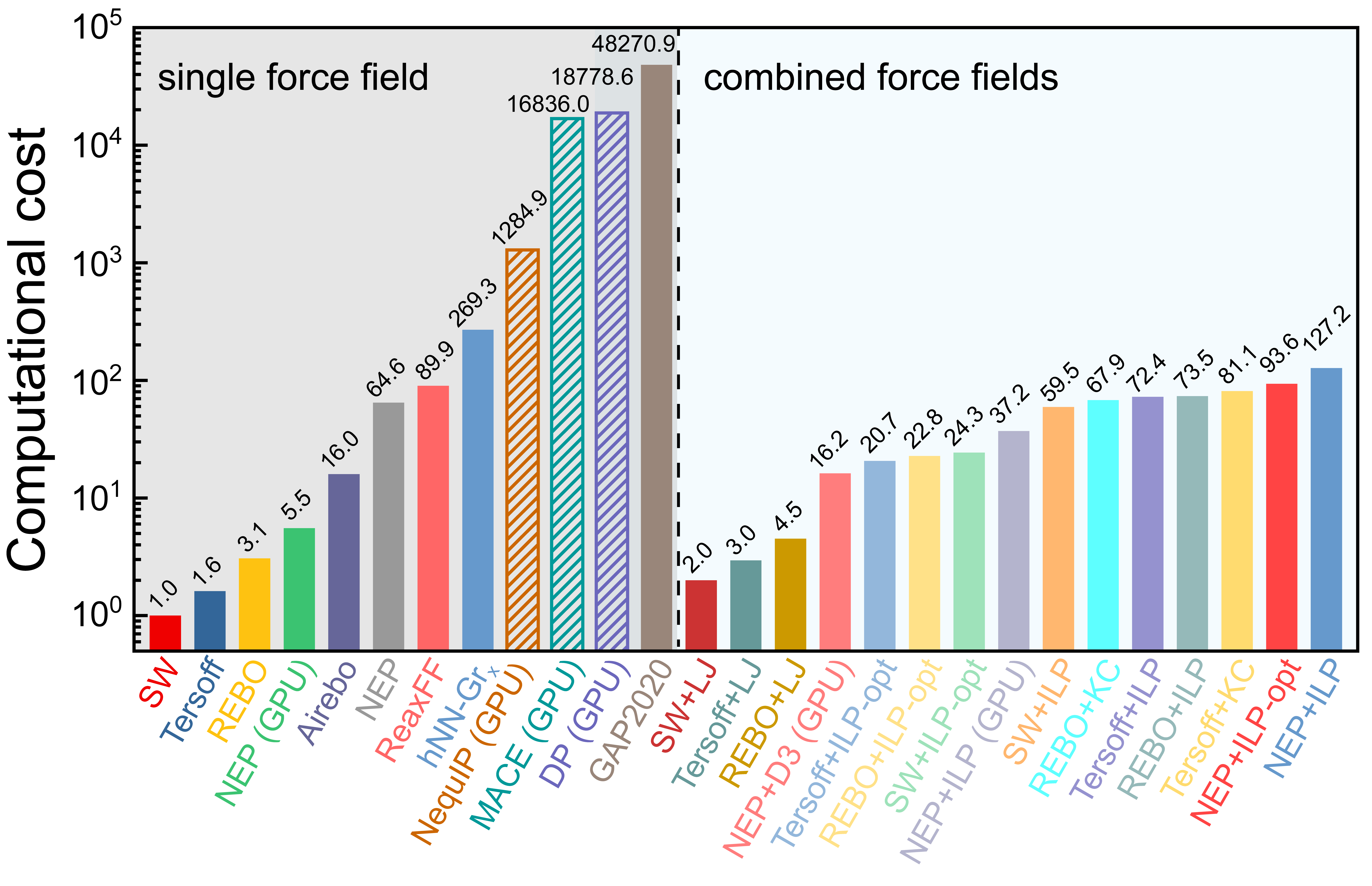}
\caption{Computational speeds of the \gls{smlp}+\gls{ilp} approach implemented in \textsc{GPUMD} using \textsc{GPU} and \textsc{LAMMPS} using \textsc{CPU}. Benchmarks were conducted on \gls{mos2} monolayer, \gls{mos2}/\gls{mos2} bilayer, \gls{gr} monolayer, and \gls{gr}/\gls{gr} bilayer, all containing 13,824 atoms, except for the \gls{nequip}, MACE and \gls{dp} tests on \textsc{GPU}, due to memory limitations. The models of the last three tests, each containing 1,024 atoms, are shown as hatched bars. The performance of other potential models and hybrid approaches is also included for comparison. The computational cost is evaluated by normalizing the per-atom, per–time-step runtime with respect to that of the \gls{sw} potential.}
\label{fig:speed}
\end{center}
\end{figure*}

\subsection{Efficiency of the \textit{s}MLP+ILP approach}
The \gls{smlp}+\gls{ilp} approach not only improves accuracy but also maintains high computational efficiency compared to the existing pure \glspl{mlp} and hybrid empirical potential + \gls{ilp} methods. In this section, we benchmark its performance against several representative approaches, including \gls{sw}, Tersoff, \gls{rebo}, \gls{airebo} \citep{stuart2000reactive}, \gls{reaxff} \citep{chenoweth2008reaxffa}, \gls{sw}+\gls{lj}, Tersoff+\gls{lj}, \gls{rebo}+\gls{lj}, Tersoff+\gls{kc} \citep{kolmogorov2005registrydependent}, \gls{rebo}+\gls{ilp}, Tersoff+\gls{ilp}, \gls{nep}+D3 \citep{ying2024combining}, and a set of existing \gls{mlp} models: \gls{nep} \citep{ying2024combining}, \gls{dp} \citep{ying2025advances}, MACE \citep{ying2025advances}, \gls{gap} \citep{bartok2010gaussian}, and hNN-Gr$_x$ \citep{wen2019hybrid}. In addition, we also trained the latest \gls{nequip} model (v0.15.0) \citep{tan2025highperformance} using an existing dataset for bilayer \gls{gr}~\citep{ying2024effect} (see \gls{sm} Section S8 for details).

To accurately capture the decay of binding energy with interlayer distance in \gls{vdw} interfaces, the \gls{ilp} model employs a cutoff of \SI{16}{\angstrom}, while the pure \gls{nep} model~\citep{ying2024combining} uses a radial cutoff of \SI{10}{\angstrom} and an angular cutoff of \SI{4.5}{\angstrom}. For the \gls{nep}+D3 approach \citep{ying2024combining}, both radial and angular cutoffs are set to \SI{4.5}{\angstrom}, with an additional \SI{16}{\angstrom} cutoff for the D3 correction. The cutoffs used in the \gls{dp} and hNN-Gr$_x$ models are all no longer than \SI{10}{\angstrom}. The MACE model uses two message passing layers with a cutoff radius of \SI{6}{\angstrom}, while the \gls{nequip} model employs four message passing layers with a cutoff radius of \SI{7}{\angstrom}.

As shown in \autoref{fig:speed}, we benchmarked these approaches on both \textsc{GPU} and \textsc{CPU} platforms, using a single NVIDIA GeForce RTX 4090 GPU and 36 Intel Xeon 6240 CPU cores, respectively. All \textsc{GPU}-based tests were conducted using \textsc{GPUMD} (version 3.9.5), except for the \gls{dp}, \gls{nequip} and MACE models, which were tested using \textsc{LAMMPS} (feature release, 10 September 2025). All \textsc{CPU}-based tests were also run using \textsc{LAMMPS} (stable release, 29 August 2024) and we used both non-OPT and OPT versions of \gls{ilp} potential in \textsc{LAMMPS} \citep{gao2021lmff}. The considered types of \gls{vdw} bilayers were \gls{gr} monolayer, \gls{mos2} monolayer, \gls{gr}/\gls{gr} homogeneous bilayer and \gls{mos2}/\gls{mos2} homogeneous bilayer. For each test, simulations were performed in the microcanonical ensemble for 1,000 steps with a time step of \SI{1}{\femto\second}. Notably, on a single desktop GPU, the computational cost of \gls{smlp}+\gls{ilp} is 37.2-fold relative to the \gls{sw} potential (normalized as shown in \autoref{fig:speed}), comparable to that of the pure \gls{nep}, \gls{nep}+D3, and Tersoff+\gls{ilp} models, and approximately two orders of magnitude faster than the \gls{nequip} model and three orders than the \gls{dp} and MACE models. On the \textsc{CPU} platform, the \gls{smlp}+\gls{ilp} performance is similarly close to Tersoff+\gls{ilp} and Tersoff+\gls{kc} models, and one order of magnitude faster than other \gls{mlp} models. Our results also highlight the high computational efficiency of the \gls{nep} approach, which is a key reason we adopted it to train the \gls{smlp} model in our hybrid \gls{smlp}+\gls{ilp} framework.

\section{Moir\'e superlattices in complex vdW heterostructures}

\subsection{Comparison with Experimental Moir\'e Patterns}
Having demonstrated the accuracy and efficiency of our \gls{smlp}+\gls{ilp} approach, we now apply it in large-scale atomistic simulations to explore structural reconstruction in complex \gls{vdw} systems. Previous studies have shown that certain multilayer \gls{vdw} materials undergo local reconstruction, forming Moir\'e out-of-plane distortions to maximize the overall areas corresponding to the energetically favorable (commensurate) stacking mode \citep{mandelli2019princess, ouyang2021parity}. The emergence of Moir\'e superlattices involves the formation of sharp domain walls of unfavorable stacking between nearly commensurate regions, which relieve excess in-plane stress induced by interlayer lattice mismatch. 

\begin{figure}[htb]
\begin{center}
\includegraphics[width=0.8\columnwidth]{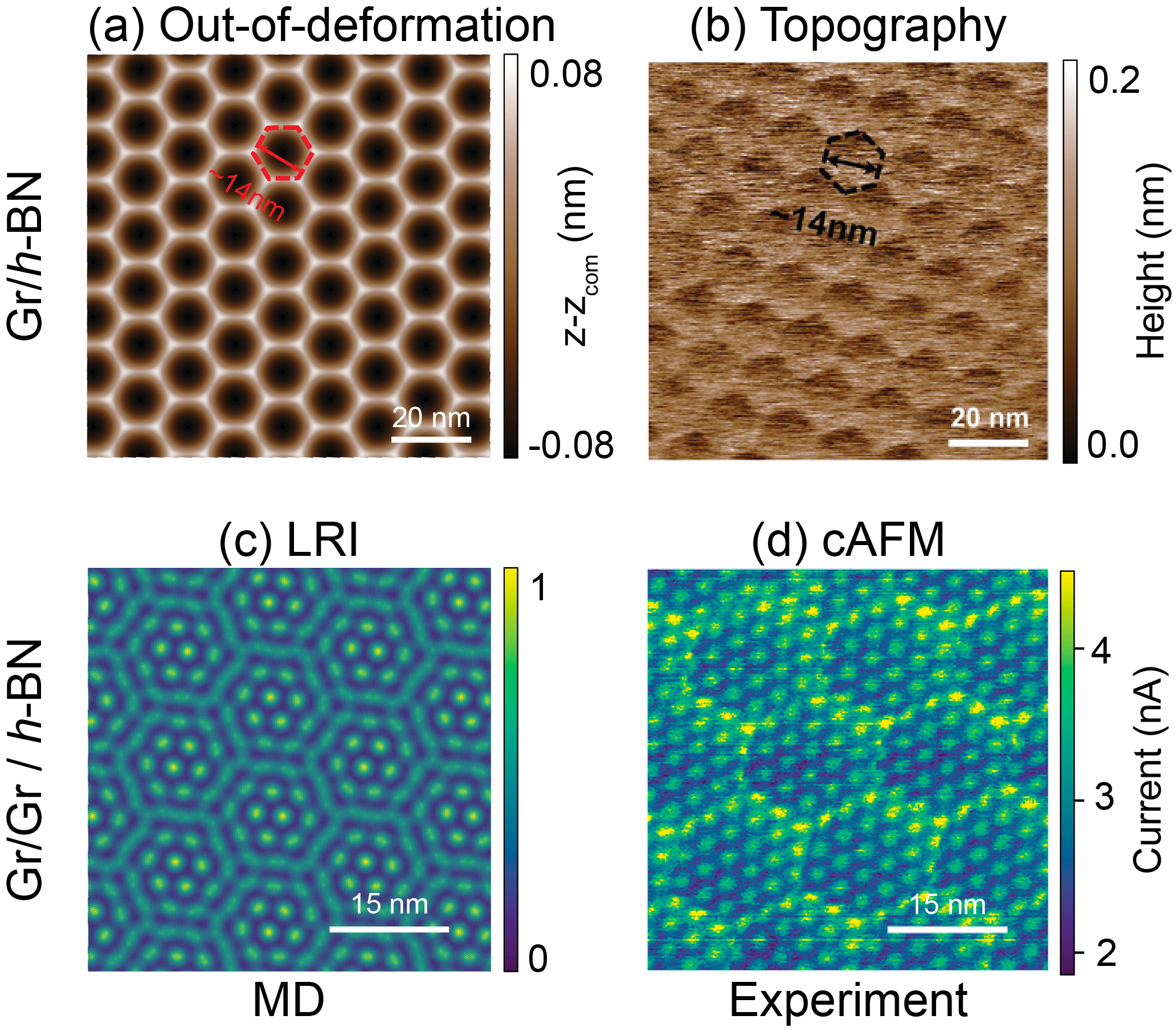}
\caption{(a) \gls{smlp}+\gls{ilp} predicted out-of-plane deformation and (b) experimentally measured topography of a relaxed \gls{gr}/\gls{hbn} heterostructure. (c) \gls{smlp}+\gls{ilp} predicted \gls{lri} map and (d) experimentally measured conductive atomic force microscopy image of a relaxed \gls{gr}/\gls{gr}/\gls{hbn} heterostructure with a \ang{4.2} twist angle between \gls{gr} layers. Panel (b) is adapted with permission from Ref.~\citep{huang2022origin}. Copyright 2022, Wiley. Panel (d) is adapted with permission from Ref.~\citep{huang2021imaging}. Copyright 2021, American Chemistry Society.}
\label{fig:moire}
\end{center}
\end{figure}

Herein, we demonstrate the advantage of \gls{smlp}+\gls{ilp} approach in modeling complex heterostructures by predicting the Moir\'e patterns in both \gls{gr}/\gls{hbn} bilayer and \gls{gr}/\gls{gr}/\gls{hbn} trilayer. For both systems, the experimental measurements of Moir\'e superlattices are available \citep{huang2022origin, huang2021imaging}, enabling direct validation of our predictions. To that end, we constructed \gls{gr} ($56\times56$)/\gls{hbn} ($55\times55$) heterojunction models (containing 36,352 atoms) to consider the 1.8\% lattice mismatch between the two monolayers. Beyond this intrinsic lattice mismatch, a twisted angle of \ang{4.2} was applied to the top two \gls{gr} layers and \ang{0.04} to the middle \gls{gr} and \gls{hbn} in the trilayer system to match the experimental setup, resulting in a \gls{gr} ($52\times59$)/\gls{gr} ($59\times52$)/\gls{hbn} ($58\times51$) heterojunction model (containing 54,858 atoms). To recreate the experimental height observation \citep{yankowitz2016pressureinduced, xue2011scanning}, we added substrates for \gls{gr}/\gls{hbn} bilayer and benchmarked the influence of different simulation setup (see \gls{sm} Section S10 for details). The substrates will not change the Moir\'e patterns but only inhibit the out-of-plane deformation.

\begin{figure*}[htb]
\begin{center}
\includegraphics[width=1.8\columnwidth]{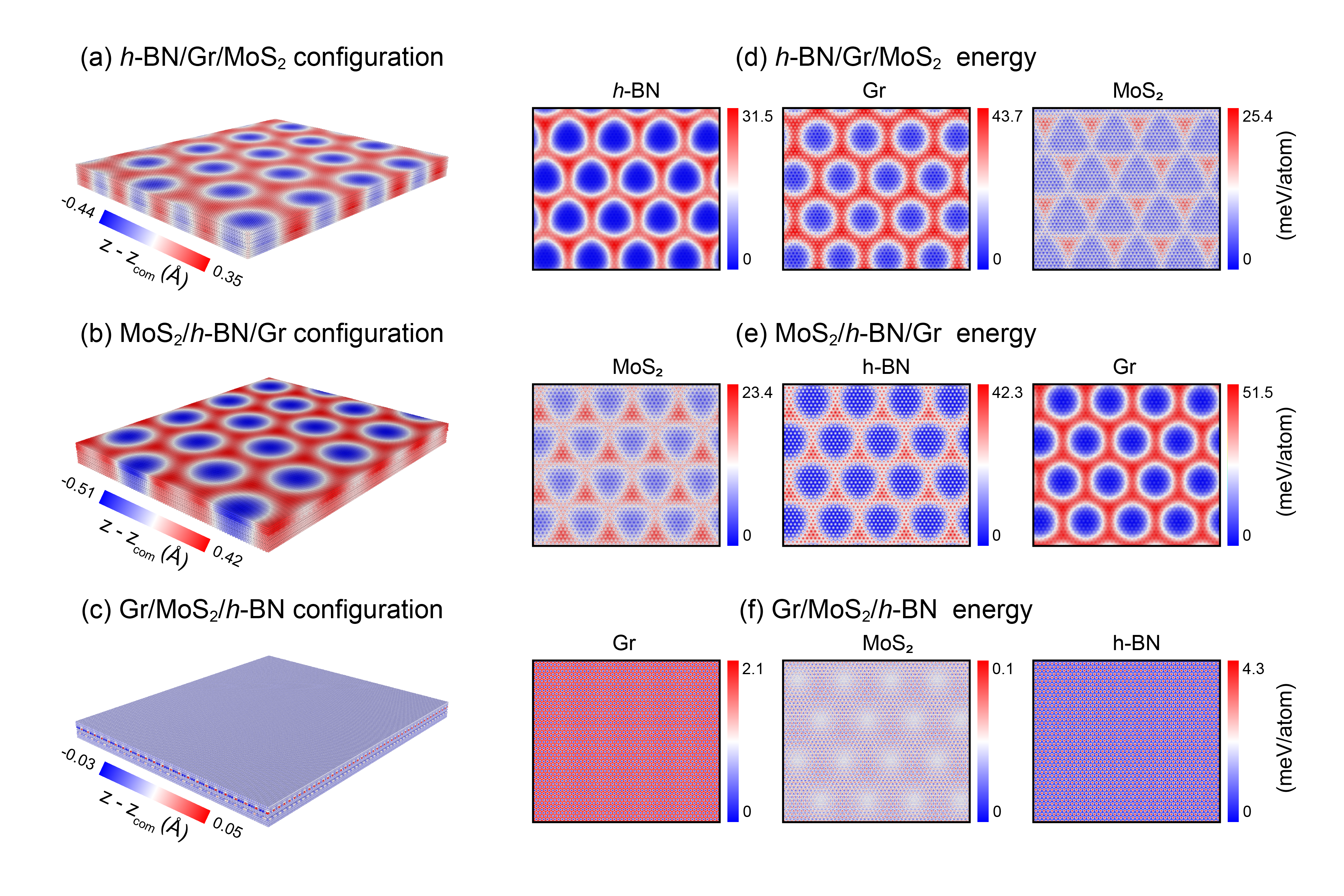}
\caption{(a)-(c) Out-of-plane deformation and (d)-(f) energy distribution of the optimized substrate-supported \textit{h}-BN/Gr/\ce{MoS2} (top), \ce{MoS2}/\textit{h}-BN/Gr (middle), and Gr/\ce{MoS2}/\textit{h}-BN (bottom) structures. Panels (d)-(f) present the weighted atomic energy maps for each component monolayer in the optimized \gls{vdw} heterostructures, with all values presented relative to the corresponding minimum energy of the respective layer.}
\label{fig:GMB}
\end{center}
\end{figure*}

Initially, the models exhibit high stress at the boundaries, necessitating relaxation. We achieve this through energy minimization described in \gls{sm} Section S1. With a force convergence criterion below \SI{1e-6}{\electronvolt\per\angstrom}, we obtained the optimized Moir\'e superlattices shown in \autoref{fig:moire}, and compared them with experimental observations. For the \gls{gr} ($56\times56$)/\gls{hbn} ($55\times55$) system with 8 flexible substrate layers, a hexagonal Moir\'e superlattice with a periodicity of \SI{14}{\nano\meter} emerges due to the intrinsic lattice mismatch between \gls{gr} and \gls{hbn} (\autoref{fig:moire}(a)), in excellent agreement with the experimentally measured topography \citep{huang2022origin} (\autoref{fig:moire}(b)). The flexible trilayer model likewise exhibits an out-of-plane Moir\'e corrugation with a characteristic wavelength of about \SI{14}{\nano\meter}. (see \gls{sm} Section S10) For the \gls{gr} ($52\times59$)/\gls{gr} ($59\times52$)/\gls{hbn} ($58\times51$) system, we employed the local registry index (\gls{lri}) method (see \gls{sm} Section S11 for details) \citep{leven2016multiwalled}, which characterizes the local stacking configurations for the optimized superlattice. As shown in \autoref{fig:moire}(c), the \gls{smlp}+\gls{ilp} model predicts a double-Moir\'e pattern: a larger \SI{14}{\nano\meter} pattern arising from the \gls{gr}/\gls{hbn} lattice mismatch, and a smaller one originating from the lattice mismatch between twisted \gls{gr}/\gls{gr} interface. This prediction closely resembles the experimental Moir\'e superlattice observed via the conductive atomic force microscopy~\citep{huang2021imaging} (\autoref{fig:moire}(d)). 

\subsection{Stacking order-dependent Moir\'e superlattices in Gr/\textit{h}-BN/\texorpdfstring{\ce{MoS2}}{MoS2} heterostructures}
Building on the successful reproduction of experimental Moir\'e superlattices in \gls{gr}/\gls{hbn} systems, we extend the \gls{smlp}+\gls{ilp} approach to predict Moir\'e patterns in more complex \gls{vdw} heterostructures composed of \gls{gr}, \gls{hbn}, and \gls{mos2}. Considering the intrinsic lattice mismatches of 1.8\% between \gls{gr} and \gls{hbn}, and 29.1\% between \gls{gr} and \gls{mos2}, we constructed \gls{gr}($111\times111$)/\gls{hbn}($109\times109$)/\gls{mos2}($86\times86$) heterojunction models (containing 423,552 atoms), exhibiting lattice mismatch in each layer reduced to below 0.05\%. To examine stacking order-dependent reconstruction behavior, we considered three configurations: \gls{gr}/\gls{mos2}/\gls{hbn}, \gls{hbn}/\gls{gr}/\gls{mos2}, and \gls{mos2}/\gls{hbn}/\gls{gr}. The initial interlayer spacing was set to the equilibrium value of the corresponding bilayers (see \gls{sm} Section S9 for details). Building on the preceding analysis of the substrate thickness, we also adopt an eight-layer substrate for the substrate-supported model in this study. Following the experimental configuration \citep{niu2025ferroelectricity}, the two ends of the supported model are further thickened by four additional substrate layers each (see \gls{sm} Section S10 for details). All trilayer heterostructure models were optimized using the same procedure as the \gls{gr}/\gls{hbn} system (see \gls{sm} Section S1 for details).

\autoref{fig:GMB}(a)-(c) show the relative out-of-plane deformations of each component monolayer in substrate-supported \gls{hbn}/\gls{gr}/\gls{mos2}, \gls{mos2}/\gls{hbn}/\gls{gr} and \gls{gr}/\gls{mos2}/\gls{hbn} heterostructures. When the \gls{gr} and \gls{hbn} are adjacent (see \autoref{fig:GMB}(a) and (b)), the deformation is dominated by their interlayer interaction, giving rise to a Moir\'e pattern with a periodicity of approximately \SI{14}{\nano\meter}. In contrast, when the \gls{mos2} layer is inserted between \gls{gr} and \gls{hbn} (see \autoref{fig:GMB}(a)), the deformation is suppressed by nearly two orders of magnitude, leading to atomically flat interfaces with negligible out-of-plane corrugation. The similar phenomenon is also found in the simulation of suspended trilayer models (see \gls{sm} Figure S21). This stacking order-dependent suppression aligns well with recent experimental observations \citep{niu2025ferroelectricity}.

To understand why different stacking orders lead to distinct Moir\'e superlattices, we calculated the weighted mean energy of each atom by summing the energies of its neighbor atoms within three nearest hexagonal rings (see \gls{sm} Section S12 for details), effectively reflecting the local stacking environments. \autoref{fig:GMB}(d)-(f) present the weighted atomic energy maps for each monolayer layer in suspended \gls{hbn}/\gls{gr}/\gls{mos2}, \gls{mos2}/\gls{hbn}/\gls{gr} and \gls{gr}/\gls{mos2}/\gls{hbn} heterostructures, respectively.

For \gls{hbn}/\gls{gr}/\gls{mos2} and \gls{mos2}/\gls{hbn}/\gls{gr} system, the energy Moir\'e patterns of \gls{gr} and \gls{hbn} are dominated by a \SI{14}{\nano\meter} hexagonal supercell. Although smaller-scale Moir\'e features appear in the layers adjacent to the \gls{mos2} layer (\gls{gr} layer in \autoref{fig:GMB}(d) and \gls{hbn} layer in \autoref{fig:GMB} (e)), their energy corrugations are substantially weaker compared to those associated with the larger \SI{14}{\nano\meter} pattern. To minimize the overall energy, energetically favorable stacking (commensurate) regions expand by compressing the incommensurate areas, thereby inducing out-of-plane deformation to alleviate the in-plane stress. In both the \gls{hbn}/\gls{gr}/\gls{mos2} and \gls{mos2}/\gls{hbn}/\gls{gr} systems, the initial energy corrugation of the \gls{mos2} monolayer is minimal, exhibiting only weak Moir\'e modulations (see \gls{sm} Figure S23). As a result, it cannot compete with the stronger interlayer interaction between \gls{gr} and \gls{hbn}, and is thus forced to conform, presenting a similar \SI{14}{\nano\meter} hexagonal Moir\'e superlattice.

In contrast, the \gls{gr}/\gls{mos2}/\gls{hbn} structure exhibits no pronounced Moir\'e pattern in the \gls{gr} and \gls{hbn} layers, as the presence of the intervening \gls{mos2} layer significantly increases their separation, weakening the interlayer interactions. As a result, during energy minimization, the high-energy stacking regions in each layer cannot contract effectively. This leads to a substantial suppression of out-of-plane deformation—nearly two orders of magnitude lower than that observed in the \gls{hbn}/\gls{gr}/\gls{mos2} and \gls{mos2}/\gls{hbn}/\gls{gr} structures.

\section{Interfacial sliding dynamics of vdW materials with edges}

\begin{figure}[htb]
\begin{center}
\includegraphics[width=\columnwidth]{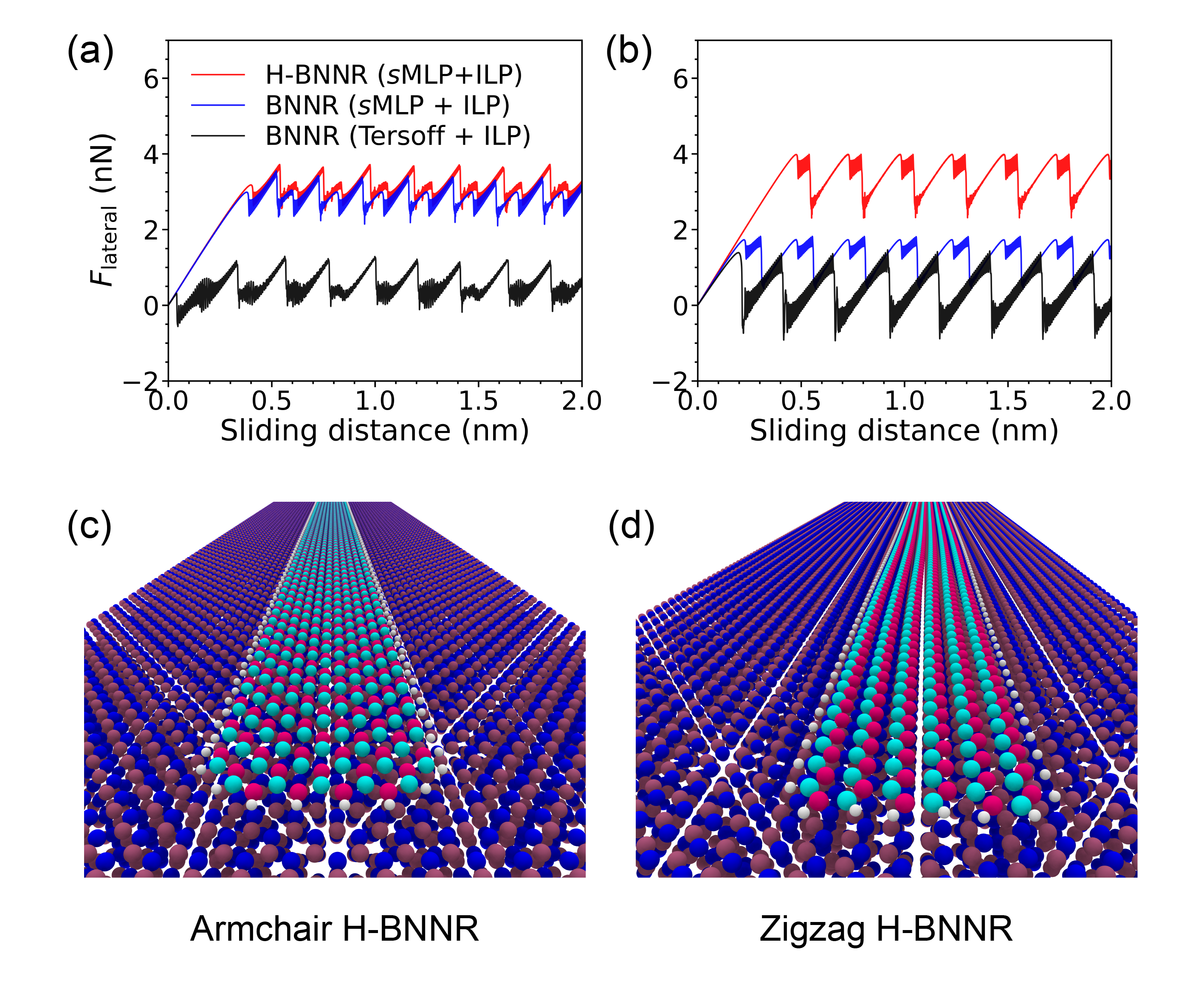}
\caption{Lateral force as a function of sliding distance for (a) armchair and (b) zigzag \glspl{bnnr} sliding on an \gls{hbn} substrate, obtained from \gls{hbnnr} (\gls{smlp}+\gls{ilp}), \gls{bnnr} (\gls{smlp}+\gls{ilp}), and \gls{bnnr} (Tersoff+\gls{ilp}) simulations. The geometries of a long armchair and zigzag \glspl{hbnnr} deposited along the armchair and zigzag axes of \gls{hbn} substrates are presented in Panels (c) and (d), respectively. Mauve, blue, pink, light blue, and white spheres represent boron (substrate), nitrogen (substrate), boron (slider), nitrogen (slider), and hydrogen atoms, respectively.}
\label{fig:hbn_friction}
\end{center}
\end{figure}

To evaluate the transferability of the \gls{smlp}+\gls{ilp} framework to tribological properties and to demonstrate the additional mechanistic insight it affords relative to existing empirical potentials and pure \glspl{mlp}, we simulated the interfacial sliding dynamics of \gls{vdw} materials with edges. Specifically, we have studied the shear dynamics of an \gls{bnnr} sliding on an \gls{hbn} substrate (\autoref{fig:hbn_friction}(c)-(d)). We have simulated both the shear dynamics of \gls{hbnnr} and hydrogen-free \gls{bnnr}. The former configuration can only be simulated with \gls{smlp}+\gls{ilp}, while the latter can be simulated both by \gls{smlp}+\gls{ilp} and Tersoff+\gls{ilp}. The system is driven at a constant sliding velocity of \SI{1}{\meter\per\second} under weak Langevin damping at effectively zero temperature, and the lateral force is obtained from the elastic force acting on the pulling spring. All simulations followed a consistent protocol (See \gls{sm} Section S14 for details).

\autoref{fig:hbn_friction} illustrates typical lateral force traces for armchair and zigzag \glspl{bnnr} sliding on an \gls{hbn} substrate. In all cases, the three force fields generate a characteristic stick--slip response with a sawtooth lateral force profile, yet the magnitude and regularity of the slips depend strongly on both edge chemistry and the choice of intralayer potential. For armchair nanoribbons, the hydrogen-passivated BNNR modeled with the \gls{smlp}+\gls{ilp} yields the largest average lateral force, \SI{3.1048\pm0.0063}{\nano\newton}, while the hydrogen-free counterpart using the same framework gives a slightly lower value of \SI{2.8557\pm0.0323}{\nano\newton}. In comparison, the hydrogen-free BNNR computed with the Tersoff+\gls{ilp} systematically underestimates the friction with an average force of \SI{0.5675\pm0.0254}{\nano\newton}. The discrepancies become even more pronounced for zigzag nanoribbons: the hydrogen-passivated BNNR with \gls{smlp}+\gls{ilp} produces the strongest stick-–slip signal with an average lateral force of \SI{3.4356\pm0.0010}{\nano\newton}, whereas the hydrogen-free counterpart within the same framework drops to \SI{1.2958\pm0.0006}{\nano\newton}, and the hydrogen-free BNNR using Tersoff+\gls{ilp} again yields to smallest lateral forces, with an average value of \SI{0.3451\pm0.0367}{\nano\newton}. This pronounced underestimation is consistent with the large force errors of the Tersoff potential at edge atoms, up to about \SI{2}{\electronvolt\per\angstrom} as quantified in the \gls{sm} Section S4, demonstrating that the employed Tersoff parametrization~\cite{lindsay2010optimized} does not provide a quantitatively reliable description of edge regions.

\begin{figure*}[htb]
\begin{center}
\includegraphics[width=1.6\columnwidth]{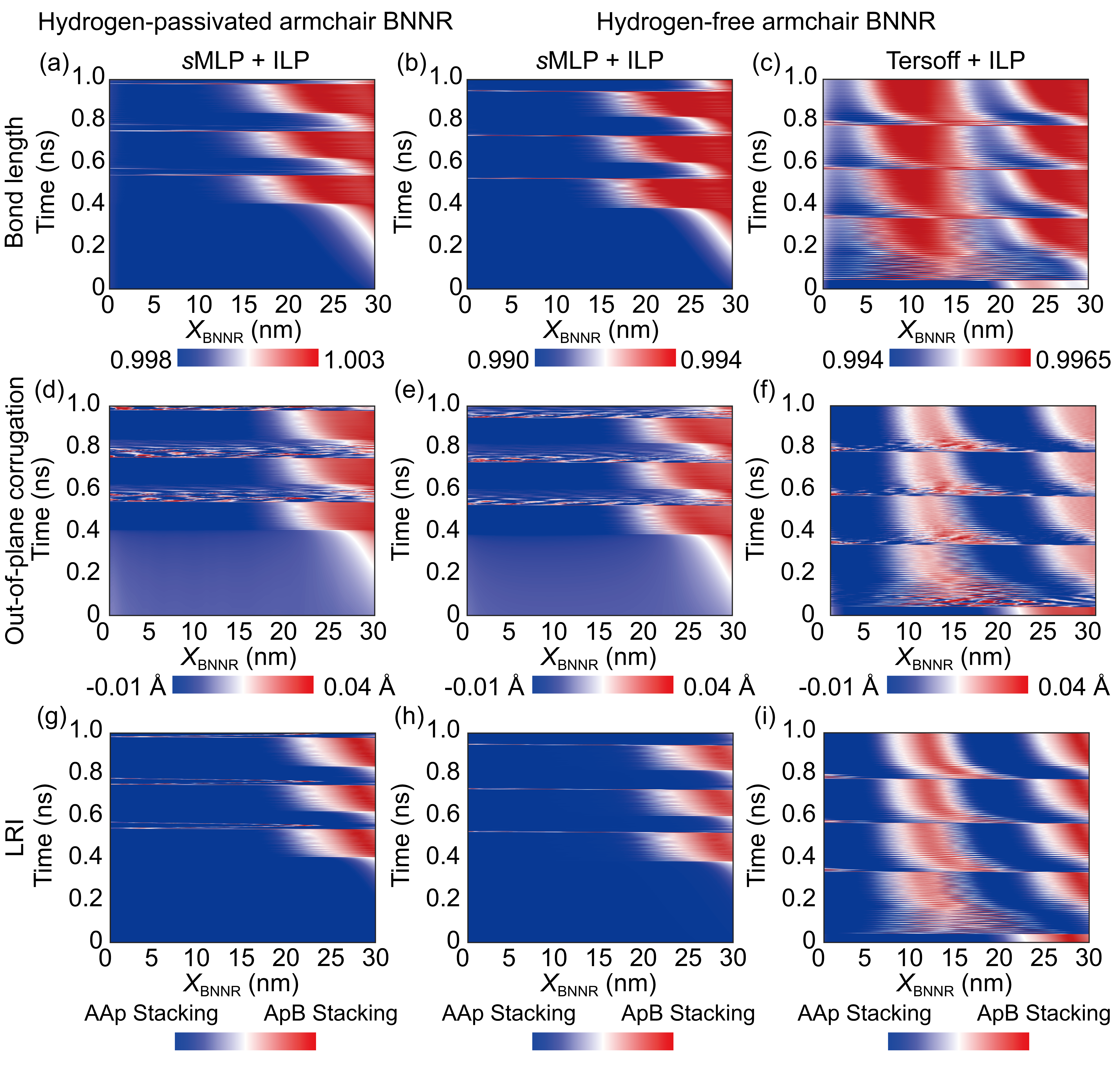}
\caption{Spatiotemporal maps of (a)–(c) bond length modulation, (d)–(f) out-of-plane corrugation, and (g)–(i) \gls{lri} for an armchair \gls{hbn} nanoribbon sliding on an \gls{hbn} substrate. Columns correspond to hydrogen-passivated armchair BNNR (\gls{smlp}+\gls{ilp}) (left), hydrogen-free armchair BNNR (\gls{smlp}+\gls{ilp}) (middle), and hydrogen-free armchair BNNR (Tersoff+\gls{ilp}) (right). The horizontal axis gives the position along the nanoribbon, $X_{\mathrm{BNNR}}$, and the vertical axis denotes simulation time.}
\label{fig:hbn_armchair_maps}
\end{center}
\end{figure*}

To clarify the microscopic origin of these trends, we analyze the spatiotemporal evolution of the bond length, out-of-plane corrugation, and \gls{lri} along the nanoribbon during sliding (\autoref{fig:hbn_armchair_maps}). The results show a consistent stick--slip mechanism across all models, yet with distinct differences in strain localization. For both hydrogen-passivated/free armchair BNNRs, adopting \gls{smlp}+\gls{ilp} force field, the bond length modulation reveals sharp slip fronts that nucleate at the driven edge and propagate coherently across the ribbon (\autoref{fig:hbn_armchair_maps}(a) and (b)). This is accompanied by simultaneous evolution on out-of-plane corrugation (\autoref{fig:hbn_armchair_maps}(d) and (e)) and a single, sharp domain wall moving smoothly in the \gls{lri} maps (\autoref{fig:hbn_armchair_maps}(g) and (h)), indicating rigid-body sliding with minimal internal distortion. In contrast, the Tersoff+\gls{ilp} model exhibits broader, more diffuse regions of bond stretching (\autoref{fig:hbn_armchair_maps}(c)) and out-of-plane undulations (\autoref{fig:hbn_armchair_maps}(f)) that extend into the ribbon interior. Notably, these deformations are accompanied by widespread fluctuations in stacking registry throughout the ribbon interior (\autoref{fig:hbn_armchair_maps}(i)), rather than distinct, uniform domains. This suggests that the Tersoff model predicts a more compliant, distributed deformation mode where the stacking order is unstable across the entire contact area, preventing the coherent collective sliding captured by the \gls{smlp} and leading to an underestimation of the lateral forces.

\begin{figure*}[htb]
\begin{center}
\includegraphics[width=1.6\columnwidth]{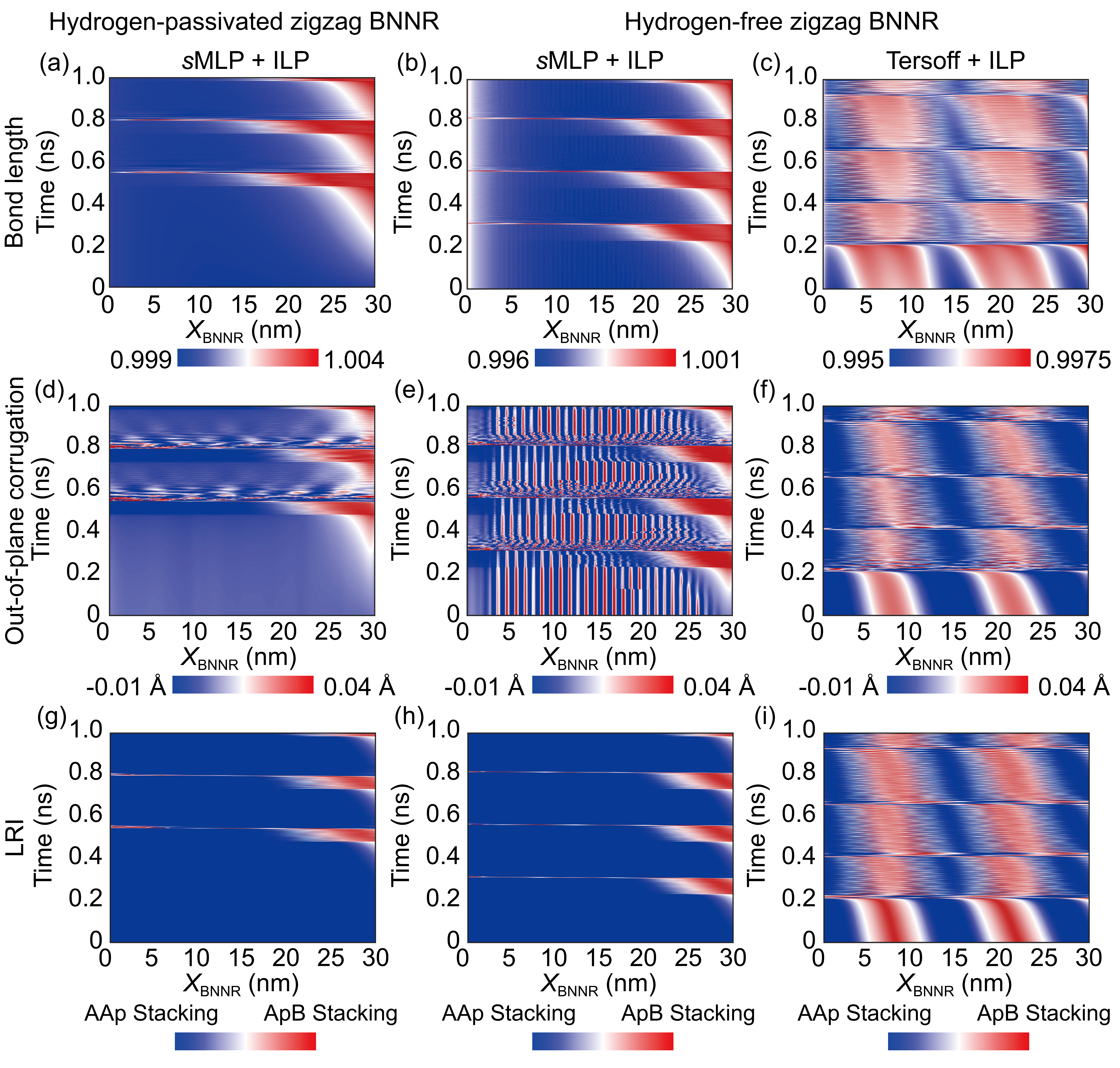}
\caption{Spatiotemporal maps of (a)–(c) bond length modulation, (d)–(f) out-of-plane corrugation, and (g)–(i) \gls{lri} for a zigzag \gls{hbn} nanoribbon sliding on an \gls{hbn} substrate. Columns correspond to hydrogen-passivated zigzag BNNR (\gls{smlp}+\gls{ilp}) (left), hydrogen-free zigzag BNNR (\gls{smlp}+\gls{ilp}) (middle), and hydrogen-free zigzag BNNR (Tersoff+\gls{ilp}) (right). Compared with the armchair case in \autoref{fig:hbn_armchair_maps}, the bare zigzag nanoribbon develops strong edge-localized buckling adopting \gls{smlp}+\gls{ilp} and pronounced mid-ribbon corrugation adopting Tersoff+\gls{ilp}, whereas hydrogen passivation for zigzag BNNR suppresses these out-of-plane distortions.}
\label{fig:hbn_zigzag_maps}
\end{center}
\end{figure*}

In contrast, zigzag nanoribbons exhibit a much more complex and edge-sensitive deformation pattern, as shown in \autoref{fig:hbn_zigzag_maps} and Figure S26 in Section S14 of the SM. For the hydrogen-free BNNR adopting \gls{smlp}+\gls{ilp} force field (middle column), the out-of-plane corrugation map (\autoref{fig:hbn_zigzag_maps}(e)) reveals that the nitrogen-terminated edge lifts from the substrate and develops high-frequency traveling buckling waves along the ribbon, whereas the boron-terminated edge remains comparatively flat. The bond-length maps (\autoref{fig:hbn_zigzag_maps}(b)) display a pronounced slip front that nucleates near the driven edge and propagates downstream, accompanied by bands of enhanced bond stretching that follow the edge buckling. The \gls{lri} field (\autoref{fig:hbn_zigzag_maps}(h)) retains a single stacking boundary that drifts smoothly with time, indicating that the local registry evolves continuously as the slip front advances. In the Tersoff+\gls{ilp} case (right column), both the bond-length (\autoref{fig:hbn_zigzag_maps}(c)) and corrugation (\autoref{fig:hbn_zigzag_maps}(f)) maps are dominated by long-wavelength, small-amplitude undulations that extend over the entire ribbon, similar to the armchair case (\autoref{fig:hbn_armchair_maps} (c) and (f)). The \gls{lri} profile varies continuously along the sliding direction without sharp domain walls, consistent with a deformation mode in which the applied shear stress is continuously redistributed through quasi-uniform corrugation rather than being stored and released in sharply localized stick--slip events. This distributed stress-relief mechanism rationalizes the systematically smaller lateral forces obtained with Tersoff+\gls{ilp} in \autoref{fig:hbn_friction}(b) and the associated underestimation of friction.

Hydrogen passivation qualitatively modifies the behavior of zigzag nanoribbons. In the hydrogen-passivated zigzag BNNR (\gls{smlp}+\gls{ilp}) simulations (left column of \autoref{fig:hbn_zigzag_maps}), saturation of the nitrogen-terminated edge strongly suppresses its tendency to buckle, leading to markedly reduced out-of-plane corrugation (\autoref{fig:hbn_zigzag_maps}(d)) and maintaining both edges in intimate contact with the substrate throughout sliding. The bond-length (\autoref{fig:hbn_zigzag_maps}(a)) and \gls{lri} (\autoref{fig:hbn_zigzag_maps}(g)) maps show that the slip front is now confined to a narrow zone near the driven edge, while the middle and trailing segments of the ribbon slide in an almost rigid-body fashion with only minor internal distortion. As a result, the effective contact area between the nanoribbon and the substrate is increased, and elastic energy can accumulate over a larger number of atoms before each slip event, giving rise to a substantial enhancement of the lateral force. This edge-stabilization effect is inherently absent in the Tersoff+\gls{ilp} model, where the inaccurate intralayer stiffness and simplified edge description favor long-wavelength corrugation and mid-ribbon stress redistribution.

Overall, these results demonstrate that the frictional response of \glspl{bnnr} is governed by an intricate coupling between interfacial registry, intralayer bending rigidity, and edge chemistry, features that are naturally resolved within the \gls{smlp}+\gls{ilp} framework. For armchair nanoribbons, where both edges remain essentially flat during sliding, hydrogen passivation exerts only a modest influence on the sliding dynamics. In strong contrast, zigzag nanoribbons are highly sensitive to edge termination: unsaturated nitrogen-terminated edges act as compliant hinges that promote stress relaxation through edge buckling that expands to the interior region of nanoribbon, thereby reducing the macroscopic friction. Hydrogen passivation stabilizes these edges, suppressing out-of-plane distortions and restoring a robust high-friction stick--slip regime. Crucially, for hydrogen-passivated zigzag BNNR, the \gls{smlp}+\gls{ilp} model captures this physical stabilization effect while avoiding the spurious, broad stress redistribution predicted by the Tersoff+\gls{ilp} potential. This further supports that the proposed \gls{smlp}+\gls{ilp} framework is well suited for quantitative simulations of complex nanotribological phenomena in \gls{vdw} materials.

\section{Conclusions}

In conclusion, we present a hybrid modeling framework, \gls{smlp}+\gls{ilp}, for simulating \gls{vdw} materials, especially complex heterostructures. In this approach, the \gls{smlp}, with a large number of parameters, is dedicated to capturing short-range intralayer interactions, while the \gls{ilp}, with minimal parameterization, efficiently describes all long-range interlayer interactions. By assigning distinct interlayer and intralayer interactions to separate potential components, this hybrid framework enhances simulation accuracy without sacrificing computational efficiency. Moreover, the \gls{smlp}+\gls{ilp} approach significantly reduces the required training dataset by at least an order of magnitude compared to fully machine-learned models, enabling faster and more accessible training for complex heterostructures. 

To realize this framework, we employed \gls{nep} as the \gls{smlp} component and enabled its seamless integration with the \gls{ilp} model in both \textsc{LAMMPS} and \textsc{GPUMD}, achieving comparable computational efficiency with empirical potentials. Notably, our framework achieves a remarkable speed of over \SI{2e6}{\atom\cdot\step\per\second} in \textsc{GPUMD} on a single NVIDIA RTX 4090 desktop GPU when simulating complex trilayer \gls{vdw} heterostructures with 423,552 atoms. Validated on monolayer \gls{gr}, \gls{hbn}, and \gls{mos2}, the trained \gls{smlp} models yield force \glspl{rmse} an order of magnitude lower than empirical potentials. Edge energy errors for monolayer \gls{gr} and \gls{hbn} are reduced to \SI{0.01}{\electronvolt\per\atom}, compared to \qtyrange{0.1}{1}{\electronvolt\per\atom} for \gls{rebo} and Tersoff. The \gls{smlp}+\gls{ilp} framework also reproduces the phonon spectra, mechanical, and thermal transport properties of graphite and bulk $h$-BN in close agreement with \gls{dft} and experimental results, confirming its accuracy, transferability, and efficiency for simulating complex \gls{vdw} heterostructures.

Finally, we applied the \gls{smlp}+\gls{ilp} framework to investigate the emergence of Moir\'e superlattices in complex \gls{vdw} heterostructures composed of \gls{gr}, \gls{hbn}, and \gls{mos2} monolayers. For both bilayer \gls{gr}/\gls{hbn} and twisted trilayer \gls{gr}/\gls{gr}/\gls{hbn} systems, the framework successfully reproduces Moir\'e patterns in excellent agreement with experimental observations. Building on this, we further explored how stacking order affects structural reconstruction in trilayer \gls{gr}/\gls{hbn}/\gls{mos2} systems with and without substrates, where interlayer interactions between \gls{gr} and \gls{hbn} layers play a key role. For the models with substrates, when \gls{gr} and \gls{hbn} layers are adjacent, the system exhibits pronounced out-of-plane hexagonal corrugation of approximately \SI{0.4}{\angstrom}, preserving the characteristic \SI{14}{\nano\meter} Moir\'e periodicity seen in bilayer \gls{gr}/\gls{hbn}, and for the models without substrates, the deformation amplitude is about \SI{3}{\angstrom}. In contrast, when the \gls{mos2} layer is inserted between \gls{gr} and \gls{hbn}, regardless of whether a substrate is present, their interaction becomes effectively decoupled, resulting in a significant suppression of deformation to less than \SI{0.1}{\angstrom}, and the Moir\'e pattern disappears. As a complementary nanotribological application, we employed the same \gls{smlp}+\gls{ilp} framework to study \gls{hbn} nanoribbons sliding on \gls{hbn} substrates. The simulations reveal distinct frictional features that can not be captured by the conventional Tersoff+ILP model.

These structural-reconstruction and nanotribology case studies demonstrate both the extensibility and the necessity of the proposed \gls{smlp}+\gls{ilp} framework. Reliably describing the coupled mechanical, thermal and frictional phenomena that emerge in realistic, large‑scale \gls{vdw} heterostructures requires a model that can simultaneously resolve intralayer elasticity, interfacial registry, and the detailed atomic structure of free edges and buried interfaces. The \gls{smlp}+\gls{ilp} approach provides precisely this integrated description.

\vspace{0.5cm} 
\noindent{\textbf{CRediT authorship contribution statement}}

\textbf{Hekai Bu:} Software, Formal analysis, Methodology, Investigation, Writing - original draft. \textbf{Wenwu Jiang:} Formal analysis, Investigation, Writing - original draft. \textbf{Penghua Ying:} Conceptualization, Methodology, Formal analysis, Validation, Investigation, Writing - review \& editing. \textbf{Ting Liang:} Investigation, Writing - review \& editing, Formal analysis. \textbf{Zheyong Fan:} Software, Supervision, Writing - review \& editing. \textbf{Wengen Ouyang:} Conceptualization, Methodology, Supervision, Resources, Funding acquisition, Project administration, Validation, Writing - review \& editing.

\vspace{0.5cm} 
\noindent{\textbf{Declaration of competing interest}}

W.O., W.J. and H.B. are inventors on a Chinese patent covering the methodology of the hybrid \gls{smlp}+\gls{ilp} framework (No. ZL 202510197024.9; assigned to Wuhan University). The authors declare no other competing interests.

\vspace{0.5cm} 
\noindent{\textbf{Acknowledgments}}

We thank Oded Hod and Michael Urbakh for many inspiring discussions on this topic. W.O. acknowledges the National Natural Science Foundation of China (12472099, U2441207), the Fundamental Research Funds for the Central Universities (No. 2042025kf0050 and 2042025kf0013), the start-up fund of Wuhan University (No. 600460100), and the GHfund B (202407023964). P.Y. is supported by the Israel Academy of Sciences and Humanities \& Council for Higher Education Excellence Fellowship Program for International Postdoctoral Researchers. T.L. gratefully acknowledges the discussions and support from Prof. Jianbin Xu, and further acknowledges support from the RGC GRF (No. 14220022) and the CUHK Ph.D. Studentship. Computations were conducted at the Supercomputing Center of Wuhan University, the National Supercomputer TianHe-1(A) Center in Tianjin and Computing Center in Xi’an.

\vspace{0.5cm} 
\noindent{\textbf{Data availability}}

Complete input and output files for the \gls{nep} training of monolayer \gls{gr}, \gls{hbn} and \gls{mos2} are freely available at \url{https://github.com/ouyang-laboratory/nep_data}. In \textsc{LAMMPS}, documentation for \gls{ilp} is available at \url{https://docs.lammps.org/pair_ilp_graphene_hbn.html} and \url{https://docs.lammps.org/pair_ilp_tmd.html}, respectively. The \gls{nep} interface to \textsc{LAMMPS} is available at \url{https://github.com/brucefan1983/NEP_CPU}. In \textsc{GPUMD}, documentation for the hybrid \gls{nep}+\gls{ilp} potential is available at \url{https://gpumd.org/potentials/nep\_ilp.html}. The hybrid \gls{nep}+\gls{ilp} potential is currently available in \textsc{GPUMD}-v4.0 (\url{https://github.com/brucefan1983/GPUMD}) and later versions.

\bibliographystyle{myaps}
\bibliography{refs}
\end{document}


\maketitle

\tableofcontents{}

\clearpage
\newpage

\section{\normalsize 1. Method}
\label{section:method}

\subsection{\normalsize 1.1. DFT calculations}
\label{subsection:dft}
All \gls{dft} calculations for the \gls{smlp} model and edge energy were performed using the Vienna Ab initio Simulation Package \citep{Kresse1996Efficient,kresse1999from} with the \gls{pbe} \citep{perdew1996generalized} exchange--correlation functional within the projector augmented wave \citep{blochl1994projector} framework. For the electronic self-consistent loop, we set a threshold of \SI{1e-7}{\electronvolt} with an energy cutoff of \SI{850}{\electronvolt}.  A Gaussian smearing width of \SI{0.1}{\electronvolt} was employed. A sufficiently large vacuum spacing for all monolayer systems was applied along the out-of-plane direction to eliminate spurious interactions between periodic images. Due to differences in the supercell sizes, distinct \textit{k}-point sampling schemes were employed. For the \gls{gr} and \gls{hbn} systems, a $3 \times 3 \times 1$ Monkhorst--Pack grid was adopted. In contrast, for the \gls{mos2} system, the Brillouin zone was sampled using a $\Gamma$-centered mesh with a reciprocal-space resolution of \SI{0.15}{\per\angstrom}.

\subsection{\normalsize 1.2. MD simulation details}
\label{subsection:md}

MD simulations for mechanical property evaluation were performed using the \textsc{LAMMPS} package \citep{thompson2022lammps}. To determine the bulk modulus of the graphite system, simulations were conducted with periodic boundary conditions in all three spatial directions, and the equations of motion were integrated with a time step of \SI{1}{\femto\second}. The system temperature was set at \SI{300}{\kelvin} using a Nos\'e--Hoover thermostat with a relaxation time of \SI{0.25}{\pico\second}. For simulations under hydrostatic pressure, a Nos\'e--Hoover barostat \citep{shinoda2004rapid,connor2015airebom} with a damping constant of \SI{1.0}{\pico\second} was employed. To construct the $P$--$V$ curves, simulations were conducted in the NPT ensemble at \SI{300}{\kelvin} under varying target pressures. Each system was equilibrated for \SI{100}{\pico\second}, followed by a \SI{100}{\pico\second} production run during which the average system volume was recorded. This procedure was repeated for hydrostatic pressures ranging from 0 to \SI{14}{\giga\pascal} to estimate the bulk modulus. All atomic visualizations were generated using the OVITO package~\cite{stukowski2009visualization}.

Uniaxial tensile simulations were performed by incrementally deforming the simulation box along either the armchair or zigzag direction to evaluate the Young’s modulus and Poisson’s ratio of the graphite system, while maintaining zero stress along the transverse direction using the NPT ensemble at \SI{300}{\kelvin}.  The strain was applied in 0.02\% increments with a time step of \SI{1}{\femto\second}. In each increment, the system was uniaxially stretched at a constant strain rate of \SI{1e-4}{\per\pico\second} to reach the target strain, followed by a \SI{0.5}{\nano\second} relaxation stage in the NPT ensemble to ensure stress equilibration in the transverse direction. Strain and stress were recorded and statistically averaged over the final \SI{0.2}{\nano\second} of each relaxation stage.

\subsection{\normalsize 1.3. Structural optimization}
\label{subsection:optimization}

All structural optimizations for \gls{vdw} heterostructures were performed in multiple steps to ensure structural relaxation and the release of residual stress in \textsc{LAMMPS}. Initially, the atomic positions and the simulation box dimensions were optimized using the \gls{cg} \citep{schwartz1997family} and FIRE algorithms \citep{bitzek2006structural}. To achieve the most optimized structures, we performed 10 iterations of the three algorithms: \gls{cg} with box relaxation \citep{parrinello1981polymorphic}, \gls{cg} with fixed box, FIRE with fixed box. In each loop, the relaxation procedure was terminated when the forces acting on each degree of freedom reduce below \SI{1e-6}{\electronvolt\per\angstrom} or the optimizing steps reach 100,000. To verify that the system does not get trapped in a local minimum, the whole system is heated to \SI{300}{\kelvin} during 100,000 time steps using the Nos\'e--Hoover thermostat with a time-step of \SI{0.25}{\femto\second}, followed by thermal equilibration at \SI{300}{\kelvin} and zero pressure for another 100,000 steps with a time-step of \SI{1}{\femto\second}, where the pressure was controlled by the Nos\'e--Hoover barostat. After these steps, the system is further optimized using the FIRE algorithm, followed by 10 more cycles of box relaxation using the \gls{cg} algorithm. Finally, the system is optimized again using the FIRE algorithm. The optimization procedure described above ensures full relaxation of residual stress in the considered structures.

\newpage

\section{\normalsize 2. Comparison of the minimum number of configurations required for training pure MLP and \textit{s}MLP+ILP approaches}
Directly comparing the number of configurations required by different approaches is challenging, as each method and \gls{mlp} variant has distinct dataset requirements depending on its intended application. In this analysis, we focus on the scenario of training a comprehensive model capable of simulating monolayer, multilayer, and bulk structures for \gls{vdw} heterostructures composed of up to $n$ components. 

\begin{figure}[htb]
\begin{center}
\includegraphics[width=0.9\columnwidth]{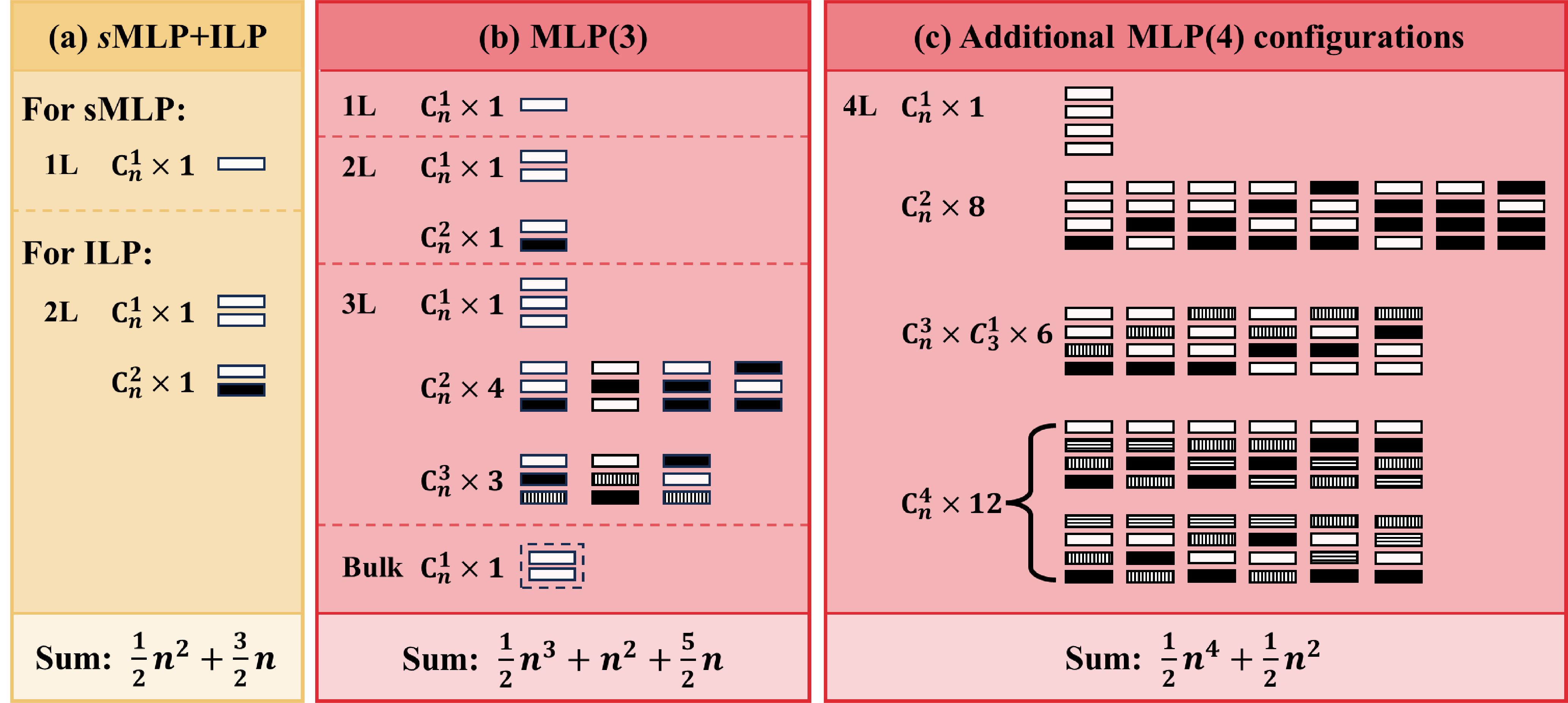}
\caption{The comparison of the minimum training configuration numbers between (a) \gls{smlp}+\gls{ilp} and (b)-(c) pure \gls{mlp} approaches for \gls{vdw} heterostructures composed of up to $n=3$ (b) and $n=4$ (c) components. Panel (b) ($\mathrm{MLP(3)}$) and panel (c) ($\mathrm{MLP(4)}$) demonstrate that as the number of layers required for accurate modeling increases, the dataset size grows substantially, highlighting the efficiency advantage of the modular \gls{smlp}+\gls{ilp} approach.
}
\label{sfig:configurations}
\end{center}
\end{figure}

For a pure \gls{mlp} approach, it is generally necessary to include at least three layers in the training set to capture both interfacial and surface environments. The resulting minimum number of configurations required is illustrated in \autoref{sfig:configurations}(b). The dataset for pure \gls{mlp} should at least include:
\begin{itemize}
  \item Monolayer configurations for each component: \( \mathrm{C}_n^1 \),
  \item Bilayer homo- and heterostructures: \( \mathrm{C}_n^1 + \mathrm{C}_n^2 \),
  \item Trilayer configurations:
  \begin{itemize}
    \item Homogeneous: \( \mathrm{C}_n^1 \),
    \item Heterogeneous (two components): \( 4 \times \mathrm{C}_n^2 \),
    \item Heterogeneous (three components): \( 3 \times \mathrm{C}_n^3 \),
  \end{itemize}
  \item Homogeneous bulk structures: \( \mathrm{C}_n^1 \),
\end{itemize}
leading to a total of
\[
\frac{1}{2}n^3 + n^2 + \frac{5}{2}n
\]
configurations for up to three-layer systems.

If up to four layers are required, additional configurations are needed, as shown in \autoref{sfig:configurations}(c):
\begin{itemize}
  \item Homogeneous four-layer structures: \( \mathrm{C}_n^1 \),
  \item Heterostructures with two components: \( 8 \times \mathrm{C}_n^2 \),
  \item Heterostructures with three components (with one component duplicated in two layers): \( \mathrm C_3^1\times6 \times \mathrm{C}_n^3 \),
  \item Heterostructures with four components: \( 12 \times \mathrm{C}_n^4 \),
\end{itemize}
resulting in an additional
\[
\frac{1}{2}n^4 + \frac{1}{2}n^2
\]
training configurations.

In contrast, the \gls{smlp}+\gls{ilp} approach requires significantly fewer configurations. Only the following are needed:
\begin{itemize}
  \item Monolayer configurations for \gls{smlp}: \( \mathrm{C}_n^1 \),
  \item Bilayer homo- and heterostructures for \gls{ilp}: \( \mathrm{C}_n^1 + \mathrm{C}_n^2 \),
\end{itemize}
yielding a total of
\[
\frac{1}{2}n^2 + \frac{3}{2}n
\]
configurations, as illustrated in \autoref{sfig:configurations}(a).

These comparisons highlight that the \gls{smlp}+\gls{ilp} approach reduces the number of required training configurations by at least an order of magnitude, significantly alleviating the burden of dataset construction for complex \gls{vdw} structures.

\newpage
\section{\normalsize 3. Reference structures for \textit{s}MLP training}
\begin{figure}[h!]
\begin{center}
\includegraphics[width=0.4\columnwidth]{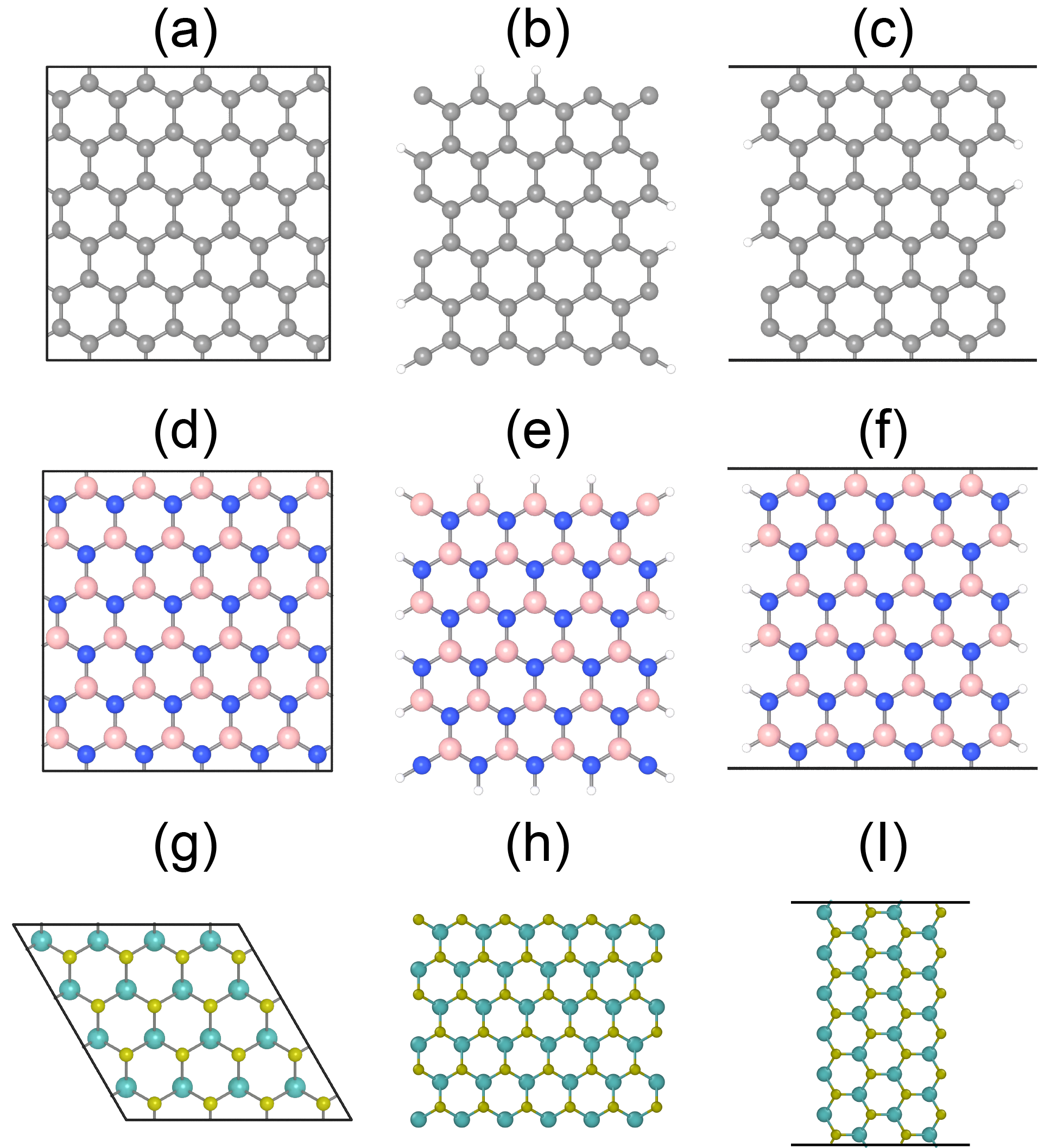}
\caption{Configurations of (a)-(c) monolayer \gls{gr}, (d)-(f) \gls{hbn} and (g)-(i) \gls{mos2}. (a), (d) and (g) are monolayers under periodic boundary conditions. (b), (e) and (h) are monolayers with edges along both armchair and zigzag directions. (c), (f) and (i) are the nanoribbons with edges along armchair or zigzag directions.}
\label{sfig:dataset}
\end{center}
\end{figure}

To construct the \gls{smlp} reference datasets, we first prepared initial atomic structures for monolayer \gls{gr}, \gls{hbn}, and \gls{mos2}. For \gls{gr}, two categories of configurations were considered, illustrated in \autoref{sfig:dataset}(a)–(c). The first includes periodic monolayer structures in a $5\times6\times1$ supercell. The second consists of non-periodic systems representing six distinct edge configurations: square and rectangular flakes containing 54 or 60 atoms, and nanoribbons with either armchair or zigzag terminations. Hydrogen atoms were introduced to passivate edge dangling bonds, with coverage ratios ranging from 0\% to 100\% in 10\% increments. The dataset for \gls{hbn} and \gls{mos2} mirrors that of \gls{gr}, encompassing periodic and non-periodic structures, both square and triangular, as shown in \autoref{sfig:dataset}(d)–(i).

Building upon these initial configurations, we generated diverse training datasets through thermal and structural perturbation sampling. For \gls{gr}, 100 periodic structures were obtained via \gls{aimd} simulations over the temperature range of \SI{100}{\kelvin} to \SI{1500}{\kelvin}, and 50 additional structures were created by applying uniform random strain ranging from −3\% to 3\%. For the non-periodic hydrogenated systems, we employed \gls{md} simulations using the MACE-MP-0 model \citep{batatia2024foundation} to sample configurations across the same temperature range. In total, 480 structures were obtained for \gls{gr}, with 384 (80\%) used for training and 96 (20\%) reserved for testing. The same protocol was applied to generate the reference datasets for \gls{hbn}, using \gls{dft}-driven molecular dynamics. For \gls{mos2}, dynamic sampling was performed using a pre-trained \gls{nep} model \citep{fredrik2023tuning}. All sampled configurations, including both thermally equilibrated and randomly strained structures, were evaluated by single-point \gls{dft} calculations to extract energies, atomic forces, and virials, which serve as training targets for the corresponding \glspl{mlp}.

\newpage
\section{\normalsize 4. \textit{s}MLP training setup and results}
We employed the \gls{nep} framework using the \texttt{nep} executable implemented in the \textsc{GPUMD} package \cite{fan2017efficient} to train the \gls{smlp} models. Based on the previously developed \gls{nep} model for bilayer \gls{gr} systems \cite{ying2024combining} and our extensive tests, we adopted the following hyperparameter settings in the \texttt{nep.in} file for the \gls{mlp} of \gls{gr}.

\begin{verbatim}
type		2 C H
version		3        
cutoff		5 5      
n_max		8 8      
basis_size	12 12   
l_max		4 2 0    
neuron		50       
lambda_1	0.05     
lambda_2	0.05   
lambda_e	1.0      
lambda_f	1.0    
lambda_v	0.1   
batch		10000   
population	50       
generation	1000000
\end{verbatim}

\begin{figure}[h!]
\begin{center}
\includegraphics[width=0.8\columnwidth]{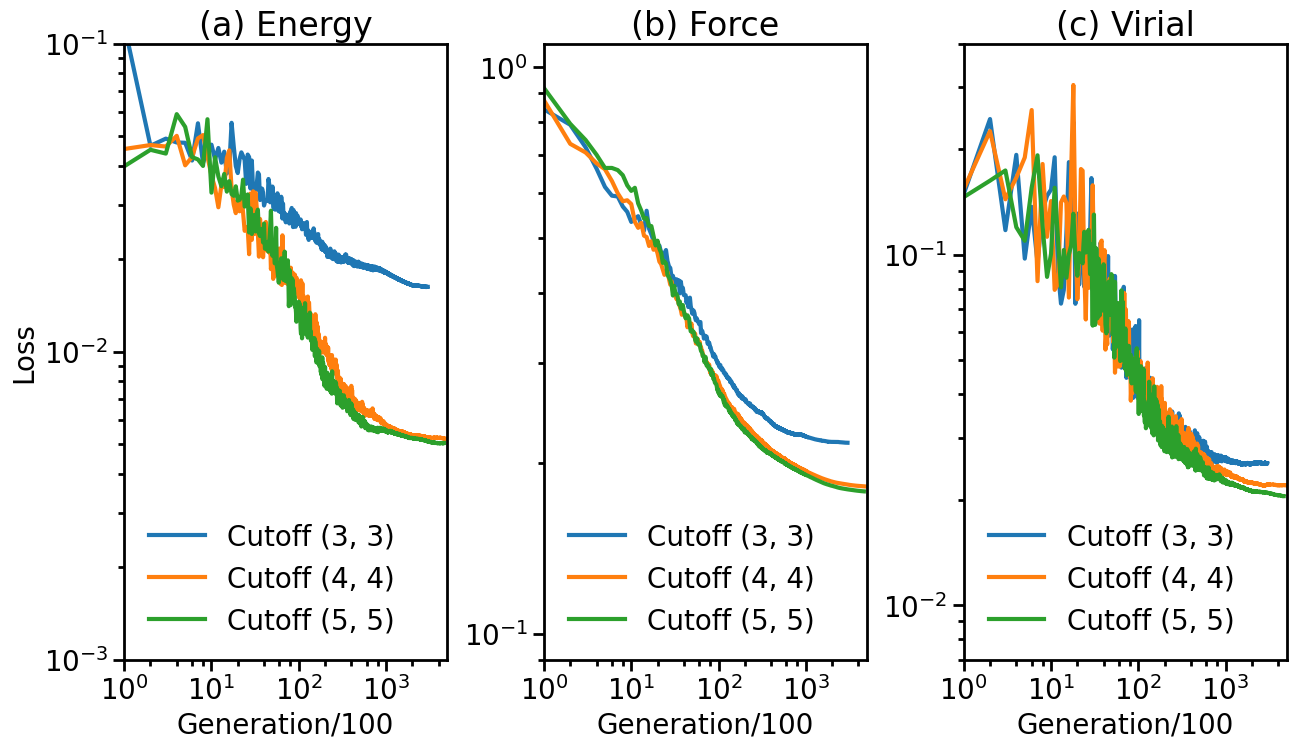}
\caption{Evolution of loss functions of \gls{smlp} models with different cutoffs for \gls{gr} monolayers during training processes.}
\label{sfig:rcut_compare}
\end{center}
\end{figure}
Specifically, the third-generation \gls{nep} framework \cite{fan2022gpumd} was used, with both radial and angular descriptor cutoffs set to \SI{5}{\angstrom}. The numbers of radial and angular descriptor components were set to 8, each employing 12 basis functions. Many-body interactions up to fourth order were included for the angular descriptors. The hidden layer of the \gls{nn} consisted of 50 neurons. Parameter optimization was carried out using the \gls{snes} algorithm with a population size of 50 to minimize the loss function. The weights for both the $\mathcal{L}_1$- and $\mathcal{L}_2$-norm regularization terms were set to 0.05. The \gls{rmse} weights for energy, force, and virial were set to 1.0, 1.0, and 0.1, respectively. For further details, refer to the \textsc{GPUMD} manual available at \hyperlink{https://gpumd.org/}{https://gpumd.org/}.

For the \gls{smlp} models of \gls{hbn} and \gls{mos2}, we used the same input hyperparameters as for \gls{gr}, except that the fourth-generation \gls{nep} framework \cite{song2024generalpurpose} was employed to improve accuracy for multi-element systems. To adapt the \texttt{nep.in} files, only the number of atom types, the list of chemical species, and the version number (set to 4) need to be modified accordingly.

\begin{figure}[h!]
\begin{center}
\includegraphics[width=0.9\columnwidth]{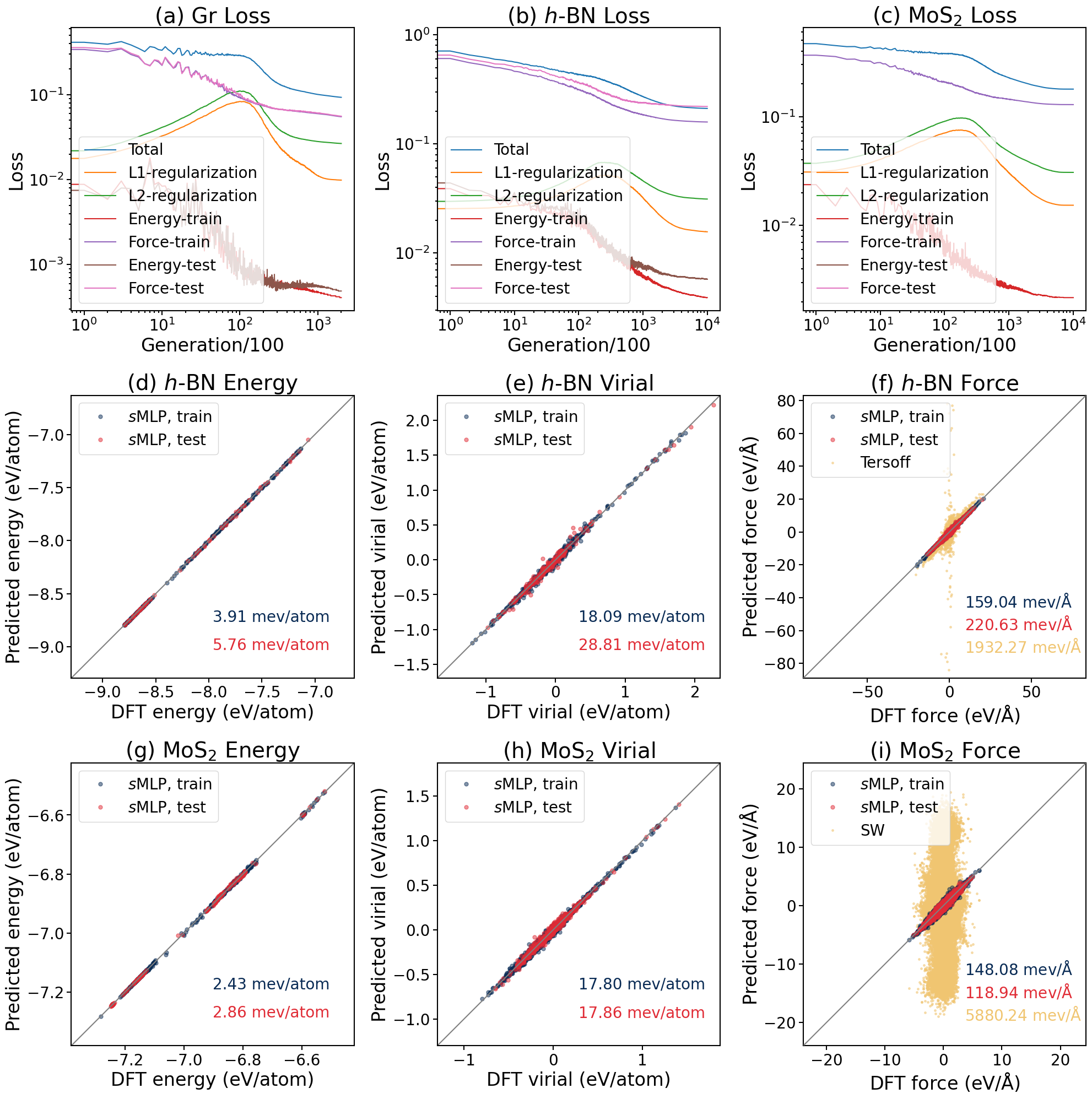}
\caption{Training performance of the \glspl{smlp} and comparison with empirical force fields. Panels (a)–(c) show the evolution of the loss components during training of the \gls{smlp} models for \gls{gr}, \gls{hbn}, and \gls{mos2}, respectively. Panels (d)–(f) present parity plots of energy, virial, and forces for \gls{hbn}, comparing \gls{smlp} predictions with \gls{dft} reference data for both the training and test sets; the diagonal lines indicate perfect agreement. Panels (g)–(i) show the corresponding parity plots for \gls{mos2}. In the force panels (f) and (i), additional markers report predictions from the empirical Tersoff potential for \gls{hbn} and the \gls{sw} potential for \gls{mos2}, respectively, evaluated on the same \gls{dft} test configurations, highlighting the accuracy gain achieved by the \glspl{smlp} over conventional empirical force fields.}
\label{sfig:train_result}
\end{center}
\end{figure}

As shown in \autoref{sfig:rcut_compare}(a)–(c), a cutoff of \SI{5}{\angstrom} was selected, as it yields higher accuracy compared to the \SI{3}{\angstrom} and \SI{4}{\angstrom} counterparts. The converged loss function values, specifically the \gls{rmse} of energy, force, and virial, are the lowest among all three \gls{smlp} models at this cutoff. In the practical training process, one million generations were used for each \gls{smlp}, allowing all loss functions to fully converge (see \autoref{sfig:train_result}(a)–(c)). \autoref{sfig:train_result}(d)–(i) demonstrate the high accuracy of the \gls{smlp} models for \gls{hbn} and \gls{mos2}, with energy \gls{rmse} values below \SI{6.0}{\milli\electronvolt\per\atom} and \SI{2.2}{\milli\electronvolt\per\atom}, respectively. As seen in panels (f) and (i), the force \gls{rmse} values of the \gls{smlp} models are more than one order of magnitude smaller than those of the Tersoff~\cite{lindsay2010optimized} and \gls{sw}~\cite{jiang2017parameterization} potentials evaluated on the same \gls{dft} train and test configurations, highlighting the clear advantage of the \gls{smlp} approach over conventional empirical force fields.

\newpage

\newpage

\section{\normalsize 5. ILP model for $h$-BN and the contribution of Coulomb interactions}
\subsection{\normalsize 5.1. ILP Potential parameters}
In this section, we extend this conclusion to \gls{hbn}. Specifically, we fitted a new \gls{ilp} model without the Coulomb term to \gls{hbn} following the same procedure as in Ref.~\cite{ouyang2018nanoserpents}. The \gls{dft} reference dataset, binding energy curves and potential energy surfaces from Ref.~\cite{ouyang2018nanoserpents}, includes three periodic structures (\gls{gr}/\gls{gr}, \gls{gr}/\gls{hbn} and \gls{hbn}/\gls{hbn}) and 10 finite structures (Benzene dimer, Borazine dimer, B$_{12}$N$_{12}$H$_{12}$ dimer, Coronene dimer, Benzene/Coronene, Borazine/B$_{12}$N$_{12}$H$_{12}$, Benzene/Borazine, Benzene/B$_{12}$N$_{12}$H$_{12}$, Borazine/Coronene and
Coronene/B$_{12}$N$_{12}$H$_{12}$), all at the Heyd–Scuseria–Ernzerhof hybrid functional in combination with the many-body dispersion method (HSE+MBD) level \cite{hermann2020density}. The parameters of the updated \gls{ilp} model are listed in \autoref{stab:bn ilp para}.

\begin{table*}[htb]
\caption{List of \gls{ilp} parameter values for \gls{hbn} based systems. The training set includes the binding energy curves and the sliding potential surfaces from Ref.~\cite{ouyang2018nanoserpents}. A value of $R_{\mathrm{cut}}=$ \SI{16}{\angstrom} is used throughout.}
\centering
\label{stab:bn ilp para}
\small
\begin{tabular}{ccccccccccc}
\hline
 & $\beta_{ij}$ & $\alpha_{ij}$ & $\gamma_{ij}$ & $\varepsilon_{ij}$ & $C_{ij}$ & $d_{ij}$ & $s_{\mathrm{R}, ij}$ & $r_{\mathrm{eff}, ij}$ & $C_{6, ij}$ \\
 & (\si{\angstrom}) &  & (\si{\angstrom}) & (\si{\milli\electronvolt}) & (\si{\milli\electronvolt}) &  &  & (\si{\angstrom}) & (\si{\electronvolt\cdot\angstrom\tothe{6}}) \\
\hline
B-B& 3.0337& 9.8083& 3.6790& 3.7641& 10.4505& 	14.9914& 	0.8408& 	3.7579& 	49.4922\\
N-N& 3.4016& 7.0667& 1.7401& 7.7003& 48.0262& 	14.9968& 	0.7828& 	3.3234& 	14.8101\\
H-H& 3.8077& 7.1711& 8.4620& 1.0787& 0.0165& 15.0419& 0.7219& 2.6903& 1.6144\\
B-N& 2.9863& 11.5725& 	4.9436& 	2.2178& 	4.6667& 	15.0014& 	0.7418& 	2.8354& 	24.6684\\
B-H& 3.5373& 14.6279& 	1.5515& 	0.0081& 	0.9719& 	15.4607& 	0.8173& 	3.6332& 	7.983\\
N-H& 2.7168& 11.4101& 	7.4659& 	0.0348& 	0.0172& 	15.0855& 	0.7148& 	2.6589& 	3.8652\\
\hline
\end{tabular}
\end{table*}

\begin{figure}[H]
\begin{center}
\includegraphics[width=1\columnwidth]{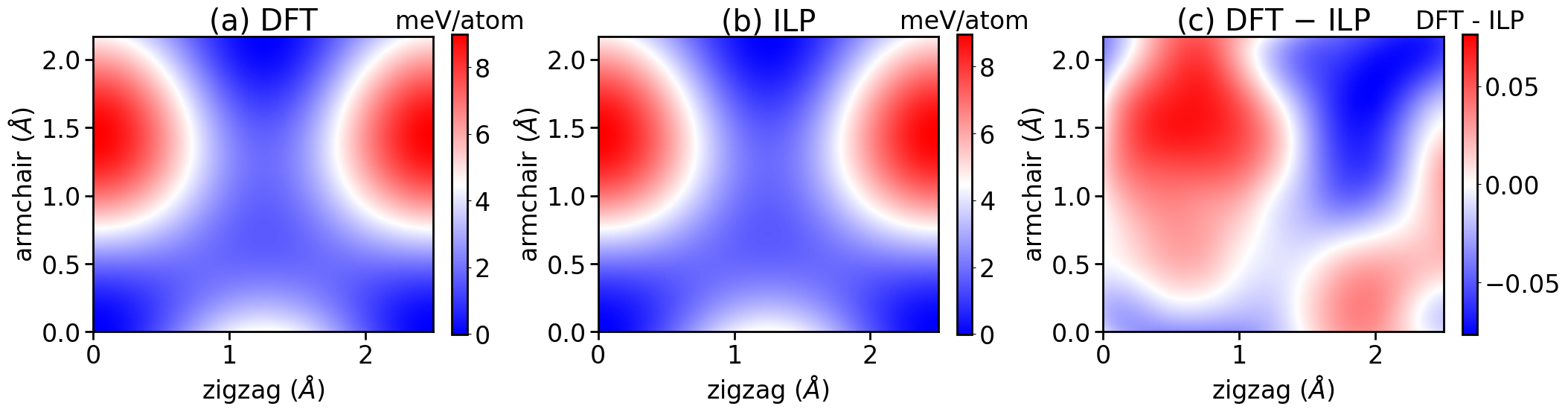}
\caption{Sliding \glspl{pes} of \gls{hbn} bilayer, calculated via rigid interlayer shifts at an interlayer distance of \SI{3.3}{\angstrom}. Panels (a)-(b) present the \gls{dft} (HSE+MBD-NL), \gls{ilp} \glspl{pes}, and their difference map, respectively. The parameters presented in \autoref{stab:bn ilp para} above are used for the \gls{ilp} calculations.
}
\label{sfig:bn AAp pes}
\end{center}
\end{figure}

\begin{figure}[hbt]
\begin{center}
\includegraphics[width=0.4\columnwidth]{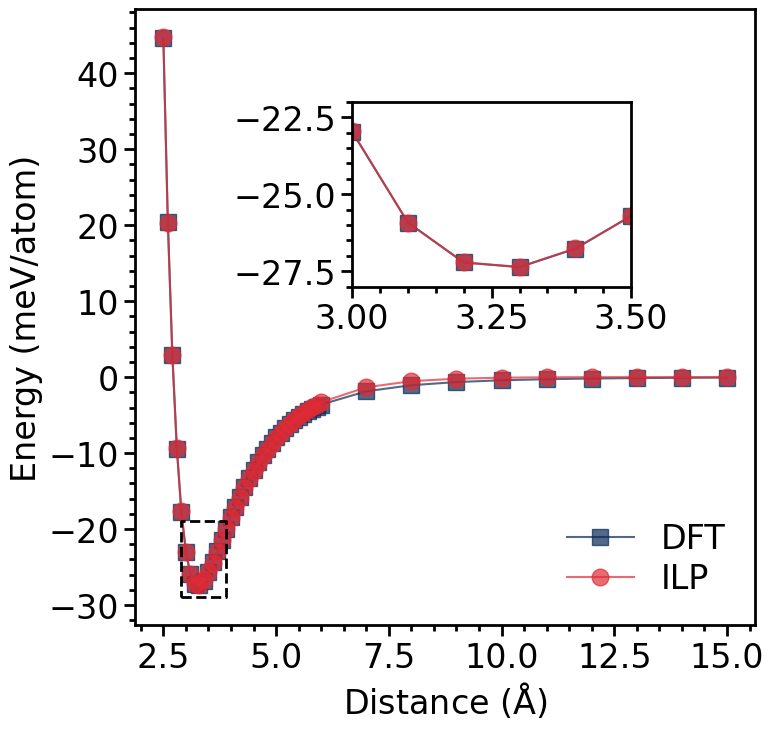}
\caption{Binding energy curves of bilayer \gls{hbn} calculated using \gls{dft}, \gls{ilp} with parameters of \autoref{stab:bn ilp para}.
The inset provides a zoom-in on the equilibrium interlayer distance region.}

\label{sfig:bn AAp be}
\end{center}
\end{figure}

\begin{figure}[H]
\begin{center}
\includegraphics[width=\columnwidth]{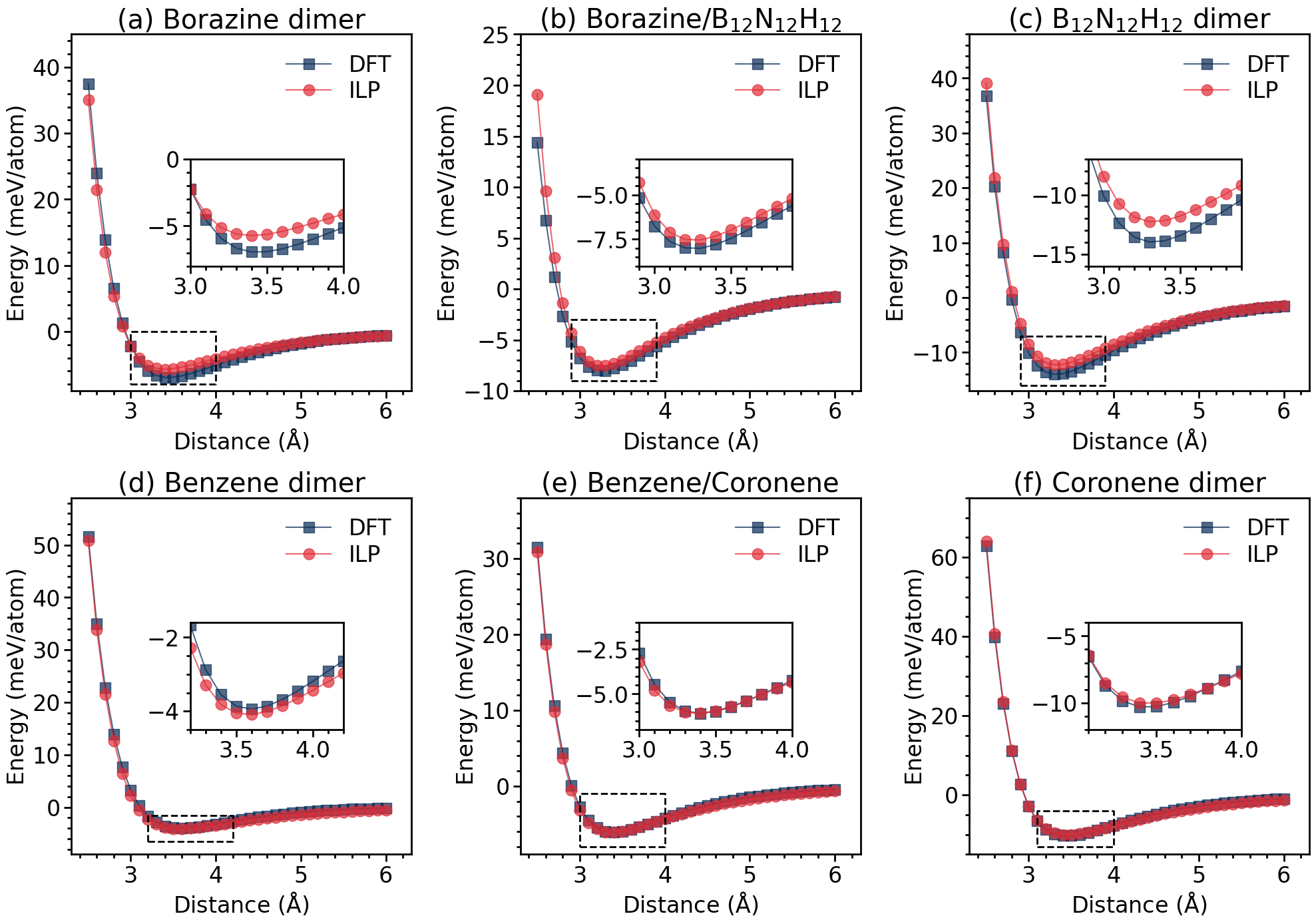}
\caption{Binding energy curves calculated for the finite homogeneous dimers of (a) Borazine, (b) Borazine/B$_{12}$N$_{12}$H$_{12}$, (c) B$_{12}$N$_{12}$H$_{12}$, (d) Benzene, (e) Benzene/Coronene, and (f) Coronene. The reported energies are measured relative to the infinitely separated dimer value and are normalized by the total number of atoms per unit-cell. The insets provide a zoom-in on the equilibrium interlayer separation region. Here, the parameters presented in \autoref{stab:bn ilp para} are used.
}
\label{sfig:bn ilp1}
\end{center}
\end{figure}

\begin{figure}[H]
\begin{center}
\includegraphics[width=0.67\columnwidth]{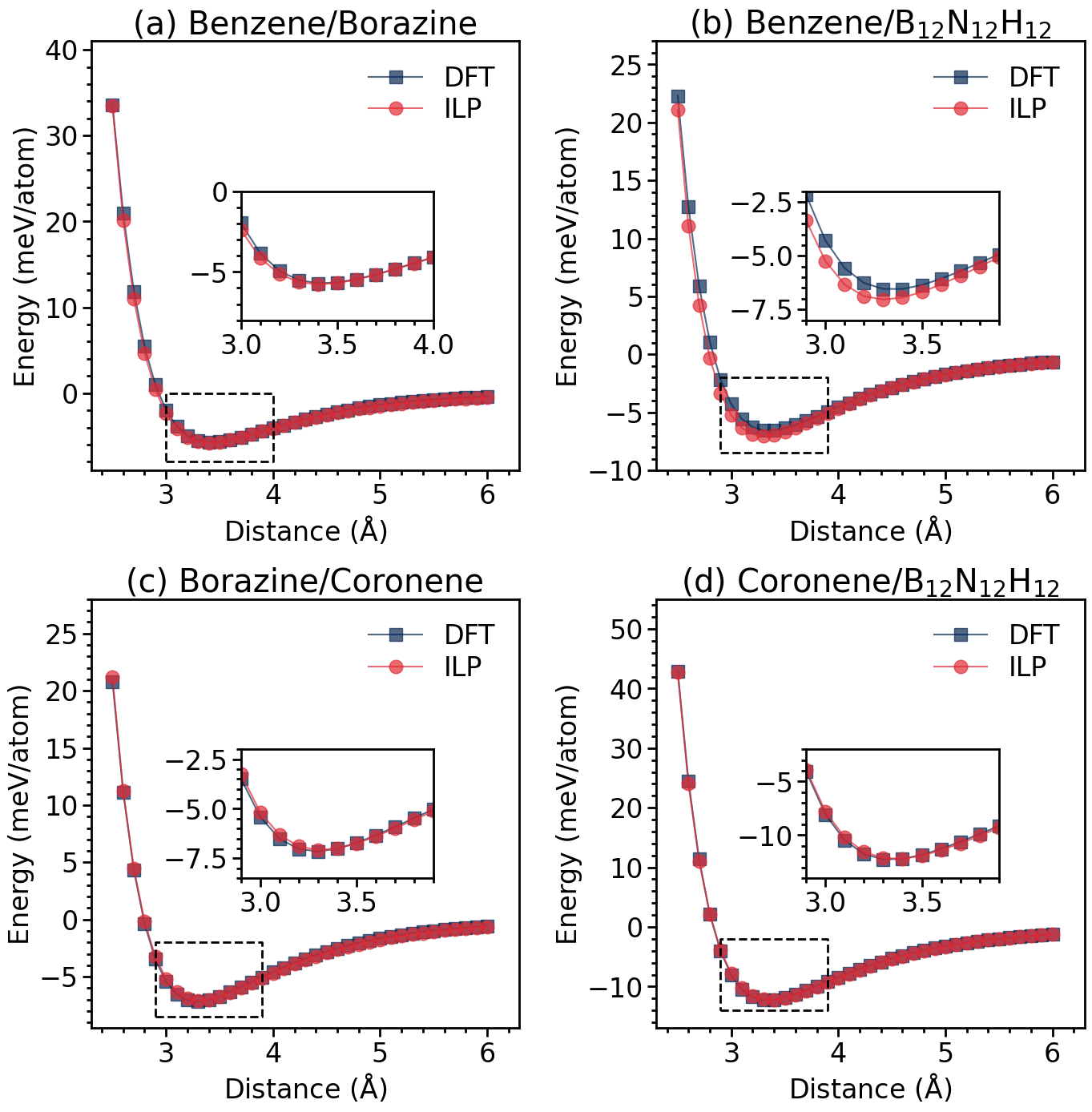}
\caption{Binding energy curves calculated for the finite heterogeneous dimers of (a) Benzene/Borazine, (b) Benzene/B$_{12}$N$_{12}$H$_{12}$, (c) Borazine/Coronene, (d) Coronene/B$_{12}$N$_{12}$H$_{12}$. The reported energies are measured relative to the infinitely separated dimer value and are normalized by the total number of atoms per unit-cell. The insets provide a zoom-in on the equilibrium interlayer separation region. Here, the parameters presented in \autoref{stab:bn ilp para} are used. 
}
\label{sfig:bn ilp2}
\end{center}
\end{figure}

\autoref{sfig:bn AAp be} and \autoref{sfig:bn AAp pes}(a)-(b) show the binding energy curves and \gls{pes} calculated for the laterally periodic bilayer \gls{hbn}, respectively, calculated using the updated \gls{ilp} model with parameters listed in \autoref{stab:bn ilp para}, compared against \gls{dft} results. The \gls{ilp} binding energy curves accurately reproduce the \gls{dft} calculations, with negligible differences near the equilibrium minimum. As shown in \autoref{sfig:bn AAp pes}(c), the differences in \glspl{pes} between \gls{dft} and \gls{ilp} are less than \SI{0.1}{\milli\electronvolt\per\atom}, demonstrating the very high accuracy of the updated \gls{ilp} model. Furthermore, \autoref{sfig:bn ilp1} and \autoref{sfig:bn ilp2} present the binding energy curves of finite homogeneous and heterogeneous dimers predicted by the updated \gls{ilp} model, with all differences relative to \gls{dft} results remaining below \SI{3}{\milli\electronvolt\per\atom}.

\newpage

\subsection{\normalsize 5.2. Contribution of Coulomb interactions }
Then we evaluated the interlayer monopolar electrostatic interactions based on a partial charge analysis of the different \gls{hbn} atoms. For bilayer \gls{hbn}, a Mulliken-charge \gls{dft} analysis, performed for the minimal unit cell at the HSE+MBD level of theory, yields partial atomic charges of +0.42\textit{e} and -0.42\textit{e} for B and N atoms, respectively. These values can then be used within the Coulomb electrostatic energy term $E=k\frac{q_iq_j}{r}, (r<r_c)$, where $k=$ \SI{14.399645}{\electronvolt\cdot\angstrom\per\square\coulomb} is the Coulomb constant, $q_i$ and $q_j$ are the effective charges of atoms $i$ and $j$ (residing in different layers), and $r_c=$ \SI{16}{\angstrom} is the cutoff. To this end, we construct a supercell consisting of a rectangular bilayer section, where each layer contains 780 B and 780 N atoms, within the \textsc{LAMMPS} package \cite{thompson2022lammps} and performed laterally periodic electrostatic calculations using the Ewald summation technique.

\begin{figure}[H]
\begin{center}
\includegraphics[width=0.4\columnwidth]{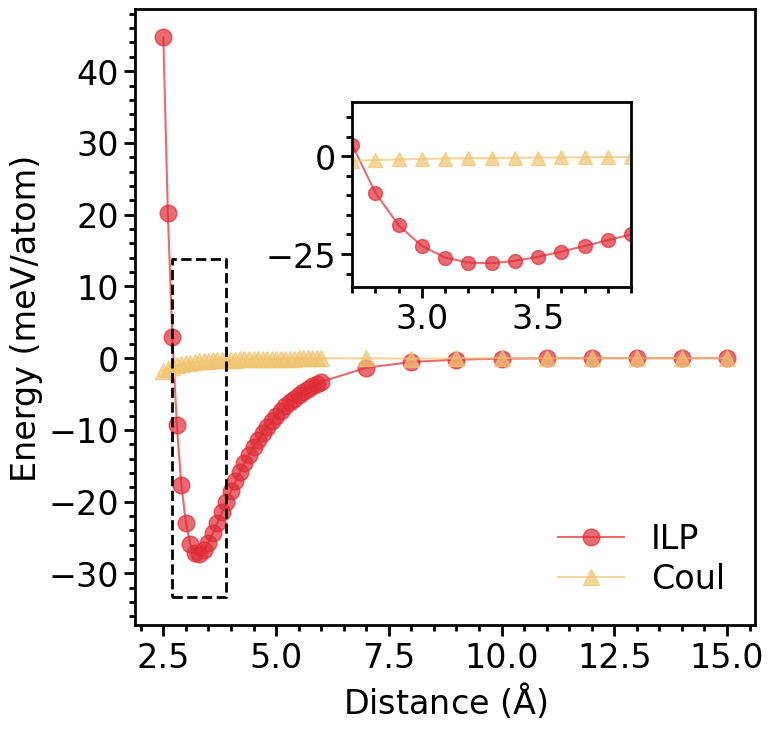}
\caption{Binding energy curves of bilayer \gls{hbn} calculated using \gls{dft}, \gls{ilp} with parameters of \autoref{stab:bn ilp para}. The monopolar electrostatic term contribution is represented by the yellow triangles. The reported energies are measured relative to the infinitely separated bilayer value and are normalized by the total number of atoms per unit-cell. The inset provides a zoom-in on the equilibrium interlayer distance region.
}
\label{sfig:bn AAp bec}
\end{center}
\end{figure}

\begin{figure}[H]
\begin{center}
\includegraphics[width=0.8\columnwidth]{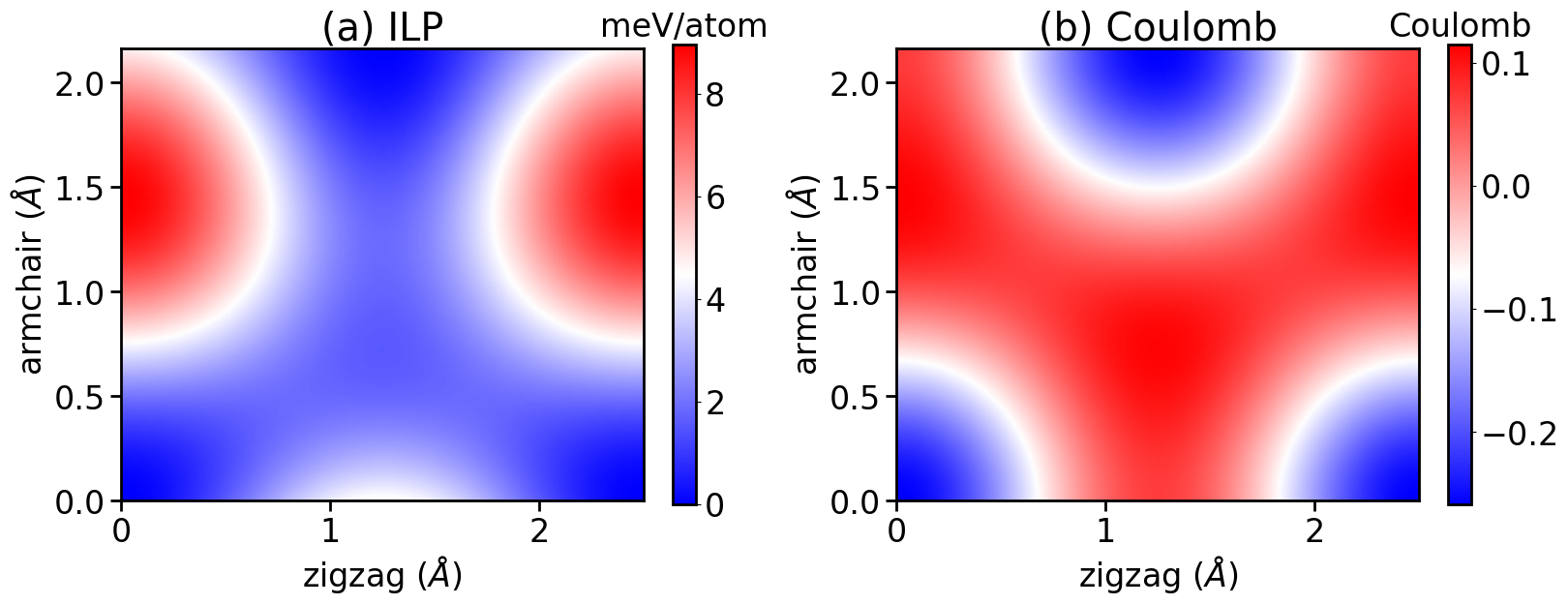}
\caption{Sliding \glspl{pes} of \gls{hbn} bilayer, calculated via rigid interlayer shifts at an interlayer distance of \SI{3.3}{\angstrom}. Panels (a)-(b) present the \gls{dft} (HSE+MBD-NL), \gls{ilp} \glspl{pes}, and their difference map, respectively. The parameters presented in \autoref{stab:bn ilp para} above are used for the \gls{ilp} calculations.
}
\label{sfig:bn AAp pesc}
\end{center}
\end{figure}

\autoref{sfig:bn AAp bec} and \autoref{sfig:bn AAp pesc} show the contributions of the Coulomb term to the calculated \gls{ilp} binding energy curves and \glspl{pes}, respectively, for the \gls{hbn} bilayer. Notably, compared to the total \gls{pes}, the Coulomb contributions account for only about 1.6\% and 3\% of the full binding energy and sliding potential energy barrier, respectively. This supports our choice to neglect monopolar electrostatic contributions in the \gls{ilp} expression to improve computational efficiency.

Previous work has shown that interlayer Coulomb interactions are significantly weaker than \gls{vdw} interactions in \gls{tmd} materials \cite{jiang2023anisotropic}. Here we also recalculated the interlayer Coulomb interactions for \gls{mos2}. For bilayer \gls{mos2}, a Mulliken-charge \gls{dft} analysis, performed for the minimal unit cell at the HSE+MBD level of theory, yields partial atomic charges of +0.52\textit{e} and -0.26\textit{e} for Mo and S atoms, respectively. To this end, we construct a supercell consisting of a rectangular bilayer section, where each layer contains 4,900 Mo and 9,800 S atoms, within the \textsc{LAMMPS} package \cite{thompson2022lammps} and also performed laterally periodic electrostatic calculations using the Ewald summation technique.

\begin{figure}[H]
\begin{center}
\includegraphics[width=0.7\columnwidth]{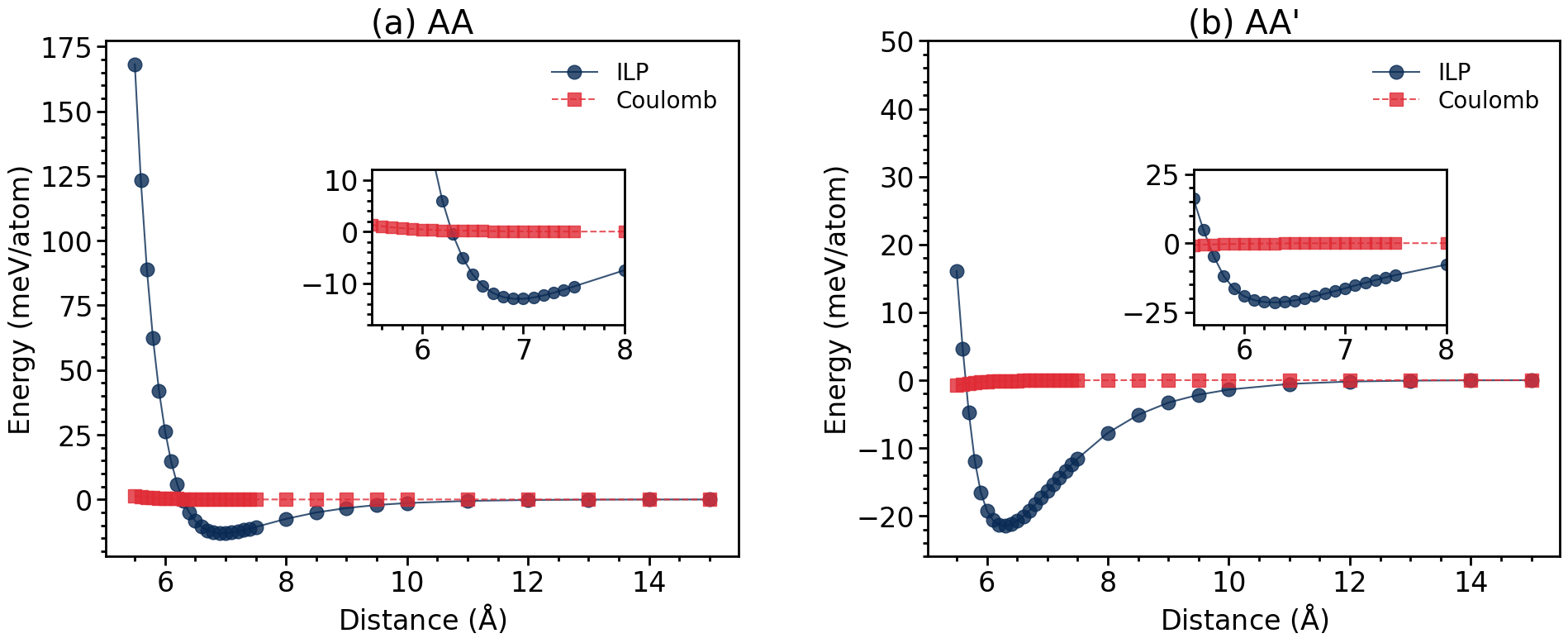}
\caption{Binding energy curves of AA (a) and AA’ (b) stacked \gls{mos2} bilayer calculated using the
ILP (blue diamond). The monopolar electrostatic term contribution
is represented by the red square. The reported energies are measured relative to the values of the
two single layers (zero for the electrostatic term) and are normalized by the total number of atoms
in the unit-cell (29,400 atoms). The insets provide a zoom-in on the equilibrium interlayer distance
region.}

\label{sfig:mos AAp bec}
\end{center}
\end{figure}

\begin{figure}[H]
\begin{center}
\includegraphics[width=0.7\columnwidth]{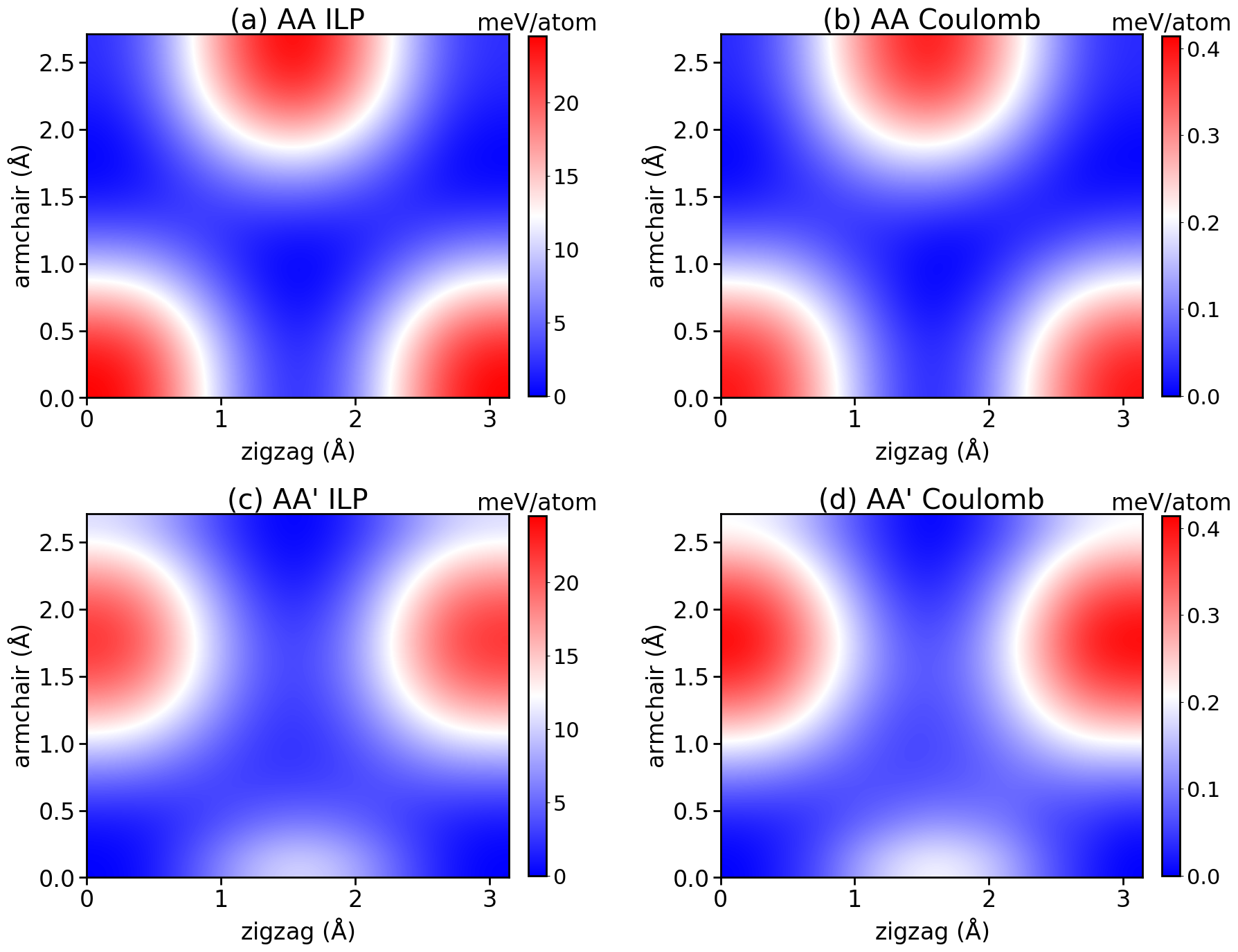}
\caption{Sliding \glspl{pes} of the bilayer \gls{mos2} homojunction, calculated at an interlayer distance of
\SI{6.24}{\angstrom} for the parallel (top panels, starting from AA-stacking mode) and antiparallel (bottom panels,
starting from AA′-stacking mode) configurations.  The reported energies are normalized by the total number of atoms
in the unit-cell (600 atoms) and measured relative to the values obtained at the AA' and AB stacking
modes for the anti-parallel and parallel configurations, respectively}

\label{sfig:mos AAp pesc}
\end{center}
\end{figure}

\autoref{sfig:mos AAp bec} and \autoref{sfig:mos AAp pesc} show the contributions of the Coulomb term to the calculated \gls{ilp} binding energy curves and \glspl{pes}, respectively, for the \gls{mos2} bilayer. Notably, compared to the total \gls{pes}, the Coulomb contributions account for only about 0.7\% and 2\% of the full binding energy and sliding potential energy barrier, respectively. This supports our choice to neglect monopolar electrostatic contributions in the \gls{ilp} expression to improve computational efficiency.

\newpage
\clearpage

\section{\normalsize 6. Coupling interaction terms in bulk vdW structures}

The \gls{smlp}+\gls{ilp} framework assumes that the total energy of a multilayer \gls{vdw} structure can be written as a superposition of intralayer and interlayer contributions. In a formal many-body description, however, the total energy of a stack containing more than two layers also includes higher-order cross-layer interaction terms that couple different interlayer pairs. Our current implementation explicitly sums individual bilayer interactions but omits these multi-body cross-couplings between distinct interlayer interfaces. In this section, we quantify the impact of the omitted coupling term by means of two complementary analyses.

\subsection{\normalsize 6.1 Comparison between bilayer and bulk ILPs}

As a first test, we consider bulk graphite as a representative single-atom-thick \gls{vdw} material composed of more than three layers. Following our previous work~\citep{ouyang2021anisotropic}, we employ two \glspl{ilp}: (i) a bilayer \gls{ilp}, fitted exclusively to the binding energy curves and sliding potential energy surfaces of isolated bilayer \gls{gr}, and (ii) a bulk \gls{ilp}, fitted to the corresponding data of bulk AB-stacked graphite. In the latter case, the effect of higher-order interlayer couplings in the infinite stack is effectively absorbed into the fitted interlayer interaction.

\begin{figure*}[htb]
  \centering
  \includegraphics[width=0.9\textwidth]{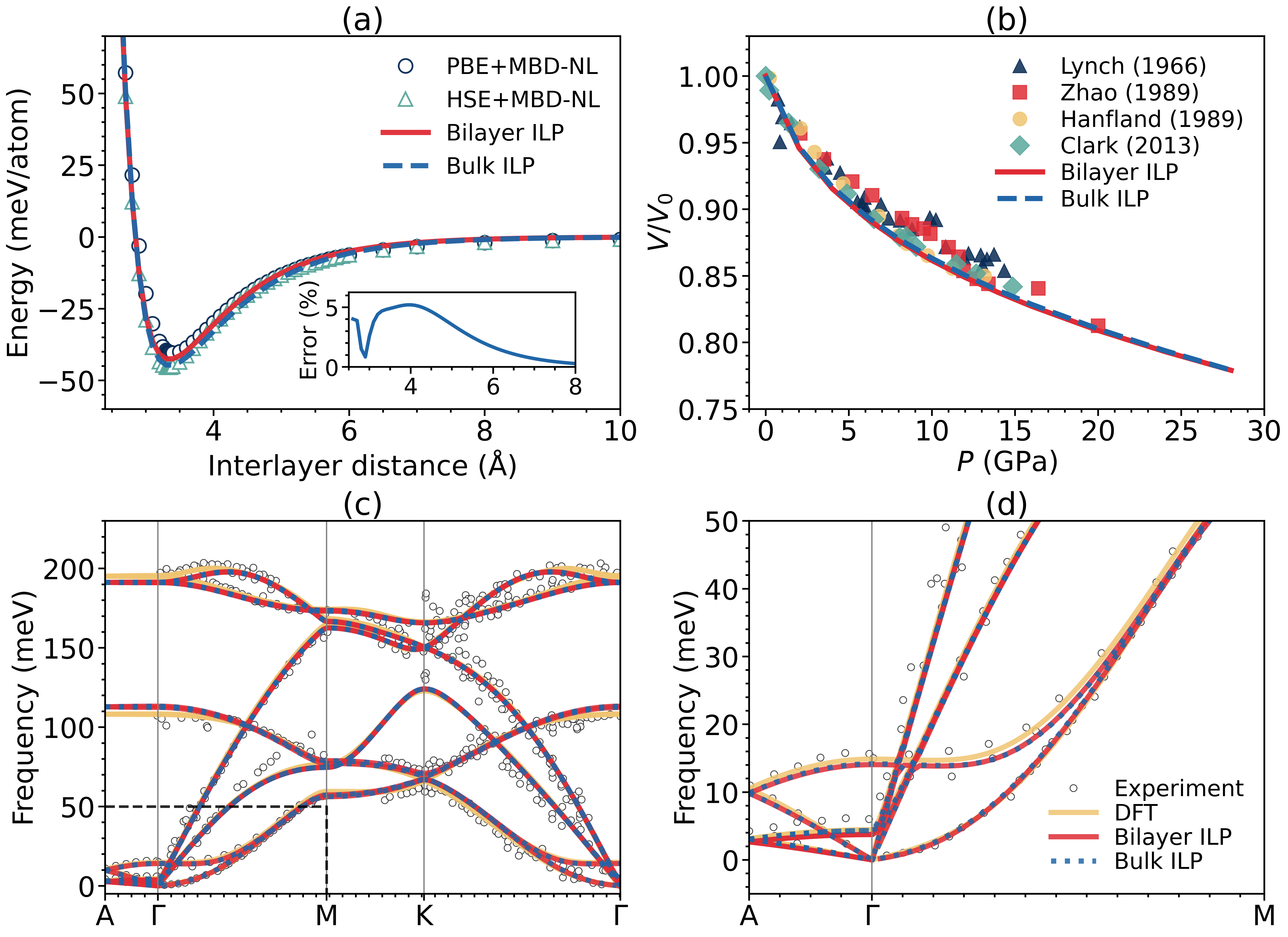}
  \caption{Validation of the bilayer and bulk \glspl{ilp} for bulk graphite.
  (a) Interlayer binding energy as a function of interlayer distance, comparing \gls{dft} benchmarks (\gls{pbe}+\gls{mbd_nl} and \gls{hse}+\gls{mbd_nl}) with predictions from the bilayer and bulk \glspl{ilp}. The inset quantifies the relative deviation between the two \glspl{ilp}, normalized by the \gls{hse}+\gls{mbd_nl} binding energy.
  (b) Equation of state showing the volume compression ratio $V/V_{0}$ versus hydrostatic pressure $P$, comparing predictions from the bilayer and bulk \glspl{ilp} with experimental data from Refs.~\citep{lynch1966effect, zhao1989xray, hanfland1989graphite, clark2013fewlayer}.
  (c) Phonon dispersion predicted by the \gls{dft} (yellow lines), bilayer \gls{ilp} (red solid lines), and bulk \glspl{ilp} (blue dashed lines), compared with experimental data (open circles) from  Ref.~\citep{wirtz2004phonon}.
  (d) Zoom-in view of dashed rectangular region in panel (c), showing the low-frequency branches along A–$\Gamma$–M , highlighting the close agreement between the two \glspl{ilp} predictions and the \gls{dft} and experimental references.}
  \label{sfig:graphite_ilp}
\end{figure*}

We re-evaluated the interlayer binding-energy curve of bulk graphite using \textsc{FHI-AIMS} with both \gls{pbe}+\gls{mbd_nl} and \gls{hse}+\gls{mbd_nl}, employing the same unit cell and computational settings as in Ref.~\citep{ouyang2021anisotropic}. As shown in \autoref{sfig:graphite_ilp}(a), both the bilayer and bulk \glspl{ilp} reproduce the \gls{dft} binding energy curve of bulk graphite with high accuracy. The inset of \autoref{sfig:graphite_ilp}(a) quantifies the relative deviation between the bilayer and bulk \glspl{ilp}, normalized by the \gls{hse}+\gls{mbd_nl} binding energy. This deviation consistently remains below $\sim5\%$ over the range of interlayer distances considered.

Using the same two \glspl{ilp}, we further evaluated the bulk modulus and phonon dispersion of bulk graphite within the \gls{smlp}+\gls{ilp} framework. \autoref{sfig:graphite_ilp}(b) demonstrates that the equation-of-state fits based on the bilayer and bulk \glspl{ilp} yield nearly indistinguishable $P$–$V$ curves. The extracted bulk moduli exhibit only minor differences, and all values fall within the spread of experimental data from X-ray diffraction; the numerical results are summarized in \autoref{stab:graphite_bulk}. The phonon spectra in \autoref{sfig:graphite_ilp}(c)–(d) likewise show that the two \glspl{ilp} produce nearly identical dispersions throughout the Brillouin zone. Minor discrepancies appear only in the low-frequency branches along the A–$\Gamma$ direction, where the bulk \gls{ilp} result is marginally closer to the \gls{dft} reference. Taken together with the binding-energy comparison above, these results indicate that, even though the bilayer \gls{ilp} is not trained on bulk configurations, neglecting explicit cross-layer terms introduces an error of at most $\sim5\%$ in the binding energy and has a negligible impact on the equilibrium structural and vibrational properties of bulk graphite.

\begin{table}[htb]
  \centering
  \caption{Bulk modulus $B$ and its zero-pressure derivative $B'$ of bulk graphite, obtained from \gls{eos} fits to \gls{md} data generated with the \gls{smlp}+\gls{ilp} model using either the bulk or bilayer \gls{ilp}, and compared with experimental values from X-ray diffraction.
  Values in parentheses indicate statistical errors. For the \gls{md} results, three different \glspl{eos}~\citep{hanfland1989graphite, murnaghan1944thecompressibility, birch1947finite, birch1952elasticity, vinet1986auniversal, vinet1987temperature} were used to fit $B$ and $B'$.}
  \label{stab:graphite_bulk}
  \begin{tabular}{lcc}
    \hline
    Method & $B$ (GPa) & $B'$ \\
    \hline
    Experiment (Ref.~\citep{lynch1966effect})   & 32(2)   & 12.3(0.7) \\
    Experiment (Ref.~\citep{hanfland1989graphite}) & 33.8(0.3) & 8.9(0.1)  \\
    Experiment (Ref.~\citep{zhao1989xray})      & 30.8(2.0) & \textemdash{} \\
    \hline
    Bulk ILP (Murnaghan)        & 32.0(2.6) &  8.9(0.7) \\
    Bulk ILP (Birch--Murn.)     & 26.8(2.0) & 15.5(1.5) \\
    Bulk ILP (Vinet)            & 31.8(2.0) & 10.6(2.1) \\
    \hline
    Bilayer ILP (Murnaghan)     & 30.8(2.3) &  9.1(0.6) \\
    Bilayer ILP (Birch--Murn.)  & 24.5(1.3) & 17.2(1.2) \\
    Bilayer ILP (Vinet)         & 30.1(1.6) & 11.0(0.5) \\
    \hline
  \end{tabular}
\end{table}

\subsection{\normalsize 6.2 Direct quantification via first-principles energy decomposition}

As a second, more direct test of the additive approximation, we performed a first-principles energy decomposition analysis to explicitly extract the cross term $\Delta E$ in multilayer systems. For bulk graphite, bulk \gls{hbn}, and the \gls{gr}/\gls{hbn} heterostructure, we computed the \gls{dft} total energy $E_{\mathrm{bulk}}(D)$ of AB-, AA$^\prime$-, and C-stacked structures, respectively, as a function of interlayer distance $D$ using \gls{pbe}+\gls{mbd_nl}.

We construct a bulk model using a fully periodic unit cell, where the size along the $c$-axis is set to twice the interlayer distance $D$. This cell contains two distinct monolayers (layer 1 and layer 2). The cross term $\Delta (D)$ is then computed as the difference between the total energy of the bulk system, $E_{\mathrm{bulk}} (D)$, and the superposition energy $E_{\mathrm{super}} (D)$, where $E_{\mathrm{super}} (D)$ is defined as the sum of the energies of the isolated monolayers plus the interlayer binding energy of each constituent layer pair:
\begin{equation}
  \Delta E(D) = E_{\mathrm{bulk}}(D) - E_{\mathrm{super}}(D),
  \label{eq:deltaE_def}
\end{equation}
\begin{equation}
  E_{\mathrm{super}}(D) = E_{\mathrm{monolayer}}^{1} + E_{\mathrm{monolayer}}^{2}
  + E_{\mathrm{interlayer}}(D),
  \label{eq:Esuper_def}
\end{equation}
where $E_{\mathrm{monolayer}}^{1}$ and $E_{\mathrm{monolayer}}^{2}$ are the energies of the isolated monolayers and $E_{\mathrm{interlayer}}(D)$ is the total accumulated interlayer potential energy of the bulk unit cell, summing interactions between layer 1 and layer 2 and all corresponding periodic images. The quantity $\Delta E(D)$ defined in \autoref{eq:deltaE_def} measures the non-additive cross-layer coupling that cannot be represented as a simple superposition of monolayer and bilayer contributions.

Within the \gls{smlp}+\gls{ilp} framework, the interlayer potential energy between any two layers can be evaluated by summing atomic site energies, which allows us to decompose the interlayer energy into layer-resolved components. This formulation allows us to decompose the interactions between any two layers into layer-resolved components:
\begin{align}
  E_{\mathrm{interlayer}}^{11}(D)
    &= E_{\mathrm{interlayer}}^{11,1}(D)
     + E_{\mathrm{interlayer}}^{11,1}(D),
    \label{eq:E11}\\
  E_{\mathrm{interlayer}}^{12}(D)
    &= E_{\mathrm{interlayer}}^{12,1}(D)
     + E_{\mathrm{interlayer}}^{12,2}(D),
    \label{eq:E12}\\
  E_{\mathrm{interlayer}}^{22}(D)
    &= E_{\mathrm{interlayer}}^{22,2}(D)
     + E_{\mathrm{interlayer}}^{22,2}(D),
    \label{eq:E22}
\end{align}
where $E_{\mathrm{interlayer}}^{ij,k}$ denotes the aggregate interlayer potential energy $E_{\mathrm{interlayer}}^{ij}$ assigned to atoms in layer $k$ arising from interactions between layers $i$ and $j$, for indices $i,j,k\in\{1,2\}$.

To reconstruct the bulk crystal properties from additive interactions, the integrated interlayer potential energy $E_{\mathrm{interlayer}} (D)$ must account for interactions between layers in the unit cell and all periodic images. For a bulk model with periodic boundary conditions along the $z$-axis, each layer interacts with image layers located in both the upper (positive $z$) and lower (negative $z$) directions. Accordingly, a constant factor of 2 is introduced in the summation to explicitly account for these bi-directional contributions at each distance $d=l\cdot D$, where $l=1,2,3\ldots$ indexes the distance multiples. Therefore, $E_{\mathrm{interlayer}} (D)$ can be represented as:
\begin{equation}
\begin{split}
  E_{\mathrm{interlayer}}(D)
  = 2 \sum_{d=(2l-1)D}
      E_{\mathrm{interlayer}}^{12,1}(d)
      + 2 \sum_{d=2l\cdot D}E_{\mathrm{interlayer}}^{11,1}(d) \\
      + 2 \sum_{d=(2l-1)D}E_{\mathrm{interlayer}}^{12,2}(d)+ 2 \sum_{d=2l\cdot D}E_{\mathrm{interlayer}}^{22,2}(d),
  \label{eq:Einter_sum_full}
  \end{split}
\end{equation}
This expression can be simplified by grouping the layer-resolved components into total bilayer interactions. Utilizing the definitions from \autoref{eq:E11}, \autoref{eq:E12} and \autoref{eq:E22}, $E_{\mathrm{interlayer}} (D)$ becomes the sum of the interlayer potential energies of distinct layer pairs:
\begin{equation}
\begin{split}
  E_{\mathrm{interlayer}}(D)
  = 2 \sum_{d=(2l-1)D}
      E_{\mathrm{interlayer}}^{12}(d)
      + \sum_{d=2l\cdot D}E_{\mathrm{interlayer}}^{11}(d) 
      + \sum_{d=2l\cdot D}E_{\mathrm{interlayer}}^{22}(d).
  \label{eq:Einter_sum_simple}
  \end{split}
\end{equation}

For a specific bilayer pair $i$–$j$ at separation $d$, the specific interaction energy can be further determined by subtracting the monolayer self-energies from the total bilayer energy:
\begin{equation}
  E_{\mathrm{interlayer}}^{ij}(d)
  = E_{\mathrm{bilayer}}^{ij}(d)
    - E_{\mathrm{monolayer}}^{i}
    - E_{\mathrm{monolayer}}^{j},
  \label{eq:Ebilayer_def}
\end{equation}
where $E_{\mathrm{bilayer}}^{ij}(d)$ is the total energy of the isolated bilayer at separation $d$, and $E_{\mathrm{monolayer}}^{i}$ and $E_{\mathrm{monolayer}}^{j}$ are  respectively the energies of the corresponding isolated monolayers $i$ and $j$.

Substituting \autoref{eq:Einter_sum_simple}, with individual interaction terms determined by \autoref{eq:Ebilayer_def}, into \autoref{eq:deltaE_def} and \autoref{eq:Esuper_def}, we computed the cross term $\Delta E(D)$ across various interlayer distances $D$. This was achieved by calculating the energy of bulk models $E_{\mathrm{bulk}} (D)$, energy of each monolayer $E_{\mathrm{monolayer}}^i$, and energy of each bilayer model $E_{\mathrm{bilayer}}^{ij}$ (d), where the separation distance $d=l\cdot D$ was summed up to a cutoff of \SI{16}{\angstrom}. All \gls{dft} calculations were performed with \textsc{FHI-AIMS} using the \gls{pbe} exchange–correlation functional~\citep{perdew1996generalized} and the \gls{mbd_nl} dispersion correction, employing the tier-2 basis set~\citep{havu2009efficient} and tight numerical settings with fully refined grid levels and a dense outer integration grid~\citep{blum2009initio}. Relativistic effects were neglected for C, B, and N atoms. For bulk models, a $19 \times 19 \times 7$ $k$-point mesh was used, whereas for monolayer and bilayer models we introduced a vacuum spacing of \SI{50}{\angstrom} and used a $19 \times 19 \times 1$ mesh.

\begin{figure*}[htb]
  \centering
  \includegraphics[width=0.95\textwidth]{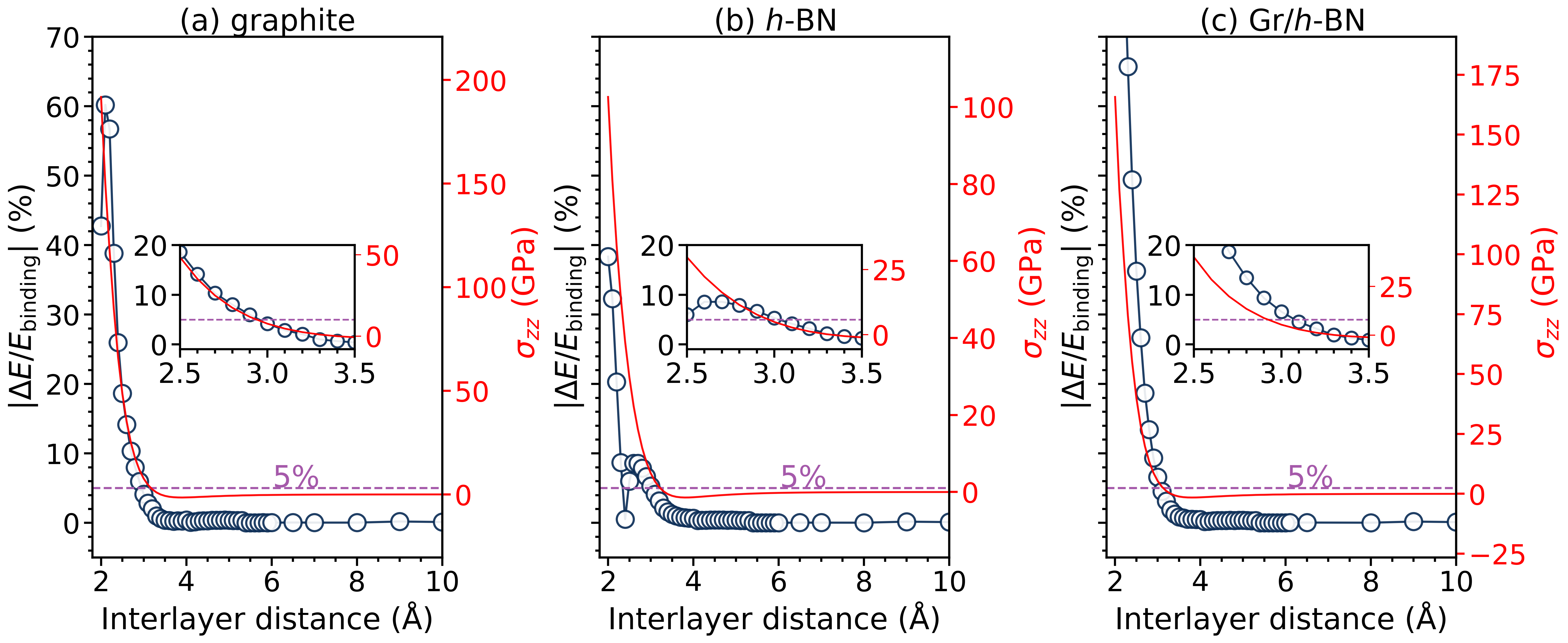}
  \caption{Quantitative assessment of the non-additive interlayer coupling term in multilayer \gls{vdw} systems. Symbols show the normalized residual energy $|\Delta E|/E_{\mathrm{binding}}$ as a function of interlayer distance for graphite (left), \gls{hbn} (middle), and the \gls{gr}/\gls{hbn} heterostructure (right), obtained from \gls{pbe}+\gls{mbd_nl} calculations. The red curve represents the normal stress $\sigma_{zz}$ as a function of the interlayer separation. The insets show magnified views of the \SIrange{2.5}{3.5}{\angstrom} interval around the equilibrium spacing. For all three systems, $|\Delta E|/E_{\mathrm{binding}}$ remains below $5\%$ once the interlayer distance exceeds about \SI{3}{\angstrom}, indicating that the neglected cross-layer many-body contribution is small in the near-equilibrium regime relevant to the present \gls{smlp}+\gls{ilp} model.
  }
  
  \label{sfig:coupling}
\end{figure*}

The results of this decomposition are presented in \autoref{sfig:coupling}. For all three systems, the normalized residual $|\Delta E|/E_{\mathrm{binding}}$ remains below $5\%$ once the interlayer distance exceeds about \SI{3}{\angstrom} in the vicinity of the equilibrium spacing. At this separation, the corresponding normal pressure $\sigma_{zz}$ is \SI{7.6}{\giga\pascal}, \SI{4.9}{\giga\pascal} and \SI{5.3}{\giga\pascal} for bulk graphite, \gls{hbn} and \gls{gr}/\gls{hbn}, respectively. This indicates that for applied normal loads below these thresholds, the neglected cross‑term in the energy decomposition is comparable to or smaller than the intrinsic fitting uncertainty between the ILP and the reference DFT data. Under stronger compression ($d$<\SI{3}{\angstrom}), both $\sigma_{zz}$ and $|\Delta E/E_{\mathrm{binding}} |$ increase sharply, indicating that higher-order cross-layer couplings become significant under such extreme loading conditions. Consequently, the combined \gls{smlp}+\gls{ilp} framework remains valid for the vast majority of experimentally relevant conditions.

Overall, these two analyses demonstrate that, for single-atom-thickness \gls{vdw} materials such as graphite and \gls{hbn}, and for the near-equilibrium configurations relevant to our thermal and mechanical property calculations, the cross term between different interlayer interactions contributes less than $\sim5\%$ of the binding energy and does not materially affect the predicted equilibrium structures, elastic properties, or phonon spectra. The linear separation of intra- and interlayer energies adopted in the \gls{smlp}+\gls{ilp} model is therefore well justified within this regime. At the same time, the energy decomposition analysis also clarifies the intrinsic limitation of this approach: under very high normal stresses or in strongly distorted multilayer configurations, the neglected cross term becomes non-negligible, and the present framework should not be used for quantitative predictions under such extreme conditions. These trends are consistent with previous studies of layered materials under pressure~\citep{ni2021stronger, martins2017raman} and with typical experimental conditions for two-dimensional materials, where normal pressures remain well below these thresholds~\citep{proctor2009highpressure, segura2021tuning}.

\newpage

\section{\normalsize 7. Benchmarks of \textit{s}MLP+ILP approach}
\subsection{\normalsize 7.1. Edge energy}

\begin{figure*}[htb]
\begin{center}
\includegraphics[width=0.8\columnwidth]{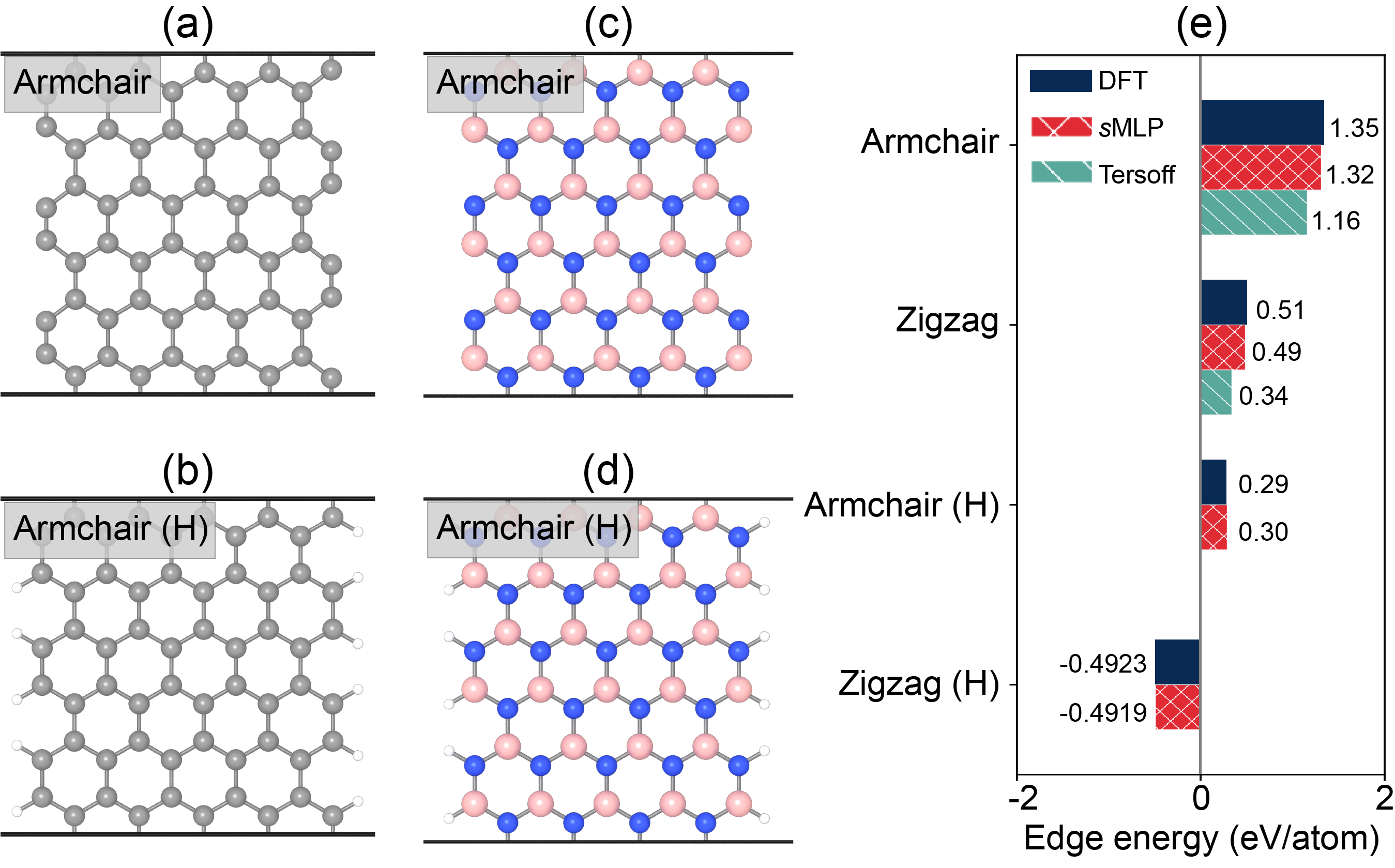}
\caption{Edge energy calculations. Panels (a)–(d) show nanoribbon structures with either armchair or zigzag edges for \gls{gr} (left) and \gls{hbn} (middle). Panels (a) and (c) depict non-hydrogenated edges, while panels (b) and (d) show hydrogenated counterparts. Panel (e) presents the edge energies of monolayer \gls{hbn} calculated using \gls{dft}, \gls{smlp}, and Tersoff potentials.
}
\label{sfig:edge energy}
\end{center}
\end{figure*}

As shown in \autoref{sfig:edge energy}(a)-(d), we calculated the edge energies of \gls{gr} and \gls{hbn} monolayers, including the configurations with edges along either the armchair or zigzag direction, and with or without hydrogenation to the total dangling edge atoms. Following the same approach used for \gls{gr} (see Eq.~14 and Fig.~2(d) of the main text), the edge energies predicted by the \gls{smlp} model for the \gls{hbn} monolayer, as shown in \autoref{sfig:edge energy}(e), closely match the \gls{dft} reference values, with deviations on the order of \SI{0.01}{\electronvolt\per\atom}, which is an order of magnitude smaller than those obtained using the Tersoff potentials.

\subsection{\normalsize 7.2. Phonon dispersions}

\begin{figure*}[htb]
\begin{center}
\includegraphics[width=0.8\columnwidth]{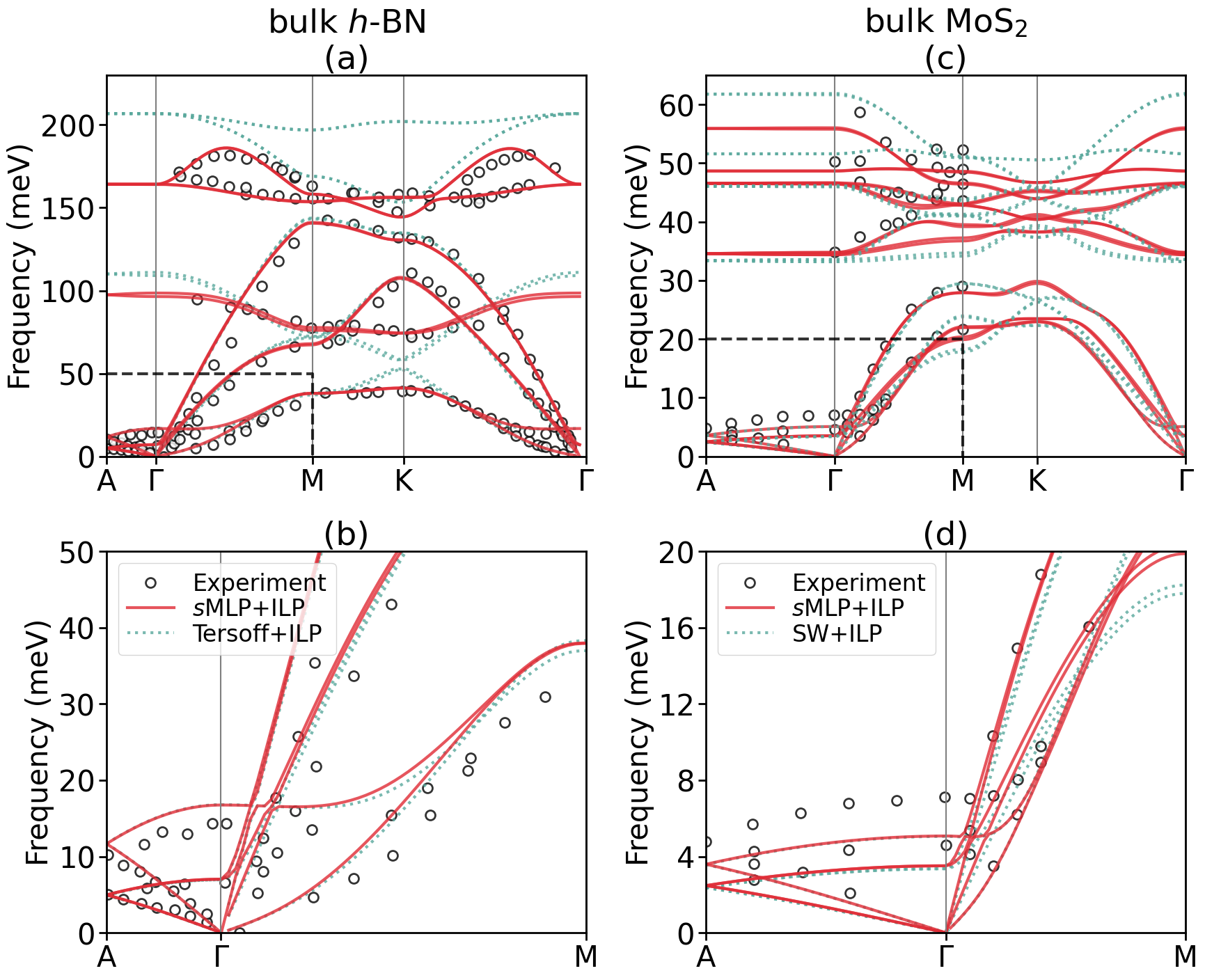}
\caption{Phonon dispersions of bulk (a)-(b) \gls{hbn} and (c)-(d) \gls{mos2}, predicted by \gls{smlp}+\gls{ilp}, empirical intralayer potential+\gls{ilp}, compared against experimental measurements. The solid red lines represent dispersion curves calculated with the \gls{smlp}+\gls{ilp} force field. The dashed-dotted green lines correspond to calculations using (a)–(b) Tersoff+\gls{ilp} for \gls{hbn} and (c)–(d) SW+\gls{ilp} for \gls{mos2}. Experimental data for bulk \gls{hbn} \cite{serrano2007vibrational} and bulk \gls{mos2} \cite{wakabayashi1975lattice} are shown as open black circles. Panels (b) and (d) provide zoom-in views of the low-energy phonon modes near the $\Gamma$-point from (a) and (c), respectively.
}
\label{sfig:phonon}
\end{center}
\end{figure*}

Phonon dispersion curves were obtained by diagonalizing the dynamical matrix within the \textsc{LAMMPS} \cite{thompson2022lammps} framework at \SI{0}{\kelvin} and zero external pressure. The dynamical matrix was constructed via numerical differentiation using a finite displacement step of \SI{1e-6}{\angstrom}, following structural relaxation under periodic boundary conditions. The phonon spectra analyzed using our homemade codes \cite{ouyang2020mechanical,ouyang2021anisotropic} were computed along high-symmetry directions in the Brillouin zone, specifically following the path A~$\rightarrow$~$\Gamma$~$\rightarrow$~M~$\rightarrow$~K~$\rightarrow$~$\Gamma$, with a total of 201 uniformly distributed \textbf{q}-points. 
\autoref{sfig:phonon}(a)-(b) show the results for \gls{hbn}, compared with Tersoff+\gls{ilp} approach and experimental measurements \cite{serrano2007vibrational}. \autoref{sfig:phonon}(c)-(d) show the results of \gls{mos2} comparing with the approach computed intralayer interactions by Stillinger-Weber (SW) potential, SW+\gls{ilp}, and experimental measurements \cite{wakabayashi1975lattice}. For both \gls{hbn} and \gls{mos2}, our \gls{smlp}+\gls{ilp} predictions closely match the experimental data across both acoustic and optical branches. In contrast, the Tersoff+\gls{ilp} model for \gls{hbn} and the SW+\gls{ilp} model for \gls{mos2} significantly overestimate the high-frequency optical modes, which mainly results from the intralayer potentials \cite{ouyang2020mechanical, ouyang2021anisotropic}. These results further demonstrate that the \gls{smlp}+\gls{ilp} approach provides superior accuracy compared to combinations of empirical intralayer potentials with \gls{ilp}.

\subsection{\normalsize 7.3. Bulk modulus}

The bulk modulus is calculated by fitting the pressure-volume ($P-V$) curves of three \gls{eos} models: the Murnaghan equation \cite{hanfland1989graphite,murnaghan1944thecompressibility},

\begin{equation}
    \frac{V}{V_0} = {[1+\frac{B'}{B}P]}^{-\frac{1}{B'}}
\label{equation:Murnaghan}
\end{equation}
the Birch-Murnaghan equation \cite{birch1947finite,birch1952elasticity},
\begin{equation}
    P = 3B\xi(1 + 2\xi)^{5/2}[1 - \frac{3}{2}(4-B')\xi]
\label{equation:BirchMurnaghan1}
\end{equation}
with
\begin{equation}
    \xi=\frac{1}{2}[(\frac{V}{V_0})^{-\frac{2}{3}} - 1]
\label{equation:BirchMurnaghan2}
\end{equation}
and the Vinet equation \cite{vinet1986auniversal,vinet1987temperature},
\begin{equation}
    P=3B\frac{1-X}{X^2}\exp{[\frac{3}{2}(B'-1)(1-X)]}, X = (\frac{V}{V_0})^{\frac{1}{3}}
\label{equation:Vinet}
\end{equation}
where, $V_0$ and $V$ are the unit cell volumes at zero and finite external pressure, respectively, while $B$ and $B'$ respectively denote the bulk modulus and its pressure derivative at zero pressure. The relationship between pressure and volume varies among the three \gls{eos} formulations: the Murnaghan, Birch–Murnaghan, and Vinet forms exhibit linear, polynomial, and exponential dependencies on pressure, respectively, when fitting the $P$–$V$ curves. 

\begin{figure*}[htb]
\begin{center}
\includegraphics[width=0.8\columnwidth]{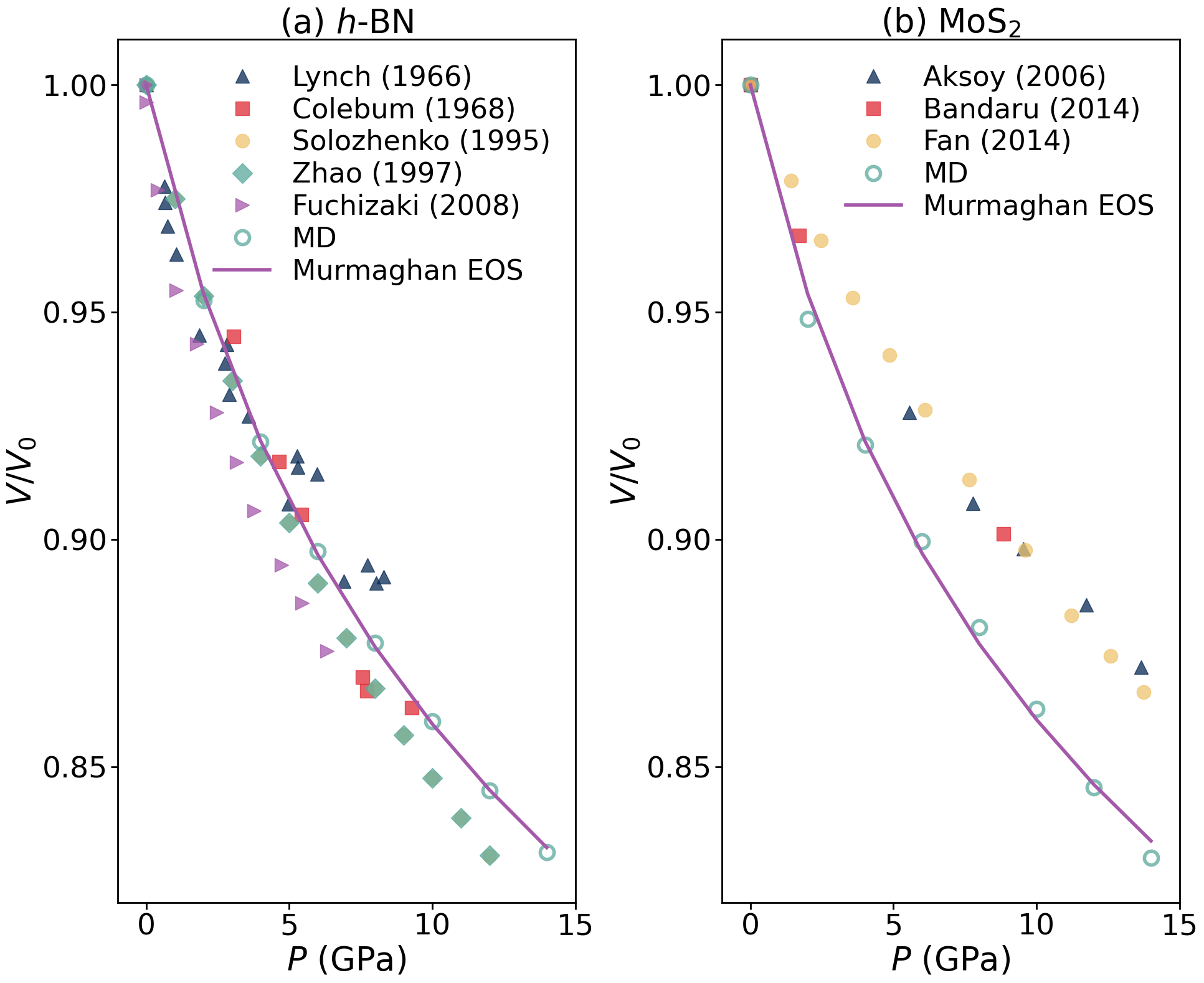}
\caption{Pressure ($P$) dependence of the normalized volume ($V/V_0$) of bulk (a) \gls{hbn} and (b) \gls{mos2}, where $V$ and $V_0$ represent the pressure-dependent and initial volumes, respectively. The green hollow circles represent \gls{md} simulation results using the \gls{smlp}+\gls{ilp} model. The solid purple line is the fitted
curve using the Murnaghan \gls{eos}; the fitted curves from the
two additional \glspl{eos} coincide closely and are therefore not
shown. In panel (a), blue triangles, red squares, yellow circles, blue diamond and purple triangles represent experimental data obtained from Ref.~\cite{lynch1966effect}, Ref.~\cite{coleburn1968irreversible}, Ref.~\cite{solozhenko1995isothermal}, Ref.~\cite{zhao1997pvtdata}, and Ref.~\cite{fuchizaki2008solid}, respectively. In panel (b), blue triangles, red squares and yellow circles represent experimental data obtained from Ref.~\cite{aksoy2006xray}, Ref.~\cite{bandaru2014effect}, Ref.~\cite{fan2014pvtequation}, respectively.
}
\label{sfig:bulk modulus}
\end{center}
\end{figure*}

As shown in \autoref{sfig:bulk modulus}, for \gls{hbn}, the computed $P$-$V$ relationship from our simulations aligns well with experimental measurements and for \gls{mos2}, the \gls{md} curve is near the experimental measurements within a little underestimation. The fitted bulk  modulus ($B$) and its zero-pressure derivative ($B'$) for \gls{hbn} and \gls{mos2} are reported in \autoref{stab:bulk_modulus_hbn} and \autoref{stab:bulk_modulus_mos2}, respectively.

\begin{table}[htb]
\setlength{\abovecaptionskip}{1.2cm}  
\caption{Bulk modulus ($B$), zero-pressure derivative ($B'$), intralayer lattice constant ($a_0$), and interlayer lattice constant ($c_0$) of bulk \gls{hbn}, calculated from \gls{md} simulations using the \gls{smlp}+\gls{ilp} model and compared with experimental data from X-ray diffraction \cite{pease1952anxray, lynch1966effect, solozhenko1995isothermal, zhao1997pvtdata, godec2000thermoelastic, alexey2006elasticity}. Values in parentheses represent statistical errors. For the \gls{md} results, three different \glspl{eos} \cite{hanfland1989graphite,murnaghan1944thecompressibility,birch1947finite,birch1952elasticity,vinet1986auniversal,vinet1987temperature} were used to fit $B$ and $B'$. $a_0$ and $c_0$ of \gls{md} are computed at \SI{300}{\kelvin}.}
\centering
\label{stab:bulk_modulus_hbn}
\begin{tabular}{lcccc}
\hline
Method & $B$ (GPa) & $B'$ & $a_0$ (\AA)  & $c_0$ (\AA) \\
\hline
Experiment (Ref.~\cite{pease1952anxray})  & --     & -- & 2.50399 (0.00005)  & 6.6612 (0.0005) \\
Experiment (Ref.~\cite{lynch1966effect})  & 22 (4)     & 18 (3) & 2.5040  & 6.6612 \\
Experiment (Ref.~\cite{solozhenko1995isothermal})  & 36.7 (0.5) & 5.6 (0.2)  & 2.504 (0.002)   & 6.660 (0.008) \\
Experiment (Ref.~\cite{zhao1997pvtdata})  & 17.6 (0.8) & 19.5 (3.4)        & 2.5043 (0.0001)  & 6.6566 (0.0006) \\
Experiment (Ref.~\cite{godec2000thermoelastic})  & 27.6 (0.5) & 10.5 (0.5)        & 2.504 (0.004)  & 6.659 (0.002) \\
Experiment (Ref.~\cite{alexey2006elasticity})  & 25.6 (0.8) &   --     &   2.506 & 6.657  \\
Experiment (Ref.~\cite{fuchizaki2008solid})  & 21 & 16        &  2.50 (0.05) &  6.66 (0.03) \\
\hline
MD (Murnaghan)       & 35.6 (1.6)  & 7.4 (0.5)  & \multirow{3}{*}{\makecell{2.5119\\(0.0003)}} &  \multirow{3}{*}{\makecell{6.4345\\(0.0046)}} \\
MD (Birch-Murn.) & 34.0 (1.1)  & 9.5 (0.5) &  \\
MD (Vinet)           & 34.9 (1.1)  & 8.6 (0.4) &  \\
\hline
\end{tabular}
\end{table}

\begin{table}[htb]
\setlength{\abovecaptionskip}{1.2cm}  
\caption{Bulk modulus ($B$), zero-pressure derivative ($B'$), intralayer lattice constant ($a_0$), and interlayer lattice constant ($c_0$) of bulk \gls{mos2}, calculated from \gls{md} simulations using the \gls{smlp}+\gls{ilp} model and compared with experimental data from X-ray diffraction \cite{aksoy2006xray, bandaru2014effect, chi2014pressure, fan2014pvtequation}. Values in parentheses represent statistical errors. For the \gls{md} results, three different \glspl{eos} \cite{hanfland1989graphite,murnaghan1944thecompressibility,birch1947finite,birch1952elasticity,vinet1986auniversal,vinet1987temperature} were used to fit $B$ and $B'$. $a_0$ and $c_0$ of \gls{md} are computed at \SI{300}{\kelvin}.}
\centering
\label{stab:bulk_modulus_mos2}
\begin{tabular}{lcccc}
\hline
Method & $B$ (GPa) & $B'$ & $a_0$ (\AA)  & $c_0$ (\AA) \\
\hline
Experiment (Ref.~\cite{aksoy2006xray})  &  53.4 (1)    & 9.2 (0.4) & --  & -- \\
Experiment (Ref.~\cite{bandaru2014effect})  & 70 (5)     & 4.5 & 3.159 (0.008) & 12.298 (0.003)\\
Experiment (Ref.~\cite{chi2014pressure})  & 47.65 (0.30) & 10.58 (0.08)  &  --  &  -- \\
Experiment (Ref.~\cite{fan2014pvtequation})  & 69 (2) & 4.7 (0.2)        & 3.163 (0.004)  & 12.341 (0.004) \\
\hline
MD (Murnaghan)       & 35.3 (5.9)  & 7.7 (1.9)  & \multirow{3}{*}{\makecell{3.1840\\(0.0005)}} &  \multirow{3}{*}{\makecell{12.556\\(0.004)}} \\
MD (Birch-Murn.) &     36.5 (6.4)  & 8.4 (2.8) &  \\
MD (Vinet)           & 37.2 (5.6)  & 7.9 (1.9) &  \\
\hline
\end{tabular}
\end{table}

\clearpage

\subsection{\normalsize 7.4. Young's modulus and Poisson's ratio}
We calculated the Young's modulus and Poisson's ratio for graphitic systems using \gls{md} simulations driven by \gls{smlp}+\gls{ilp} hybrid model. The \gls{md} details for tensile simulations are provided in the Methods section of the main text. Specifically, the Young’s modulus $E_1$ and Poisson’s ratio $\nu_{12}$ were defined as:
\begin{equation}
    E_1=\frac{\sigma_1}{\varepsilon_1}
\end{equation}
\begin{equation}
    \nu_{12}=-\frac{\varepsilon_2}{\varepsilon_1}
\end{equation}
where $\sigma_1$ is the tensile stress, $\varepsilon_1$ is the corresponding strain component along tensile direction, and $\varepsilon_2$ is the strain component along the lateral direction.

In \autoref{stab:YM and PR}, we compare the Young's modulus and Poisson's ratio of monolayer, bilayer, and bulk \gls{gr} with results obtained from other approaches, including empirical force fields, pure \glspl{mlp}, \gls{dft}, and experimental measurements. The thickness of monolayer \gls{gr} is typically assumed to be between \qtyrange{3.34}{3.4}{\angstrom}, most commonly \SI{3.35}{\angstrom}, based on experimental data \cite{lee2008measurement}, except in the case of Neek-Amal et al.  \cite{neek-amal2010nanoindentation}, who used \SI{4.5}{\angstrom} for bilayer \gls{gr}. For multilayer \gls{gr}, the thickness is defined as the number of layers multiplied by the monolayer thickness. In this study, we define the thickness of monolayer \gls{gr} as half of the interlayer lattice constant ($c_0$) of bulk graphite, i.e., \SI{3.36}{\angstrom}. As shown in \autoref{stab:YM and PR}, the Young's modulus and Poisson's ratio predicted by our \gls{smlp}+\gls{ilp} approach fall within the experimental error margins \cite{lee2008measurement, blakslee1970elastic}.

\begin{table}[ht]
\caption{Comparison of the thickness per layer $t$ (\unit{\angstrom}), Young’s modulus $E$ (\unit{\tera\pascal}), and Poisson’s ratio $\nu$ of graphite systems obtained from different methods. "Exp." and "MM" refer to experimental measurements and molecular mechanics, respectively. "MG" and "BG" denote monolayer and bilayer \gls{gr}, respectively. Values in parentheses represent statistical errors, while "zig" and "arm" indicate the zigzag and armchair directions, respectively.}
\begin{center}
\begin{tabular}{llllll}
\toprule
      & Method & Model & $t$(\unit{\angstrom}) & $E$(\unit{\tera\pascal}) & $\nu$\\
\midrule
     Exp. & \makecell[l]{Nanoindentation\\ \cite{lee2008measurement}} & MG & 3.35 & 1.0 (0.1) & 0.165 \\
     & \makecell[l]{Static tension\\ \cite{blakslee1970elastic}} & Bulk & - & 1.02 (0.03) & 0.16 (0.06) \\
\midrule
     DFT & LDA \cite{liu2007initio} & MG & 3.34 & 1.05 & 0.186 \\
     & PBE \cite{lebedeva2019elastic} & MG & 3.34 & 1.01 & 0.203 \\
     & PBE \cite{zhang2013mechanical}  &  BG & 3.41 & 0.977 & - \\
     & HF \cite{vanlier2000initio} & MG & 3.4 & 1.11 & - \\
\midrule
     MD & GAP \cite{ouyang2020accuracy} & MG & 3.4 & 0.98 & 0.193 \\
     & ANN \cite{singh2023reliable} & MG & - & 0.975 (0.080) & -\\
     & springs+LJ \cite{tsai2010characterizing} & MG & 3.4 & 0.912 & 0.261 \\
     & &  Bulk & - & 0.795 & 0.272 \\
     & REBO \cite{zhang2011mechanical} & MG & 3.35 & 0.907 & - \\
     & Tersoff \cite{ni2010anisotropic} & MG & 3.35 & \makecell[l]{1.13 (arm)\\ 1.05 (zig)} & - \\
     & Tersoff \cite{bu2009atomistic} & MG & 3.35 & 1.24 (arm) & - \\
     & Tersoff \cite{lebedeva2019elastic} & MG & 3.34 & 1.22 & -0.158 \\
     & REBO-2000 &  & & 0.7273 & 0.398 \\
     & AIREBO & & &  0.828 & 0.367 \\
     & LCBOP & & & 0.9433 & 0.221 \\
     & LJ + REBO \cite{neek-amal2010nanoindentation} & BG & 2.25 & 0.8 & - \\
\midrule
     MM & LJ + Morse \cite{cho2007mechanical} & Bulk & 3.35 & 1.153 & 0.195 \\
\midrule
     MD & \gls{smlp}+\gls{ilp} & MG & 3.36 & \makecell[l]{0.928 (0.003) (zig) \\ 0.911 (0.010) (arm)} & \makecell[l]{0.195 (0.010) (zig) \\ 0.161 (0.005) (arm)} \\
     & &  BG & & \makecell[l]{0.950 (0.001) (zig) \\ 0.943 (0.001) (arm)} & \makecell[l]{0.175 (0.002) (zig) \\ 0.179 (0.003) (arm)} \\
     & &  Bulk &  & \makecell{0.962 (0.002) (zig)\\0.956 (0.002) (arm)} & \makecell{0.186 (0.001) (zig)\\0.188 (0.001) (arm)} \\
\bottomrule
\end{tabular}
\label{stab:YM and PR}
\end{center}
\end{table}

\clearpage

\section{\normalsize 8. NequIP model for bilayer Gr}

To facilitate a fair comparison of computational efficiency with other \glspl{mlp}, we adopted the models reported in the most recent study \citep{ying2025advances}, except for the \gls{nequip} model \citep{batzner2022e3equivariant}, which has since been updated \citep{tan2025highperformance}. Accordingly, we retrained a \gls{nequip} model using the bilayer dataset from Ref.~\citep{ying2024effect}. The network employs four message-passing layers with a cutoff radius of \SI{7}{\angstrom}, 32 latent features, 8 trainable Bessel basis functions, and a maximum angular momentum of $l_{\mathrm{max}} = 1$. Training was carried out for  1000 epochs with an initial learning rate of 0.005 using \texttt{nequip} v0.15.0. The resulting \glspl{rmse} for energy and forces are \SI{2.01}{\milli\electronvolt\per\atom} and \SI{0.07}{\milli\electronvolt\per\angstrom}, respectively, attributed to the relatively simple, defect-free bilayer configurations in the training set.

\vspace{3cm}

\section{\normalsize 9. Construction of vdW heterostructures}

In this study, we constructed five \gls{vdw} heterostructures: \gls{gr}/\gls{hbn}; \gls{gr}/\gls{gr}/\gls{hbn} with an approximate \ang{4.2} twist angle between the \glspl{gr}; and three \gls{gr}/\gls{hbn}/\gls{mos2} structures with different stacking orders. The optimized lattice constants of monolayer \gls{gr}, \gls{hbn}, and \gls{mos2} are summarized in \autoref{stab:hetero detail}. The lattice mismatches between optimized \gls{gr} and \gls{hbn}, and between \gls{gr} and \gls{mos2}, are 1.83\% and 29.1\%, respectively. To minimize pre-strain, we maintained these mismatches during heterostructure construction, which led to models with a large number of atoms. The \gls{gr}/\gls{hbn} and trilayer \gls{gr}/\gls{gr}/\gls{hbn} models contain 36,352 and 54,858 atoms, respectively. Each of the three \gls{gr}/\gls{hbn}/\gls{mos2} heterostructures contains 423,552 atoms, with pre-strain kept below 0.05\% within all layers.

\begin{table*}[htb]
\caption{Lattice constants (LC) and lattice mismatches of the \gls{gr}/\gls{hbn}/\gls{mos2} heterostructure models. The second and third columns list the optimized monolayer lattice constants and the lattice constants used in the trilayer heterostructures, respectively. The \gls{hbn} and \gls{mos2} layers are slightly stretched, as indicated in the fourth column, to preserve lattice mismatches close to their optimized values. The fifth column shows the resulting lattice mismatches of the constructed models relative to \gls{gr}.}
\begin{center}
\begin{tabular}{cccccc}
\toprule
    Layer & LC (opt) & LC (set) & Strain (\%) & Mismatch (\%) & Unit cell\\
\midrule
    \gls{gr} & 2.4678  & 2.4678 & 0 & - & 111$\times$111\\
\midrule
    \gls{hbn} & 2.5129 & 2.5131 & 0.008 & 1.83 &109$\times$109 \\
\midrule
    \gls{mos2} & 3.1838 & 3.1852 & 0.044 & 29.07 & 86$\times$86\\
    \bottomrule
\end{tabular}
\label{stab:hetero detail}
\end{center}
\end{table*}

\newpage

\section{\normalsize 10. Substrate setup for heterostructures}

\subsection{\normalsize 10.1. Gr/\textit{h}-BN heterostructures}

In experimental setups, \gls{2d} materials are typically supported on bulk substrates or embedded in multilayer stacks rather than being suspended as few-layer membranes. Measurements on Gr/\gls{hbn} generally report sub-\unit{\angstrom} out-of-plane corrugation amplitudes \citep{xue2011scanning, yankowitz2016pressureinduced}. In contrast, fully suspended few-layer \gls{vdw} stacks can sustain much larger corrugations because they are not mechanically clamped by a thick substrate, reducing their effective bending rigidity. Such suspended membranes and heterostructures are of considerable interest, for example, in nano-electromechanical resonators and strain-engineering studies~\cite{zhang2010free, kang2017layer}, and therefore serve as useful upper-bound cases for the corrugation amplitude.

\begin{figure*}[htb]
  \centering
  \includegraphics[width=1\columnwidth]{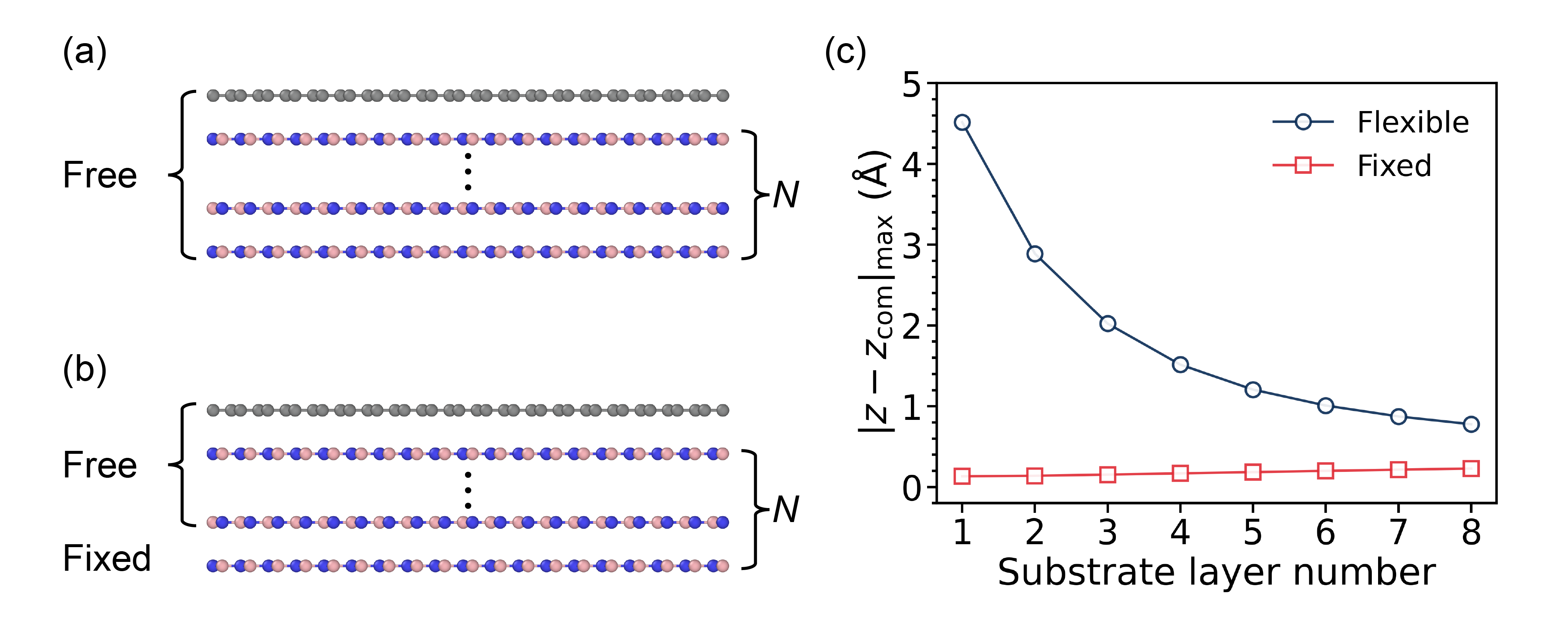}
  \caption{Effect of substrate constraints on the out-of-plane corrugation of \gls{gr}/\gls{hbn}. Panels (a) and (b) illustrate two setup strategies with N substrate layers. (a) A flexible substrate model, where all \gls{hbn} layers are allowed to fully relax. (b) A fixed bottom substrate model, in which only the lowest \gls{hbn} layer is fully fixed while the others are free to relax. (c) Maximum out-of-plane corrugation amplitude, $|z-z_{\mathrm{com}}|_{\max}$, of the \gls{gr} layer as a function of the number of \gls{hbn} substrate layers for the two models, showing that increasing the substrate thickness and constraining the bottom layer both lead to strong suppression of the corrugation. Mauve, blue, and gray spheres represent boron, nitrogen, and carbon atoms, respectively.}
  \label{sfig:substrate}
\end{figure*}

To reconcile these distinct regimes, we explicitly distinguish between suspended and substrate-supported geometries and quantify the dependence of the corrugation amplitude $|z-z_{\mathrm{com}}|_{\max}$ on the substrate model. We first consider the prototypical \gls{gr}/\gls{hbn} bilayer supported by AA$^\prime$-stacked \gls{hbn} acting as the supporting material. Two limiting substrate models are examined: a flexible substrate, in which all \gls{hbn} layers are allowed to relax, and a fixed-bottom substrate, in which only the bottom \gls{hbn} layer is frozen while the upper layers remain fully flexible. In both cases, the substrate is constructed by stacking additional \gls{hbn} layers beneath the \gls{gr}/\gls{hbn} interface.

\begin{figure*}[htb]
  \centering
  \includegraphics[trim={0 100 0 0}, clip, width=0.6\columnwidth]{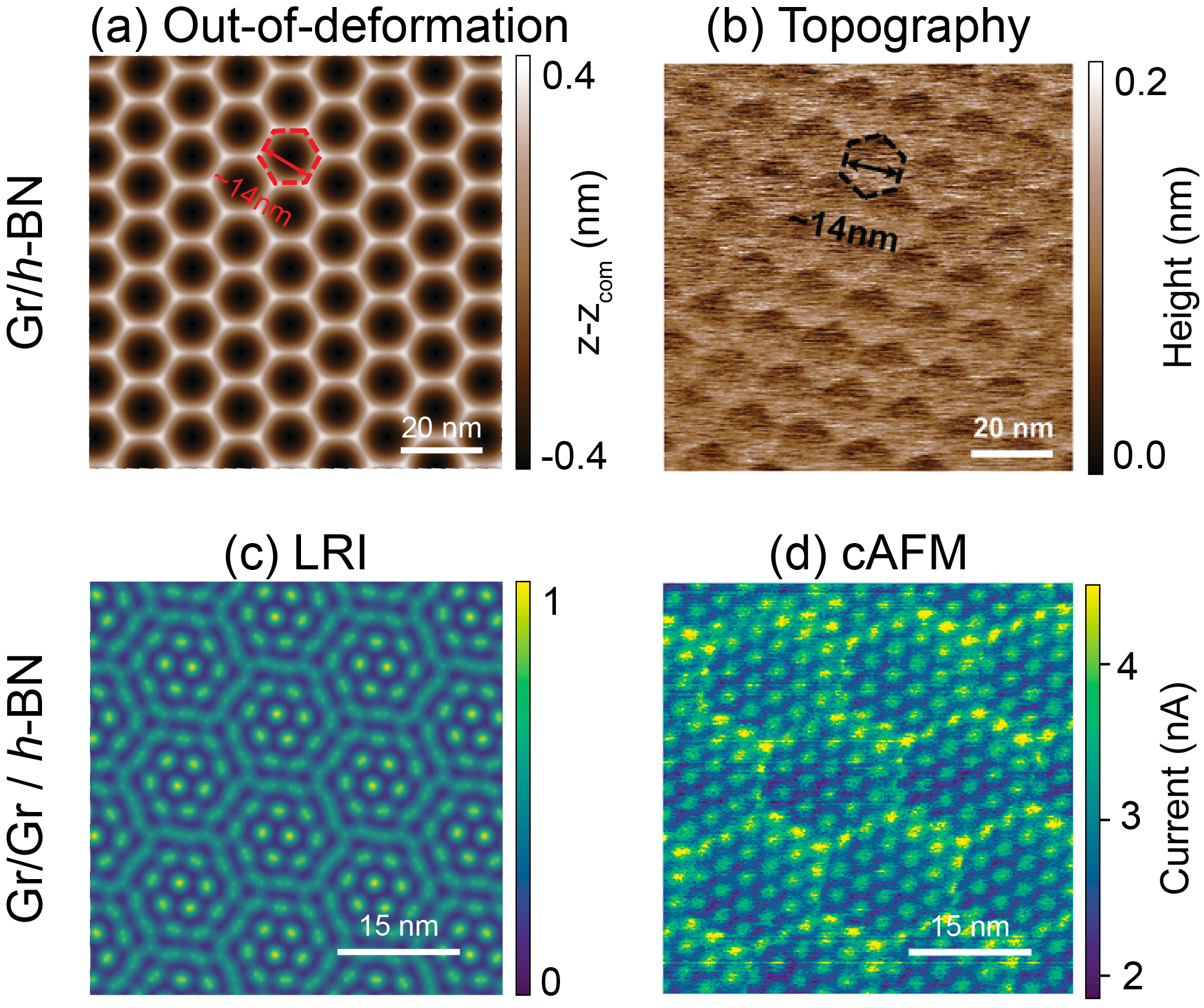}
  \caption{Comparison between simulation and experiment for \gls{gr}/\gls{hbn}-based heterostructures. (a) Out-of-plane deformation of a relaxed \gls{gr}/\gls{hbn} heterostructure predicted by \gls{smlp}+\gls{ilp}. (b) Experimentally measured topography of \gls{gr}/\gls{hbn}. Both simulation and experiment display a Moir\'e periodicity of $\sim$\SI{14}{\nano\meter} and exhibit similar contrast patterns. Panel (b) is adapted from Refs.~\citep{huang2022origin}.}
  \label{sfig:moire-2L}
\end{figure*}

\autoref{sfig:substrate} shows $|z-z_{\mathrm{com}}|_{\max}$ of the \gls{gr} layer as a function of the number of \gls{hbn} substrate layers for these two models. Owing to the strong constraint imposed on the lowest layer, the fixed-bottom model yields artificially small corrugations when the substrate is thin. By contrast, in the fully flexible model $|z-z_{\mathrm{com}}|_{\max}$ decreases rapidly with increasing substrate thickness and converges once the substrate exceeds approximately $7$--$8$ layers. At this limit, the corrugation amplitude approaches the sub-\AA{} range, in good agreement with experimental measurements on bulk-supported \gls{gr}/\gls{hbn}~\citep{xue2011scanning, yankowitz2016pressureinduced}. Importantly, adding homogeneous \gls{hbn} layers underneath the interface does not modify the Moir\'e periodicity or pattern. The substrate primarily reduces the amplitude of the height modulation. Consequently, all out-of-plane deformation results for \gls{gr}/\gls{hbn} reported in the main text are therefore obtained using an eight-layer flexible \gls{hbn} substrate. Given that suspended \gls{vdw} materials are also of considerable interest, we further evaluated the out-of-plane corrugation of a suspended bilayer \gls{gr}/\gls{hbn} heterostructure, as shown in \autoref{sfig:moire-2L}. The periodicity of the corrugation and the Moir\'e pattern are consistent with those obtained for the eight-layer-supported system in the main text, while the primary difference lies in the significantly larger corrugation amplitude in the suspended case.

\begin{figure*}[htb]
  \centering
  \includegraphics[width=0.9\textwidth]{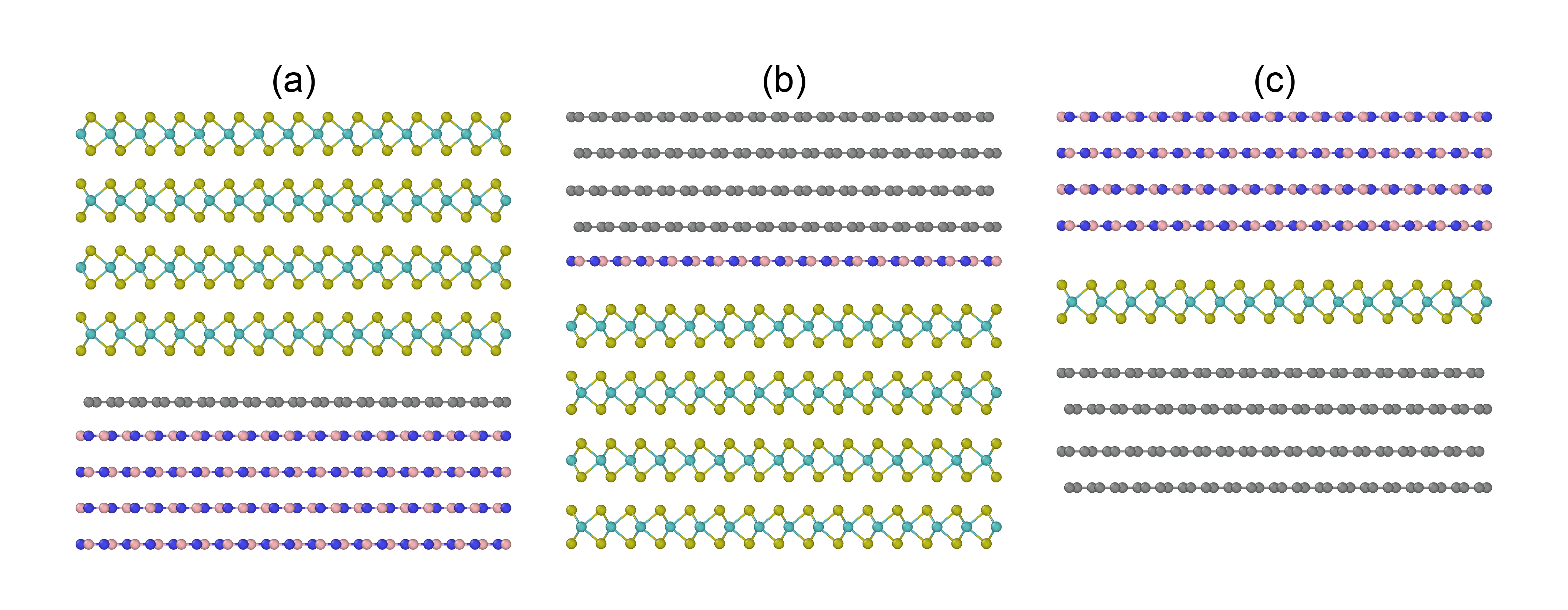}
  \caption{Substrate-supported \gls{gr}/\gls{hbn}/\gls{mos2} trilayer heterostructures used for corrugation analysis. (a) \gls{hbn}/\gls{gr}/\gls{mos2} stacking, where additional \gls{mos2} layers (top) and \gls{hbn} layers (bottom) are added as substrates. (b) \gls{mos2}/\gls{hbn}/\gls{gr} stacking with \gls{mos2} (bottom) and \gls{gr} (top) substrates. (c) \gls{gr}/\gls{mos2}/\gls{hbn} stacking with \gls{gr} (bottom) and \gls{hbn} (top) substrates. In all cases, the three central monolayers constitute the active trilayer, while the outer layers provide mechanical support mimicking experimental substrates. Mauve, blue, gray, light blue, and green spheres represent boron, nitrogen, carbon, molybdenum, and sulfur atoms, respectively.}
  \label{sfig:substrate-bgm}
\end{figure*}

\subsection{\normalsize 10.2. Gr/\textit{h}-BN/\texorpdfstring{\ce{MoS2}}{MoS2} heterostructures}

We next extend this analysis to the \gls{gr}/\gls{hbn}/\gls{mos2} trilayer heterostructures discussed in the main text. Three stacking sequences are considered: \gls{hbn}/\gls{gr}/\gls{mos2}, \gls{mos2}/\gls{hbn}/\gls{gr}, and \gls{gr}/\gls{mos2}/\gls{hbn}. To mimic the experimental geometry, the outer materials are thickened by four additional layers on each side, which act as mechanical supports \citep{niu2025ferroelectricity}. The \gls{gr}, \gls{hbn}, and \gls{mos2} substrates are constructed using AB, AA$^\prime$, and AA$^\prime$ stacking, respectively. The substrate-supported geometries are illustrated in \autoref{sfig:substrate-bgm}. In all cases, the added outer layers provide mechanical clamping while preserving the intrinsic stacking relationship within the central trilayer.

\begin{figure*}[hbt]
  \centering
  \includegraphics[width=1\textwidth]{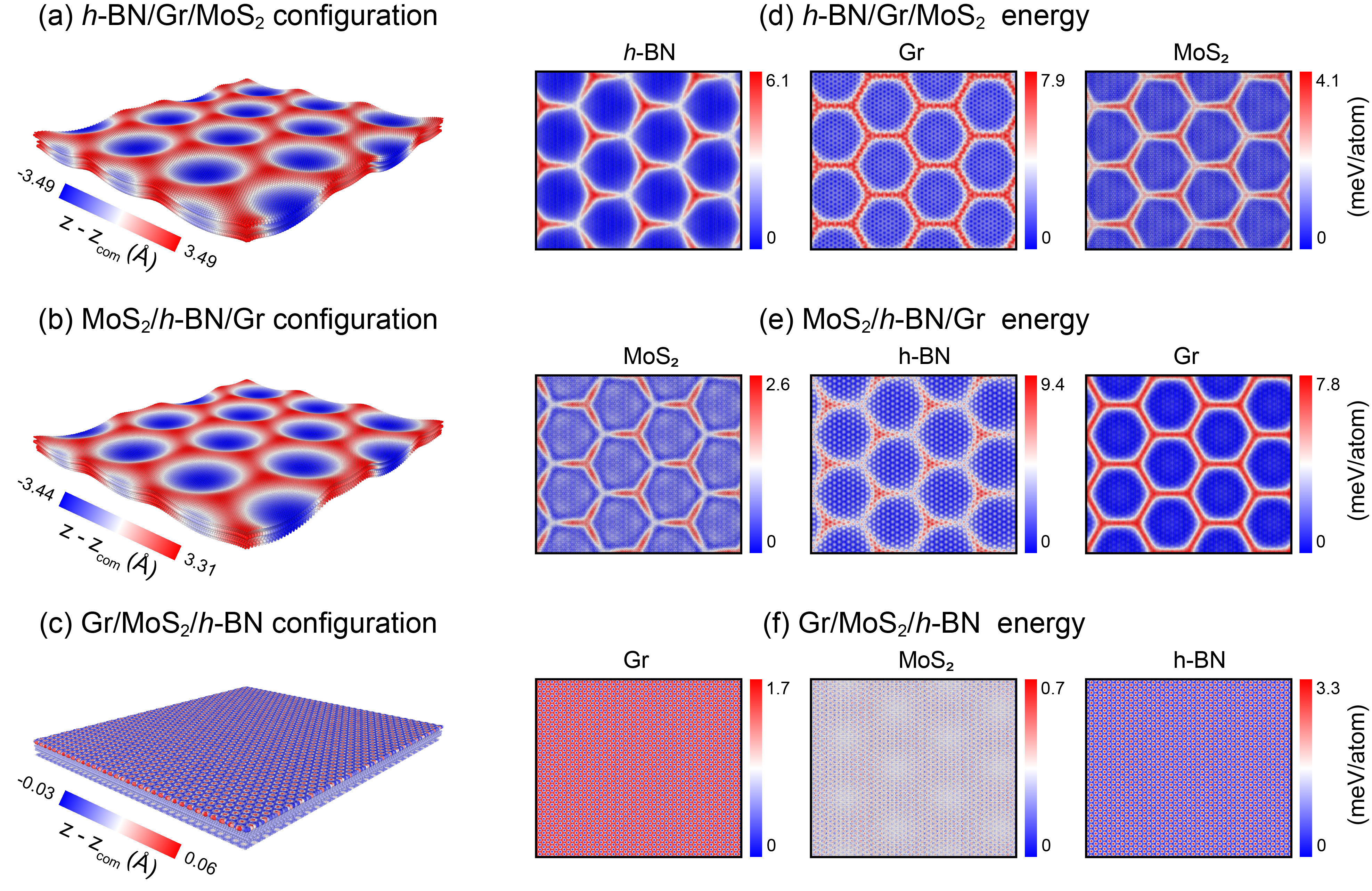}
  \caption{Fully suspended \gls{gr}/\gls{hbn}/\gls{mos2} trilayer heterostructures. (a)–(c) Out-of-plane deformation maps of the relaxed \gls{hbn}/\gls{gr}/\gls{mos2}, \gls{mos2}/\gls{hbn}/\gls{gr}, and \gls{gr}/\gls{mos2}/\gls{hbn} configurations, respectively, expressed in terms of $z-z_{\mathrm{com}}$. (d)–(f) Corresponding layer-resolved weighted atomic-energy maps for each component monolayer, relative to the minimum energy of the respective layer. When \gls{gr} and \gls{hbn} are adjacent, a long-period Moir\'e pattern with large corrugation amplitudes emerges, whereas inserting \gls{mos2} between \gls{gr} and \gls{hbn} effectively suppresses both the Moir\'e contrast and the out-of-plane deformation.}
  \label{sfig:3L-suspended}
\end{figure*}

For completeness, we also analyze fully suspended \gls{gr}/\gls{hbn}/\gls{mos2} trilayers, where only in-plane periodic boundary conditions are imposed and all atoms are allowed to relax. The corresponding out-of-plane deformation and layer-resolved weighted atomic-energy distributions are shown in \autoref{sfig:3L-suspended}. In configurations where \gls{gr} and \gls{hbn} are adjacent (\gls{hbn}/\gls{gr}/\gls{mos2} and \gls{mos2}/\gls{hbn}/\gls{gr}), a long-period (\(\sim\)\SI{14}{\nano\meter}) Moir\'e pattern develops and the corrugation amplitude $|z-z_{\mathrm{com}}|_{\max}$ of the \gls{gr} layer can reach \SI{3}{\angstrom}–\SI{5}{\angstrom}. These values constitute an upper bound on the corrugation achievable in mechanically unconstrained samples. Conversely, when \gls{mos2} is inserted between \gls{gr} and \gls{hbn} (\gls{gr}/\gls{mos2}/\gls{hbn}), their direct interaction is strongly reduced; consequently, no pronounced Moir\'e pattern emerges, and the out-of-plane corrugation becomes negligible even in the suspended geometry. While the same stacking-order dependence persists in the substrate-supported trilayers used in the main text, the corrugation amplitude is reduced by nearly an order of magnitude, consistent with available experimental observations \citep{niu2025ferroelectricity}.

\clearpage
\newpage

\section{\normalsize 11. Local registry index computation method of trilayer Gr/Gr/\textit{h}-BN heterostructure}

In the main text, we calculated the \gls{lri} for the trilayer \gls{gr}/\gls{gr}/\gls{hbn} heterostructure to distinguish different stacking configurations. Here, we briefly summarize the technical details of the \gls{lri} method; a comprehensive description is available in Ref.~\cite{leven2016multiwalled, oded2013theregistry}.

The \gls{lri} is a tool developed to quantify the local degree of lattice commensurability between adjacent layers \cite{leven2016multiwalled}. It builds upon the global registry index (GRI) approach, which measures the overall interlayer registry for different stacking configurations of rigid interfaces \cite{leven2016interlayer}. The \gls{lri} is defined as the average GRI value over the local environment of a given atom on one layer, considering its interactions with all atoms in the adjacent layer. Each atom is assigned a value between 0 (indicating optimal registry) and 1 (indicating the poorest registry), reflecting its local stacking quality.

Similar to the GRI, the \gls{lri} is evaluated via the calculation of projected overlaps between circles assigned to atoms on adjacent layers. For flat surfaces the projected circle overlap area of atoms $i$ and $n$ located on adjacent layers, $S_{in}(\rho_{in})$ , is a simple function of their lateral distance, $\rho_{in}$. 

For the trilayer \gls{gr}/\gls{gr}/\gls{hbn} heterostructure, we calculated the \gls{lri} of atoms on the middle \gls{gr} layer, $\mathrm C_1$, including the projected overlaps with both atoms on the \gls{hbn} layer and the top \gls{gr} layer, $\mathrm C_2$. With this, the \gls{ilp} of a given atom $i$ on the middle layer is defined as a function of the atomic projected overlaps, $S^{\mathrm{C_1}M}$:

\begin{equation}
LRI^{\mathrm{C_1}}_i = \frac{1}{3} \sum_{m=j,k,l} 
\frac{\sum_{M=\mathrm{C_2,B,N}}\left[\left(S^{\mathrm{C_1}M}_i + S^{\mathrm{C_1}M}_m\right) - \left(S^{\mathrm{C_1}M,\mathrm{O}}_i + S^{\mathrm{C_1}M,\mathrm{O}}_m\right)\right]}{\sum_{M=\mathrm{C_2,B,N}}\left[\left(S^{\mathrm{C_1}M,\mathrm{W}}_i + S^{\mathrm{C_1}M,\mathrm{W}}_m\right) - \left(S^{\mathrm{C_1}M,\mathrm{O}}_i + S^{\mathrm{C_1}M,\mathrm{O}}_m\right)\right]}
\end{equation}
where the sum is taken over the three nearest neighbors ($j$, $k$, $l$) of atom $i$. $S^{\mathrm{C_1}M,\mathrm{O}}$, and $S^{\mathrm{C_1}M,\mathrm{W}}$ are the sums of pair overlaps of the circle associated with a carbon atom ($i$, $j$, $k$, $l$) on the middle \gls{gr} layer and all circles of type $M$ ($\mathrm C_2$, B, or N) on the adjacent layers obtained at the optimal and worst local stacking mode, respectively.

In the present calculations, we adopt the circle radii of the GRI method \cite{leven2013robust} where
 $r_{\mathrm C_1}=r_{\mathrm C_2}=0.5~a_{\mathrm CC}$,  $r_{\mathrm B}=0.15~a_{\mathrm{BN}}$, and $r_{\mathrm N}=0.5~a_{\mathrm{BN}}$, and $a_{\mathrm{CC}}=1.42$ \unit{\angstrom} and $a_{\mathrm{BN}}=1.44$ \unit{\angstrom} are the CC and BN intralayer covalent bond lengths, respectively.

\newpage
\section{\normalsize 12. Energy analysis of trilayer Gr/\textit{h}-BN/\texorpdfstring{\ce{MoS2}}{MoS2} heterostructures}

\begin{figure*}[htb]
\begin{center}
\includegraphics[width=0.9\columnwidth]{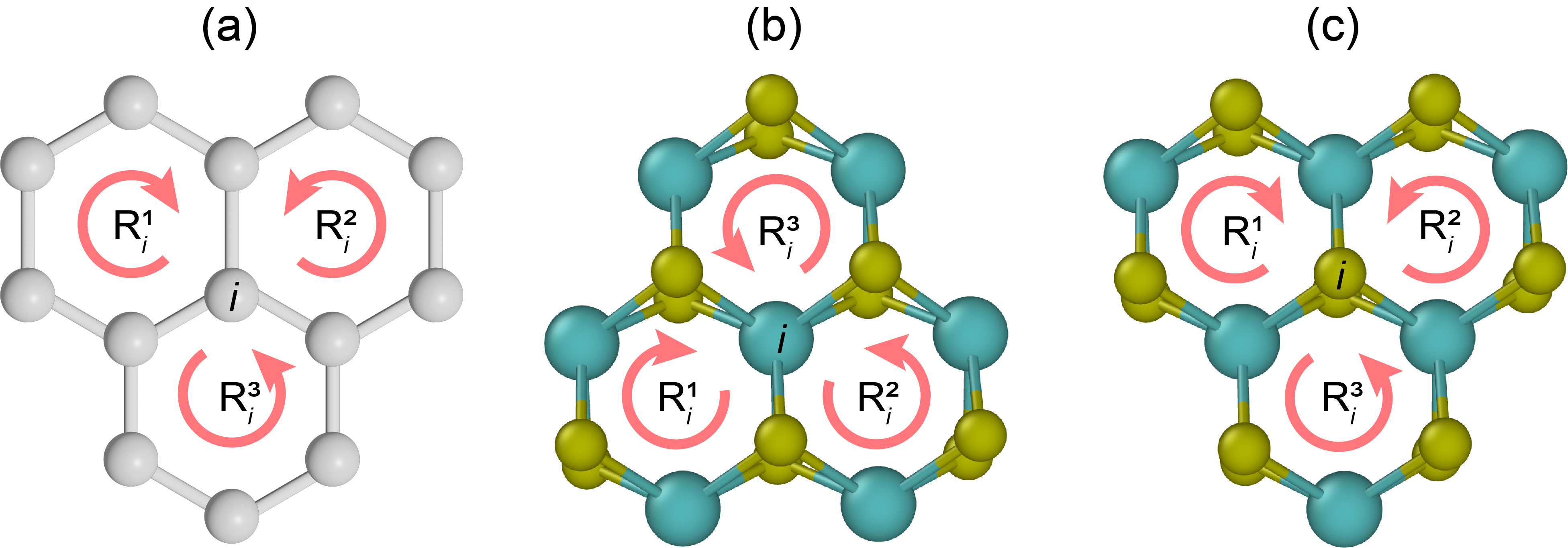}
\caption{Computation method of weighted mean energy for the central atom. The weighted mean energy of an atom $i$ is calculated by averaging atomic energies of the surrounding three adjacent hexagonal rings ($R^1_i$, $R^2_i$ and $R^3_i$). Panel (a) represents the computation method for \gls{gr}, which also works for \gls{hbn}. Panels (b) and (c) show the computation method for the Mo atoms and S atoms of \gls{mos2} layer, respectively.
}
\label{sfig:energy calc method}
\end{center}
\end{figure*}

To analyze the Moir\'e patterns and out-of-plane deformations across different stacking orders, we use a weighted averaging approach to enhance computational stability and represent the stacking energy more robustly. Specifically, the energy of each atom is first calculated. Then, for a given atom $i$, its surrounding three adjacent hexagonal rings are selected ($R^1_i$, $R^2_i$ and $R^3_i$). For \gls{gr}, \gls{hbn} and \gls{mos2}, each hexagonal ring consists of six, six and nine atoms (denoted as $R_i^{1,j}$,$R_i^{2,j}$  and $R_i^{3,j}$), respectively, and the energy of these atoms within each ring is averaged to obtain the energy of that ring. Since three hexagonal rings are considered, three energies ($E_i^1$, $E_i^2$ and $E_i^3$) are obtained for the given atom $i$, which are then averaged (dividing by 18 atoms for \gls{gr} and \gls{hbn}, and by 27 atoms for \gls{mos2}) to determine its final averaged energy  (see \autoref{sfig:energy calc method}), ensuring a consistent and meaningful representation of different stacking configurations. The energy of all atoms in the layer is presented relative to the corresponding minimum energy point for better visualization of energy corrugations.

\newpage

\begin{figure*}[htb]
\begin{center}
\includegraphics[width=0.8\columnwidth]{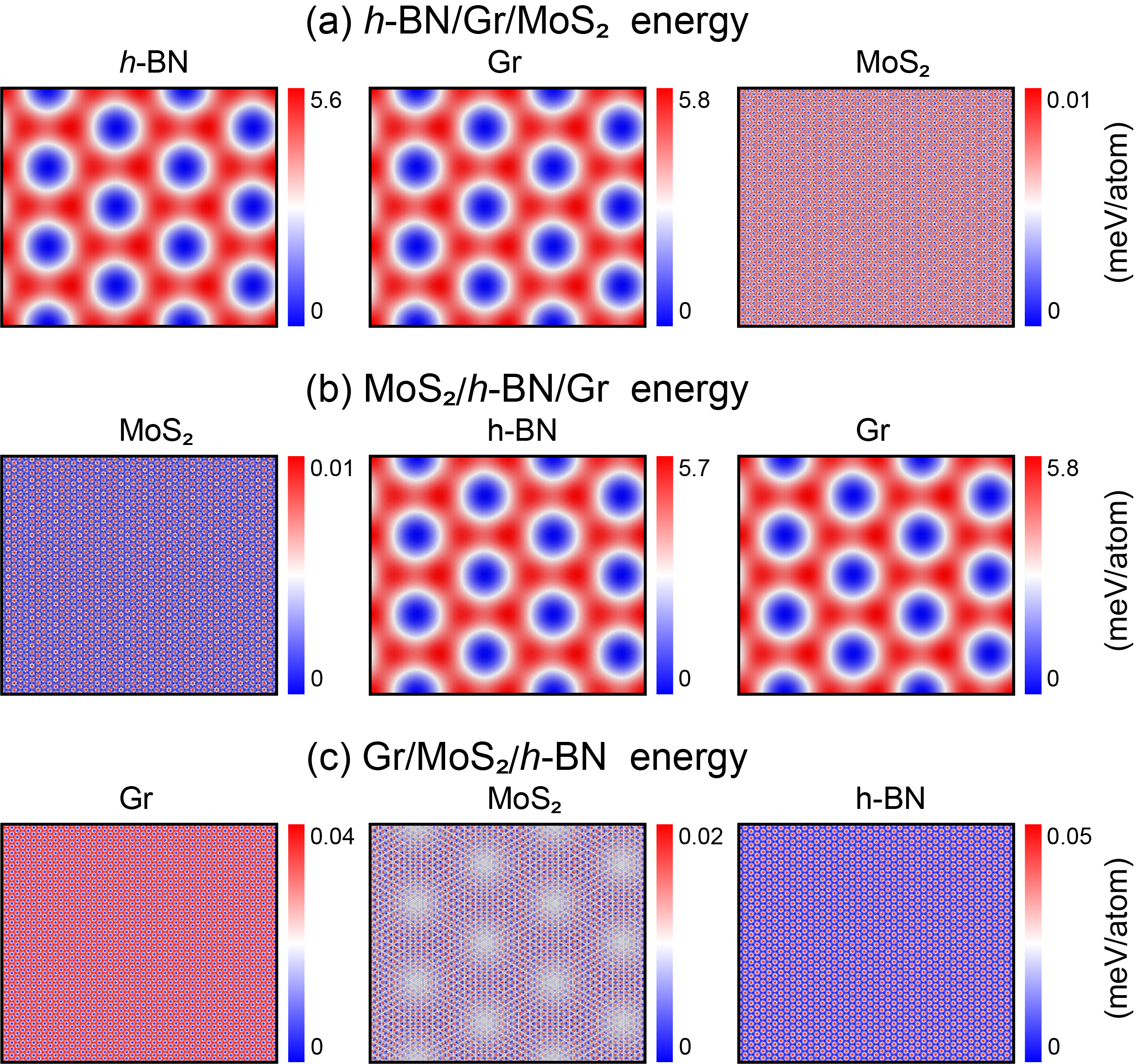}
\caption{Energy distribution of the non-optimized (a) \gls{hbn}/\gls{gr}/\gls{mos2}, (b) \gls{mos2}/\gls{hbn}/\gls{gr} and (c) \gls{gr}/\gls{mos2}/\gls{hbn} structures. Panels (a-c) present the weighted atomic energy maps for each component monolayer in the initial \gls{vdw} heterostructures, with all values presented relative to the corresponding minimum energy of the respective layer.
}
\label{sfig:G-M-B raw energy}
\end{center}
\end{figure*}

\autoref{sfig:G-M-B raw energy} (a), (b), and (c) show the weighted average energy distributions of the initial \gls{hbn}/\gls{gr}/\gls{mos2}, \gls{mos2}/\gls{hbn}/\gls{gr}, and \gls{gr}/\gls{mos2}/\gls{hbn} structures, respectively. In both the \gls{hbn}/\gls{gr}/\gls{mos2} and \gls{mos2}/\gls{hbn}/\gls{gr} configurations, where \gls{hbn} and \gls{gr} are adjacent, the \gls{mos2} monolayer exhibits minimal energy corrugation with only weak Moir\'e modulations, while the \gls{gr} and \gls{hbn} layers form a \SI{14}{\nano\meter} Moir\'e superlattice. In contrast, in the \gls{gr}/\gls{mos2}/\gls{hbn} system, where \gls{gr} and \gls{hbn} are separated by an inserted \gls{mos2} layer, all three monolayers show only weak Moir\'e modulations. In the former two configurations, low-energy regions expand to reduce the total system energy, primarily driven by the strong interaction between adjacent \gls{gr} and \gls{hbn} layers. However, in the \gls{gr}/\gls{mos2}/\gls{hbn} configuration, high-energy regions remain largely unrelaxed due to short-wavelength Moir\'e patterns across all three layers. Consequently, inserting a \gls{mos2} layer between \gls{gr} and \gls{hbn} effectively suppresses out-of-plane deformation.

\newpage

\section{\normalsize 13. Thermal conductivity of bulk Gr and \textit{h}-BN}

\subsection{\normalsize 13.1. HNEMD theory}

Within the \gls{hnemd} framework~\cite{fan2019homogeneous}, the system is driven out of equilibrium by an external force defined as~\cite{gabourie2021spectral}:
\begin{equation}
  \boldsymbol{F}_i^{\mathrm{ext}}=E_i\boldsymbol F_e + \boldsymbol F_e\cdot \boldsymbol W_i,
  \label{eq:hnemd2}
\end{equation}
where $E_i$ and $\boldsymbol W_i$ are the total energy and virial tensor of atom $i$, respectively, and $\boldsymbol F_e$ is a driving force parameter with units of inverse length. This applied force induces a non-equilibrium heat current $\boldsymbol J$, whose ensemble average is linearly proportional to the driving force via~\cite{fan2019homogeneous}:
\begin{equation}
\frac{\langle J^{\alpha} \rangle}{TV}=\sum_{\beta} \kappa^{\alpha\beta}  F_e^\beta.
\label{eq:hnemd1}
\end{equation}
Here, $T$ and $V$ denote the system temperature and volume, respectively, and $\kappa^{\alpha\beta}$ is the thermal conductivity tensor. In this study, we focus on the diagonal component in the out-of-plane $z$ direction, computing the thermal conductivity as:
\begin{equation}
\kappa^{zz}=\frac{\langle J^z \rangle}{T V F_e^z }.
\label{eq:hnemd3}
\end{equation}
The driving force parameter is set to $F_e^z=9\times10^{-6}/$\unit{\angstrom}. This magnitude has been validated to maintain the system within the linear response regime while generating a measurable heat current with a satisfactory signal-to-noise ratio for the out-of-plane thermal conductivity calculation of bulk graphite and \gls{hbn}.

A notable advantage of the \gls{hnemd} approach is its capability to provide a spectral decomposition of thermal conductivity~\cite{fan2017thermal}. Specifically, the conductivity along the $z$ direction can be expressed as:
\begin{equation}
\kappa^{zz}=\int_0^{\infty} \frac{\mathrm d\omega}{2\pi} \kappa^{zz} (\omega),
\label{eq:hnemd4}
\end{equation}
where
\begin{equation}
\kappa^{zz} (\omega)=\frac{2}{TVF_e^z} \int_{-\infty}^{\infty}\mathrm dt e^{i\omega t} K^z (t)	.
\label{eq:hnemd5}
\end{equation}
Here, $K^z (t)$ is the $z$-component of the virial-velocity correlation function~\cite{gabourie2021spectral}:
\begin{equation}
\boldsymbol K (t)=\sum_i \langle\boldsymbol W_i (0)\cdot \boldsymbol v_i (t)\rangle
\label{eq:hnemd6}
\end{equation}

\subsection{\normalsize 13.2. Cross-plane thermal conductivity}

To further assess the transferability of the combined \gls{smlp}+\gls{ilp} framework to generic \gls{vdw} crystals, we evaluated the cross-plane thermal conductivity of bulk \gls{gr} and \gls{hbn} by performing \gls{hnemd} simulations at \SI{300}{\kelvin}. For each material, we constructed three multilayer stacks containing 20, 40, and 60 atomic layers, with heat transport imposed along the crystallographic $c$ axis. All simulations were carried out with a time step of \SI{1}{\femto\second} in the canonical ensemble using a Nosé–Hoover thermostat at \SI{300}{\kelvin}. Each system was first equilibrated for \SI{2}{\nano\second} before switching on the homogeneous driving field. The subsequent production stage lasted \SI{20}{\nano\second}, during which the heat current was recorded every 1000 steps to compute the running averages of $\kappa$. For every thickness, we ran three independent simulations with different initial velocity seeds to quantify statistical uncertainties.

\begin{figure*}[htb]
\begin{center}
\includegraphics[width=1\columnwidth]{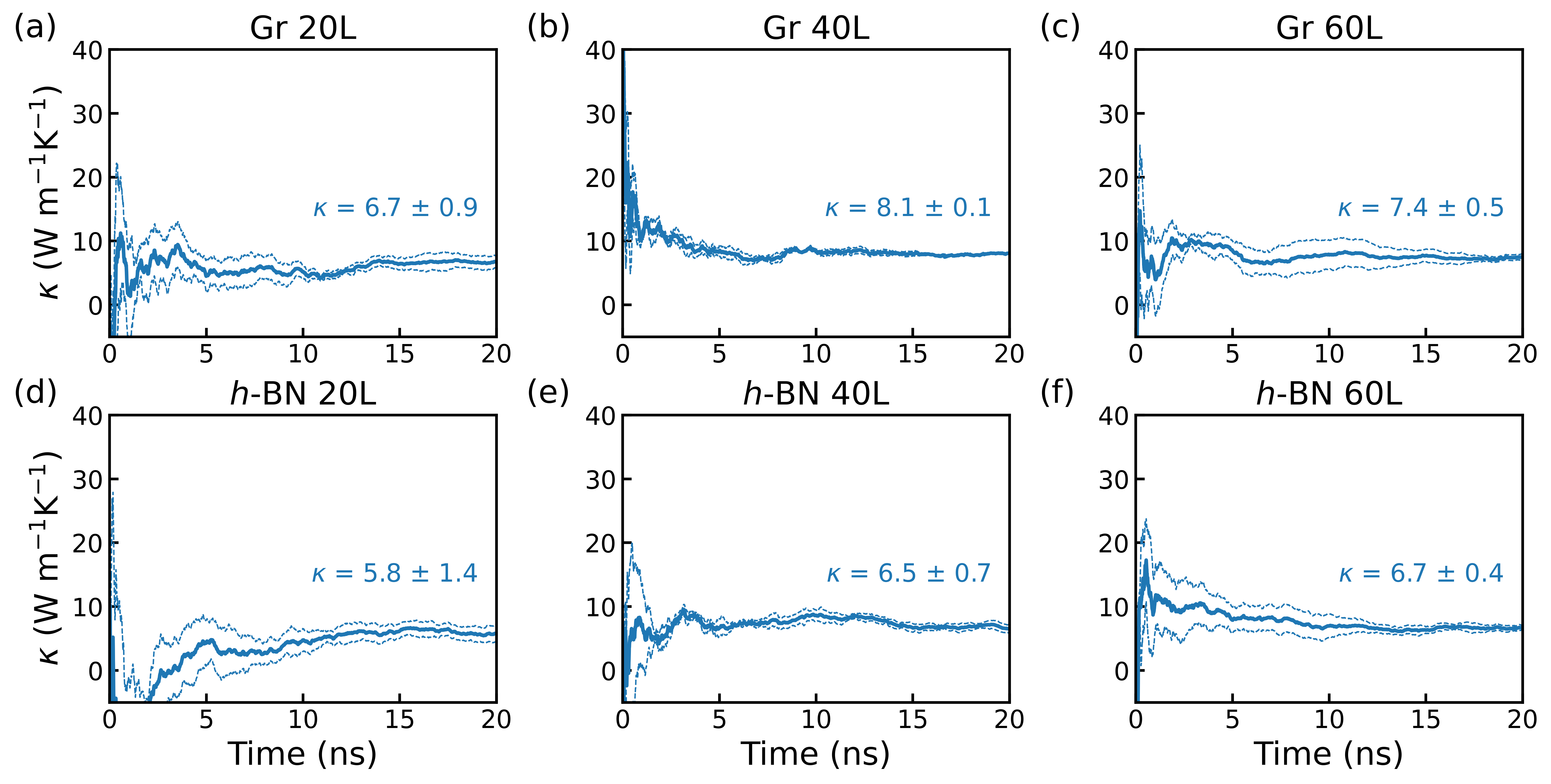}
\caption{Time evolution of the running average of the cross-plane thermal conductivity obtained from \gls{hnemd} simulations for multilayer \gls{gr} and \gls{hbn} at \SI{300}{\kelvin}. Panels (a)–(c) correspond to \gls{gr} with 20, 40, and 60 layers, respectively, while panels (d)–(f) show results for the corresponding \gls{hbn} systems.}

\label{sfig:TC}
\end{center}
\end{figure*}

The temporal evolution of the running averages of the cross-plane thermal conductivity $\kappa$ is presented in \autoref{sfig:TC}. For all multilayer \gls{gr} and \gls{hbn} systems, the cumulative averages converge within the first $\sim\SI{10}{\nano\second}$ and subsequently exhibit only small fluctuations around the steady-state values. The resulting thermal conductivities fall in the range of $6.7 \pm 0.9$ to $8.1 \pm 0.1$~\unit{\watt\per\meter\per\kelvin} for \gls{gr} and $5.8 \pm 1.4$ to $6.7 \pm 0.4$~\unit{\watt\per\meter\per\kelvin} for \gls{hbn}. The results indicate that the 20–60 layer models already provide a good approximation to the bulk cross-plane limit within the statistical uncertainty of our simulations.

These \gls{md} results align well with available experimental measurements~\cite{yang2025ultrahigh,han2018submm,jaffe2023thicknessdependent,yuan2019modulating,jiang2018anisotropic,salihoglu2020energy} of the $c$-axis thermal conductivity of highly ordered graphite and bulk \gls{hbn} at room temperature, with variation mainly arising from differences in sample quality, defect concentration, and measurement techniques. The close agreement between our \gls{hnemd} predictions and these experimental benchmarks validates the present force field parametrization and simulation protocol for describing thermal properties in bulk \gls{gr} and \gls{hbn}.

\newpage

\section{\normalsize 14. Friction behavior of \textit{h}-BN nanoribbons}

We consider a nanoribbon system in which a \gls{bnnr} slides on a multilayer \gls{hbn} substrate. The interlayer interactions are consistently described by the same \gls{ilp} used in the previous sections, whereas the intralayer interactions are modeled either by the \gls{smlp} or by the Tersoff potential. We have simulated both the shear dynamics of \gls{hbnnr} and hydrogen-free \gls{bnnr}. However, since the available Tersoff parametrization for \gls{hbn} does not include hydrogen, the \glspl{hbnnr} are simulated exclusively using the \gls{smlp}+\gls{ilp} force fields.

   \begin{figure*}[htb]
\begin{center}
  \includegraphics[width=0.9\textwidth]{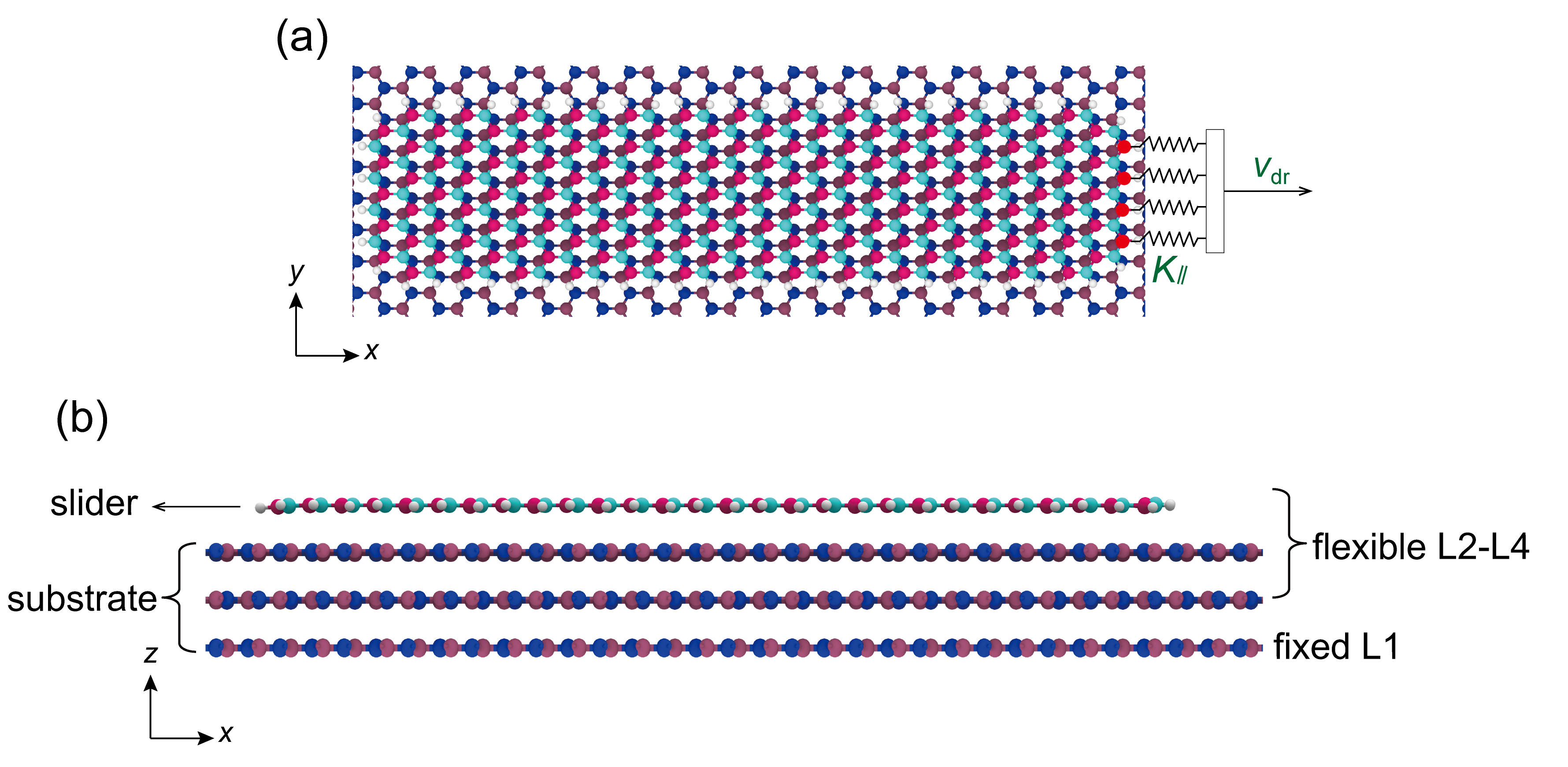}
  \caption{Simulation setup for the \gls{hbn} nanoribbon sliding calculations.
  (a) Top view of an \gls{hbn} nanoribbon (slider) resting on a multilayer \gls{hbn} substrate.  
  The slider is pulled along the $x$ direction via a harmonic spring of stiffness $k_{\parallel}$ attached to a stage moving at a constant drift velocity $V_{\mathrm{dr}}$.  
  (b) Side view in the $x$–$z$ plane. The top \gls{hbn} nanoribbon L4 serves as the slider, while the underlying multilayer \gls{hbn} stack (L1-L3) forms the substrate.  
  The bottom substrate layer L1 is held fixed, whereas the upper substrate layers L2-L3 and slider L4 are flexible and allowed to relax during sliding. Mauve, blue, pink, light blue, and white spheres represent boron (substrate), nitrogen (substrate), boron (slider), nitrogen (slider), and hydrogen atoms, respectively.}
  \label{fig:hbn_friction_set}
\end{center}
\end{figure*}

All sliding simulations were performed in LAMMPS package~\cite{thompson2022lammps}, following the same protocol as illustrated in \autoref{fig:hbn_friction_set}. The initial configurations of the \gls{bnnr} and \gls{hbnnr} sliders are first obtained by geometry optimization using the FIRE algorithm~\cite{bitzek2006structural},  with a maximum residual force threshold of $10^{-6}\,\mathrm{eV/\text{\AA}}$. Subsequently, damped sliding friction simulations were carried out under the NVE ensemble with a time step of \SI{1}{\femto\second}. The rightmost atoms of the \gls{bnnr} were connected to a driving stage via parallel lateral springs, where the stage moving at a constant velocity of $V_x$ = \SI{1}{\meter\per\second} along the $x$ direction (see \autoref{fig:hbn_friction_set}(a)). The overall spring stiffness was fixed at $K_x$=10 N/m \citep{szlufarska2008recent}, yielding a spring constant of $k_x=K_x/n$ for each dragged spring, where $n$ is the total number of dragged atoms. The contact geometry consists of a \gls{bnnr} sliding on a three-layer \gls{hbn} substrate (layers L1–L3), shown in \autoref{fig:hbn_friction_set}(b). The bottom substrate layer (L1) was kept frozen to mimic a rigid support, while the atoms in the middle layer (L2) were coupled to a Langevin thermostat with damping coefficient ($\eta=1 ~\mathrm{ps}^{-1}$) \citep{ouyang2018nanoserpents}. The lateral spring force on each dragged atom is given by:
\begin{equation}
F_i (t)=k_x [x_i (t)-X^{\mathrm{stage}} (t)]	
\end{equation}
where $X^{\mathrm{stage}} (t)$ and $x_i (t)$ are the instantaneous position of the driving stage and the $i^{\mathrm{th}}$ boron and nitrogen atoms, respectively. The total lateral force acting on the \gls{bnnr} is then calculated as $F_L =\sum_{i=1}^n F_i(t)$.

\begin{figure*}[hbt]
\begin{center}
\includegraphics[width=1\columnwidth]{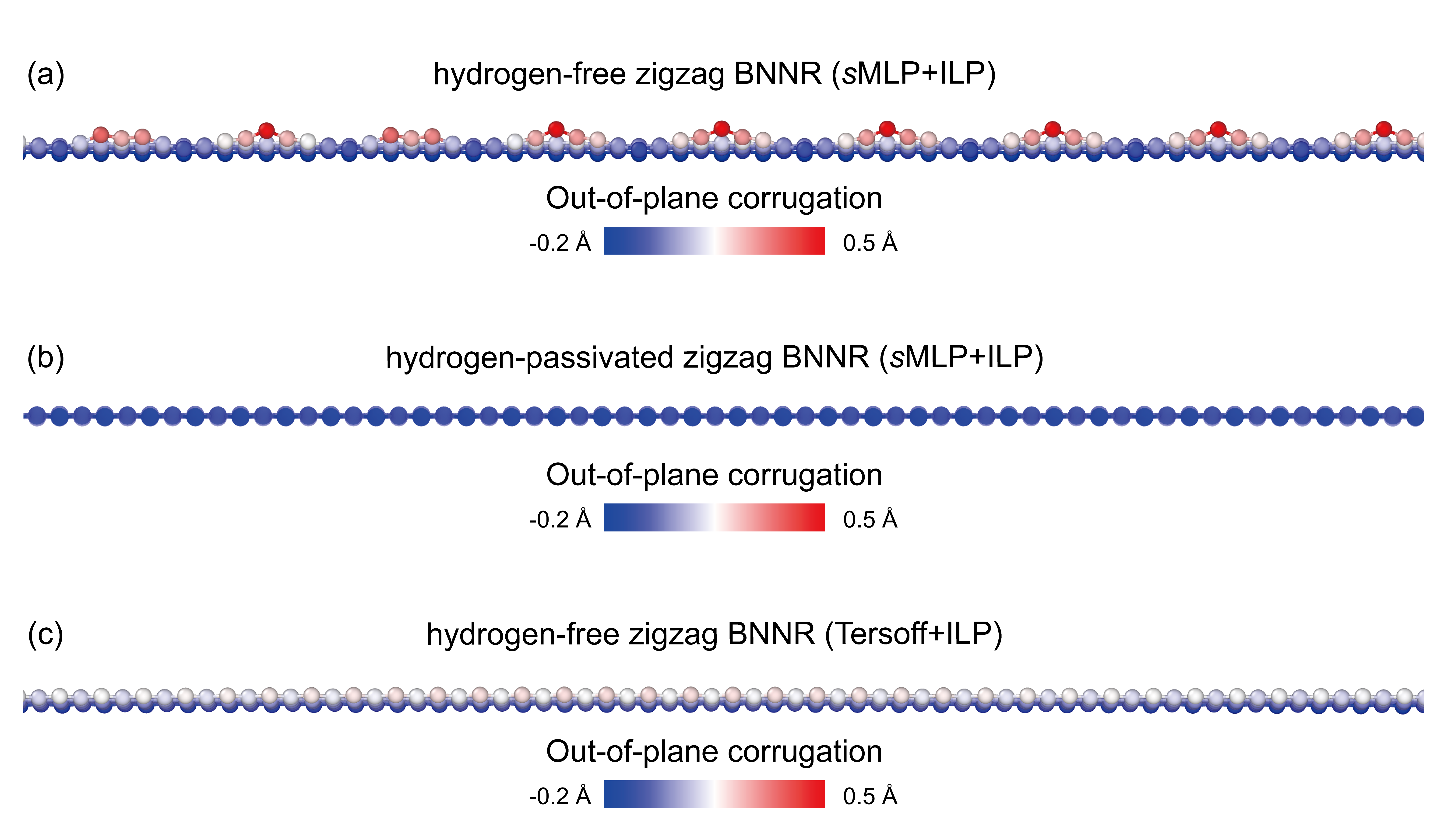}
\caption{Representative simulation snapshots of zigzag \gls{bnnr} with and without hydrogen passivation.  
  (a) Hydrogen–free zigzag \gls{bnnr} adopting \gls{smlp}+\gls{ilp} force field, where the out–of–plane corrugation of the edge atoms is color–coded according to the height scale shown below.  
  (b) Hydrogen–passivated zigzag \gls{bnnr} adopting \gls{smlp}+\gls{ilp} force field, which remains nearly flat with strongly suppressed out–of–plane corrugation.
  (c) Hydrogen–free zigzag \gls{bnnr} adopting Tersoff+\gls{ilp} force field.
}
\label{sfig:hyd}
\end{center}
\end{figure*}

As shown in \autoref{sfig:hyd}, in sharp contrast to the essentially flat armchair \glspl{bnnr}, the hydrogen-free zigzag \gls{bnnr} simulated using \gls{smlp}+\gls{ilp} force field exhibits pronounced out-of-plane buckling. The corrugation is mainly localized at the nitrogen-terminated edge, where the undercoordinated nitrogen atoms bend toward the substrate. Conversely, hydrogen passivation suppresses this edge buckling, resulting in an almost perfectly flat zigzag \gls{bnnr}.

\clearpage
\newpage

\phantomsection

\bibliographystyle{myaps.bst}
\bibliography{refs}